\pdfoutput=1
\documentclass[conference,compsoc]{IEEEtran}

\usepackage[nocompress]{cite}
\usepackage[dvipsnames,table]{xcolor}
\usepackage{amsmath}
\usepackage{amssymb}
\usepackage{amsthm}
\usepackage{mathtools}
\usepackage{bm}
\usepackage{array}
\usepackage{booktabs}
\usepackage{multirow}
\usepackage{graphicx}
\usepackage{algorithm}
\usepackage{algpseudocode}
\usepackage{mleftright}
\usepackage{refcount}
\usepackage[olditem,oldenum]{paralist}
\newenvironment{enumerate*}[1][]{\begin{inparaenum}[(1)]}{\end{inparaenum}}
\usepackage{tcolorbox}
\usepackage{xurl}
\usepackage[hypertexnames=false]{hyperref}
\hypersetup{
	breaklinks=true,
	colorlinks=true,
	citecolor=ForestGreen,
	linkcolor=MidnightBlue,
	urlcolor=Maroon,
	pdfauthor={Muhammad Arzaki, Muhammed F. Esgin, Markku-Juhani O. Saarinen, Amin Sakzad, Ron Steinfeld},
    pdftitle={LaTSEC: Lattice-Based Threshold Signatures with Native Non-Interactive Identifiable Abort and Compact Per-Party Storage},
}
\usepackage{cleveref}
\usepackage[
n,
advantage,
operators,
sets,
adversary,
landau,
probability,
notions,
logic,
ff,
mm,
primitives,
events,
complexity,
asymptotics,
keys]{cryptocode}

\newcommand{\fullversionurl}{\url{https://github.com/latsec-ts/latsec}}

\algnewcommand\algorithmicto{\textbf{to}}
\algnewcommand{\LineComment}[1]{\Statex \(\triangleright\) \textcolor{gray}{#1}}

\tcbset{
  gamebox/.style={
    colback=white,
    colframe=black,
    boxrule=0.5pt,
    sharp corners,
    width=\linewidth,
    top=1mm,
    bottom=1mm,
    left=1mm,
    right=1mm,
  }
}

\theoremstyle{plain}
\newtheorem{theorem}{Theorem}
\newtheorem{lemma}{Lemma}
\newtheorem{corollary}{Corollary}

\theoremstyle{definition}
\newtheorem{definition}{Definition}

\theoremstyle{remark}
\newtheorem{remark}{Remark}

\newcommand{\randgets}{\stackrel{\$}{\gets}}

\crefname{item}{Items}{Items}
\Crefname{item}{Item}{Items}

\begin{document}

\title{LaTSEC: Lattice-Based Threshold Signatures with Native Non-Interactive Identifiable Abort and Compact Per-Party Storage}

\author{%
\IEEEauthorblockN{Muhammad Arzaki\IEEEauthorrefmark{1},
Muhammed F. Esgin\IEEEauthorrefmark{1},
Markku-Juhani O. Saarinen\IEEEauthorrefmark{2},
Amin Sakzad\IEEEauthorrefmark{1}, and
Ron Steinfeld\IEEEauthorrefmark{1}}
\IEEEauthorblockA{\IEEEauthorrefmark{1}Faculty of Information Technology, Monash University, Melbourne, Australia\\
\{muhammad.arzaki, muhammed.esgin, amin.sakzad, ron.steinfeld\}@monash.edu}
\IEEEauthorblockA{\IEEEauthorrefmark{2}Faculty of Information Technology and Communication Sciences, Tampere University, Tampere, Finland\\
markku-juhani.saarinen@tuni.fi}}
\maketitle

\begin{abstract}
    \boldmath
    We present \(\mathsf{LaTSEC}\), a three-round lattice-based threshold signature scheme with native non-interactive identifiable abort.
    Each party holds one small-norm signing share and is associated with one public abort key, against which its partial signature is verified directly, without zero-knowledge proofs or an additional interactive protocol.
    Prior lattice-based threshold signatures that identify misbehaving parties rely on one of three mechanisms: zero-knowledge proofs, an additional interactive protocol, or an everywhere-short secret sharing scheme in which each party holds multiple signing shares, each with its own public abort key.
    Compared with schemes that use the third mechanism, \(\mathsf{LaTSEC}\) substantially reduces per-party storage.
    The construction uses \(\mathsf{SEC}\), a recent ramp everywhere-short secret sharing scheme based on lossless expanders that gives each party a \textit{single} short share.
    We prove correctness, identifiable-abort security, and threshold unforgeability under the standard \(\mathsf{MLWE}\) and \(\mathsf{MSIS}\) assumptions.
    The unforgeability proof introduces the \textit{Repeated-\(\bm{A}\) Matrix Hint-MLWE} problem, a new variant of \(\mathsf{MLWE}\) with hints that captures the complete tuple of correlated public abort keys.
    We rigorously prove that this new variant admits a reduction from the standard \(\mathsf{MLWE}\) problem.
    For \(60\) parties at \(128\)-bit security, \(\mathsf{LaTSEC}\) tolerates up to \(6\) corrupted parties, and, empirically, a randomly chosen set of at least \(46\) parties can sign with probability at least \(0.99\).
    Each party holds one signing share of about \(14.3\) KiB and is associated with one \(24\)-KiB public abort key.
\end{abstract}

\section{Introduction}\label{Section:Introduction}

A threshold signature \cite{desmedt1989threshold} distributes a signing key among \(n\) parties so that sufficiently large sets can sign jointly, while smaller coalitions cannot forge.
Threshold signatures have been built from RSA \cite{gennaro2008threshold}, Schnorr \cite{komlo2021frost,ruffing2022roast,crites2023fully}, BLS \cite{bacho2022adaptive}, and ECDSA \cite{canetti2020uc,castagnos2023bandwidth,doerner2024threshold}.
FROST \cite{komlo2021frost} is a noteworthy example that also provides \textit{native non-interactive identifiable abort}: each partial signature is checked directly against a participant-specific public key, so a failed session exposes the misbehaving parties without further interaction or proofs.
Several threshold ECDSA schemes also provide identifiable abort \cite{canetti2020uc,castagnos2023bandwidth}, and ROAST turns FROST into a robust, asynchronous threshold signature \cite{ruffing2022roast}.
However, these schemes rely on mathematical problems that are vulnerable to large-scale quantum computers, and there is a need to develop \textit{post-quantum} threshold signatures with practical efficiency.
The NIST call for multi-party threshold schemes \cite{brandao2026nist} has made this a concrete target.

\paragraph{\textbf{Lattice-based threshold signatures}}

Lattice-based threshold and distributed signatures use trapdoor sampling \cite{bendlin2013share}, hash-and-sign with noise flooding \cite{espitau2024flood}, generic multi-party computation (MPC) \cite{tang2023efficient}, threshold homomorphic encryption (THE) \cite{boneh2018threshold,agrawal2022round,gur2024two}, or Fiat-Shamir signatures, either with aborts in distributed and multi-signature settings \cite{damgaard2021two,boschini2022musig,chen2023efficient,alkadri2024practical} or without aborts in the Raccoon style \cite{del2024raccoon,del2024threshold}.
Trapdoor sampling and noise flooding impose strong parameter constraints that make the resulting schemes costly for large thresholds.
For example, the per-party communication of Pelican grows linearly with the threshold \cite{espitau2024flood}.
Generic MPC and THE incur high communication costs, reaching \(3\) MB per party in the THE-based GKS scheme of Gur et al.\ \cite{gur2024two}.
In distributed Fiat-Shamir signatures with aborts, a rejection by any signer restarts the signing protocol.
Raccoon removes this rejection step \cite{del2024raccoon}, and Threshold Raccoon builds on it to obtain practical threshold signatures from standard lattice assumptions \cite{del2024threshold}.
Our scheme \(\mathsf{LaTSEC}\) follows this Raccoon-style approach.
A separate line targets compatibility with \(\mathsf{ML\text{-}DSA}\), either with more signing rounds \cite{bienstock2026quorus} or for small committees \cite{celi2026efficient}.
For practical lattice-based threshold signatures with low communication and security under standard assumptions, a central remaining question is how to handle malicious signers.

\paragraph{\textbf{Handling malicious signers}}

In Raccoon-style signatures, the response \(\bm{z}=c\bm{s}+\bm{r}\) must stay short, that is, have small norm \cite{del2024raccoon}---we use \textit{short} in this sense throughout.
A malicious signer can cause signing to fail, and how a scheme identifies it depends on its secret sharing: existing schemes use either uniform shares, mostly from Shamir sharing, or everywhere-short sharing, which we discuss next and build on.

With Shamir sharing \cite{shamir1979share}, a partial response \(c\lambda_i\bm{s}_i+\bm{r}_i\) involves a uniform share \(\bm{s}_i\) and a possibly large Lagrange coefficient \(\lambda_i\), so it is neither short nor safe to reveal \cite{del2024threshold,borin2026hermine}.
Threshold Raccoon \cite{del2024threshold} hides it with one-time additive masks, an approach adopted by later schemes with adaptive security \cite{katsumata2024adaptively}, partially non-interactive two-round signing \cite{espitau2025two,zhu2025algebraic}, or two-round signing under standard \(\mathsf{MLWE}\) \cite{boschini2025ringtail}.
Chairattana-Apirom et al.\ instead use Benaloh-Leichter sharing \cite{benaloh1990generalized} with coefficients in \(\{0,\pm1\}\), but still with uniform shares \cite{chairattana2024partially}.
In all of these schemes, partial responses cannot be checked directly against public data, so identifying misbehaving signers requires an additional mechanism.

TRaccoon-IA runs an interactive protocol after a failed aggregation \cite{del2025unmasking}, whereas GKS and Olingo attach non-interactive zero-knowledge proofs to their signing messages, with per-party communication of \(3\) MB and \(852\) kB, respectively \cite{gur2024two,gur2025olingo}.
Identifiable abort also yields robustness, since the honest signers can exclude the identified parties and retry \cite{ruffing2022roast}.
\textit{Robust} schemes instead guarantee that the protocol directly outputs a signature whenever sufficiently many honest signers participate: Bendlin et al.\ give robust protocols for trapdoor sampling \cite{bendlin2013share}, while Pelican, built on Plover, relies on an approximate shortness proof in the honest-majority setting \cite{espitau2024flood,esgin2024plover}.

\paragraph{\textbf{Everywhere-short sharing and its cost}}

Native non-interactive identifiable abort over lattices becomes possible with \textit{everywhere-short} secret sharing, in which the secret, the shares, and the reconstruction coefficients are all short: each partial response is then short and can be verified against participant-specific public material \cite{del2024shortly,borin2026hermine}.
The sDKG framework of del Pino et al.\ instantiates this idea using replicated and random-assignment ramp sharing \cite{del2024shortly}, whereas Hermine uses Vandermonde sharing and adds partially non-interactive signing and proactive security \cite{desmedt1994multiplicative,battagliola2025enhancing,borin2026hermine}.
For \(n\) participants, signing threshold \(t\), and security parameter \(\kappa\), these constructions pay for shortness through the number of short sub-shares held by each participant: \(\binom{n-1}{t-1}\) under replicated sharing, one from each of \(\mu=\kappa\) parallel sharings under the ramp construction, and \(O\bigl(\tbinom{t-1+\lceil\lg n\rceil}{t-1}\bigr)\) under Vandermonde sharing \cite{battagliola2025enhancing}.
Because native non-interactive identifiable abort associates public verification material with every sub-share, this multiplicity inflates both the long-term secret-share storage of each participant and the participant-specific public material required to verify its partial responses.

\paragraph{\textbf{Why storage matters}}

Post-quantum cryptography must also run on constrained platforms with limited memory, energy, and bandwidth, where use-case-specific memory budgets matter \cite{puch2026sok}.
In the IETF classification of constrained devices \cite{bormann2014rfc}, Class~1 and Class~2 devices have about \(10\) and \(50\) KiB of RAM and \(100\) and \(250\) KiB of flash, respectively; Java Card smart cards typically have tens of KiB of persistent memory and a few KiB of RAM \cite{van2018java}; and the \textsc{pqm4} benchmarking board has \(196\) KiB of RAM and \(1\) MiB of flash \cite{kannwischer2019pqm4}.
Threshold signatures have already been implemented on such platforms, including FROST on Java Card smart cards \cite{dufka2023enabling} and the lattice-based \(n\)-out-of-\(n\) scheme DS2 \cite{damgaard2021two} on a microcontroller with \(192\) KiB of RAM \cite{ricci2024lattice}. Post-quantum protocols on banking smart cards must likewise fit larger objects into limited memory \cite{bettale2022post}.

In a threshold signature with native non-interactive identifiable abort, each participant keeps its secret-share material for the lifetime of the key, while every verifier of partial responses keeps the authenticated participant-specific public material.
Thus, a per-participant share of hundreds of KiB or more, as in the everywhere-short baselines of \Cref{Table:Intro-TS-Comparison}, already exceeds the flash capacity of a Class~2 device and the persistent memory of a typical Java Card smart card.
To the best of our knowledge, prior lattice-based threshold-signature evaluations have not treated these sizes as primary metrics; we report both per participant.

\subsection{Our Contributions}\label{Section:Contributions}

\begin{table*}[!t]
	\caption{Lattice-based threshold signatures with mechanisms for handling malicious signers at \(\kappa=128\)-bit security, in chronological order of first appearance}
	\label{Table:Intro-TS-Comparison}
	\centering
	\footnotesize
	\setlength{\tabcolsep}{3.3pt}
	\begin{tabular}{@{}l l l c r r r r r l@{}}
		\toprule
		Scheme & Access & Sharing layer & Short partial & \multicolumn{2}{c}{Storage at \(n=60\)\({}^{a}\)} & \multicolumn{3}{c}{Reported\({}^{b}\)} & Malicious signers \\
		\cmidrule(lr){5-6}\cmidrule(lr){7-9}
		& & & responses & \(|\mathsf{sk}_i|\) (KiB) & \(|\mathsf{ak}_i|\) (KiB) & \(n\) & \(|\mathsf{sig}|\) (kB) & \(\mathsf{Comm}_i\) (kB) & \\
		\midrule
		GKS \cite{gur2024two} & threshold & Shamir & \(\times\) & n/r & n/r & \(5\) & \(46.6^{c}\) & \(3000^{c}\) & proof IA\({}^{c}\) \\
		Pelican \cite{espitau2024flood} & threshold & Shamir & \(\times\) & n/r & n/r & \(48\) & \(12.3\) & \(26.8{+}56t\) & robust \\
		sDKG, replicated \cite{del2024shortly} & threshold & replicated & \(\checkmark\) & \(\geq2.71{\times}10^{8}\) & \(\geq5.06{\times}10^{8}\) & \(16\) & \(10.3\) & \(20.0\) & NI-IA \\
		sDKG, ramp \cite{del2024shortly} & ramp\({}^{d}\) & rand.-assign.\ ramp & \(\checkmark\) & \(\geq963\) & \(\geq1920\) & \(1024\) & \(11.9\) & \(25.0\) & NI-IA \\
		TRaccoon-IA \cite{del2025unmasking} & threshold & Shamir & \(\times\) & n/r & n/r & \(1024\) & \(12.7\) & \(88.2{+}6.4t\) & int.\ IA \\
		Olingo \cite{gur2025olingo} & threshold & Shamir & \(\times\) & n/r & n/a\({}^{e}\) & \(1024\) & \(11.3\) & \(852\) & proof IA \\
		Hermine \cite{borin2026hermine} & threshold & Vandermonde & \(\checkmark\) & \(\geq699\) & \(\geq2040\) & \(16^{f}\) & \(11.3\) & \(73.2\) & NI-IA \\
		\rowcolor{lightgray}
		\(\mathsf{LaTSEC}\) (this work) & ramp & \(\mathsf{SEC}\) & \(\checkmark\) & \(14.3\) & \(24.0\) & \(60\) & \(43.6\) & \(61.6\) & NI-IA \\
		\bottomrule
	\end{tabular}
	\par\smallskip
	\parbox{\textwidth}{\scriptsize
	KiB and kB: \(1024\) and \(1000\) bytes.
	Threshold and ramp: one threshold (\(t\)), or distinct thresholds, for unforgeability and signing.
	Short partial responses: partial responses are short by construction, since the sharing layer is everywhere-short.
	Robust: the protocol directly outputs a valid signature whenever sufficiently many honest signers participate.
	NI-IA, int.\ IA, and proof IA: partial responses verified directly against participant-specific public material, an additional protocol after a failed aggregation, and non-interactive zero-knowledge proofs, respectively.
	\(|\mathsf{sk}_i|\) and \(|\mathsf{ak}_i|\): total long-term secret-share storage held by participant \(i\) and total public verification material associated with that participant.
    \(|\mathsf{sig}|\) and \(\mathsf{Comm}_i\): aggregate-signature size and per-party signing communication.
	n/r and n/a: not reported and not applicable.
    \({}^{a}\)Our estimates (\Cref{Table:LaTSEC-Selected-Sizes} for \(\mathsf{LaTSEC}\)); prior schemes are extrapolated to \(n=60\) with threshold \(t=7\), matching \(\mathsf{LaTSEC}\)'s tolerance of up to \(t_\mathsf{uf}=k_p=6\) corrupted participants. These values lie outside the sources' parameter ranges; lower bounds are rounded down, and for Hermine the smallest per-participant value is given; see Appendix~\ref{Appendix:Storage-Comparison}.
	\({}^{b}\)From the sources, with \(n\) the number of participants covered by the parameter set behind the entries; for \(\mathsf{LaTSEC}\), \Cref{Table:LaTSEC-Selected-Sizes} converted from KiB.
	\({}^{c}\)As reported in \cite[Table 1]{gur2025olingo}, which notes identification without a formal IA proof.
	\({}^{d}\)Thresholds hold with overwhelming probability over a public seed.
	\({}^{e}\)Proof-based IA has no counterpart to \(|\mathsf{ak}_i|\).
	\({}^{f}\)Parameters only for \(n\leq16\), \(t\leq8\) \cite[Section 7]{borin2026hermine}; benchmarks up to \(n=t=64\).}
\end{table*}

\paragraph{\(\bm{\mathsf{LaTSEC}}\)}

We construct \(\mathsf{LaTSEC}\), a three-round lattice-based threshold signature with native non-interactive identifiable abort (\Cref{Section:LaTSEC-Main}).
Its sharing layer is \(\mathsf{SEC}\), the ramp everywhere-short secret sharing scheme of Arzaki et al.~\cite{arzaki2026SEC} built from lossless expanders, which gives each participant a single short share.
Each participant therefore holds one short signing share and is associated with one public abort key, against which its partial response is verified directly.
To the best of our knowledge, \(\mathsf{LaTSEC}\) is the \textit{first} lattice-based threshold signature with native non-interactive identifiable abort in which this holds, with the number of signing shares and public abort keys per participant independent of \(n\), the thresholds, and \(\kappa\).
The ramp access structure of \(\mathsf{SEC}\), with privacy threshold \(k_p\) and empirical reconstruction threshold \(k_c\), induces distinct unforgeability and signing thresholds \(t_\mathsf{uf}=k_p<t_\mathsf{sig}=k_c\): up to \(t_\mathsf{uf}\) corrupted participants cannot forge, while a uniformly random set of at least \(t_\mathsf{sig}\) participants can sign except with a small probability.
\Cref{Table:Intro-TS-Comparison} compares \(\mathsf{LaTSEC}\) with prior lattice-based schemes that handle malicious signers.

\paragraph{\textbf{Security under standard assumptions}}

We prove correctness, identifiable abort, and threshold unforgeability against static corruptions in the classical random oracle model under standard \(\mathsf{MLWE}\) and \(\mathsf{MSIS}\) assumptions (\Cref{Section:LaTSEC-Correctness-Security,Section:LaTSEC-Abort-Key-Simulation,Section:LaTSEC-Unforgeability}).
The unforgeability reduction in our case is particularly challenging since the underlying \(\mathsf{SEC}\) relies on \textit{computational} privacy, and the analysis must account for the additional leakage from threshold signing: every signing session reveals noisy linear functions \(c\bm{s}_i+\bm{r}_i\) of the shares.
In particular, we must simulate the complete tuple of public abort keys, which are images of correlated shares under one public matrix \(\bm{A}\), jointly with this leakage.
To handle this, we introduce Repeated-\(\bm{A}\) Matrix Hint-MLWE (\(\mathsf{RA\text{-}MHMLWE}\)), in which several \(\mathsf{MLWE}\) targets share the same \(\bm{A}\) matrix and the hints act jointly on all of their secrets.
We prove that \(\mathsf{RA\text{-}MHMLWE}\) admits a reduction from the standard \(\mathsf{MLWE}\), so our security rests on standard assumptions only.
This new problem may also be useful for other threshold schemes whose public keys are images of correlated secrets under one matrix.
Our proof thereby extends the multi-query privacy of \(\mathsf{SEC}\) \cite{arzaki2026SEC}, under which the target \(\bm{A}\bm{s}\) stays pseudorandom given the corrupted shares and the partial responses of all sessions, to the complete abort-key tuple.

\paragraph{\textbf{Concrete evaluation}}

We instantiate \(\mathsf{LaTSEC}\) for a participant set size of \(n\in\{40,60\}\) and security parameter \(\kappa\in\{128,192,256\}\), and evaluate its security and practical performance (\Cref{Section:LaTSEC-Concrete-Instantiation-Evaluation,Section:Implementation-Evaluation}).
At \(n=60\) and \(\kappa=128\), each participant stores only one \(14.3\)-KiB signing share and is associated with one \(24.0\)-KiB public abort key; the signature is \(42.6\) KiB, and the per-party signing communication is \(60.1\) KiB.
Against the everywhere-short baselines extrapolated to \(n=60\) in \Cref{Table:Intro-TS-Comparison}, under the size model described in Appendix~\ref{Appendix:Storage-Comparison}, this reduces per-participant secret-share storage by factors of at least \(48.9\) (Hermine) and \(67.5\) (ramp sDKG), and per-participant public abort-key storage by factors of at least \(85.0\) and \(80.0\), respectively.
The interactive and proof-based approaches keep shares small but pay in communication instead.
In \(\mathsf{LaTSEC}\), each partial response is checked directly against its abort key, so identifying misbehaving signers requires no further messages, and the per-party communication stays at \(60.1\) KiB (\(61.6\) kB) even when signing fails.
The reported per-party communication with identification is \(88.2+6.4t\) kB for TRaccoon-IA, \(852\) kB for Olingo, and \(3000\) kB for GKS (\Cref{Table:Intro-TS-Comparison}), each exceeding \(61.6\) kB for every threshold \(t\).
Our Rust prototype (\Cref{Section:Implementation-Evaluation}) corroborates these estimates: at \(n=60\) and \(\kappa=128\), its encoded signing shares average \(14.3\) KiB, its signatures average \(37.8\) KiB, below the worst-case \(42.6\)-KiB estimate because the encoding adapts to the actual width of the aggregate response, and one partial verification takes \(1.3\)~ms.

\paragraph{\textbf{Trade-offs}}

At \(n=60\) and \(\kappa=128\), the \(42.6\)-KiB (\(43.6\)-kB) signature of \(\mathsf{LaTSEC}\) is larger than those of the more recent schemes in \Cref{Table:Intro-TS-Comparison} for two reasons.
First, we do not currently apply bit dropping, which substantially reduces signature size and signing communication in Raccoon-style schemes \cite{del2024raccoon,del2024threshold}.
A preliminary size calculation suggests that adapting this technique to \(\mathsf{LaTSEC}\) could reduce the dominant transmitted vectors, and hence its signature and \(60.1\)-KiB per-party signing communication, by up to a factor of two.
Realizing this saving is challenging: the resulting rounding errors and hints would further complicate partial-response verification, aggregation, and the identifiable-abort and unforgeability analyses, so we leave it to future work.
Second, the abort-key simulation requires wider Gaussians and hence a larger modulus and dimension in our setting.
The signature of \(\mathsf{LaTSEC}\) is nevertheless smaller than the \(46.6\)-kB signature of GKS and the \(144.5\)-kB signature of the two-round line of \cite{chairattana2024partially,zhu2025algebraic}, as reported in \cite[Table 1]{gur2025olingo}.

Because \(\mathsf{SEC}\) is a ramp scheme, the signing threshold exceeds the unforgeability threshold: at \(n=60\), we have \(t_\mathsf{uf}=6\) and \(t_\mathsf{sig}=46\), whereas the threshold schemes in \Cref{Table:Intro-TS-Comparison} use a single threshold.
Here, \(t_\mathsf{sig}\) is the minimum signing-set size the protocol enforces, not a bound on which coalitions can reconstruct the key (\Cref{Section:Implementation-Correctness-IA}).
\(\mathsf{LaTSEC}\) also uses three message-dependent rounds and a trusted dealer.
Its concrete instantiations currently go up to \(n=60\), the largest party count evaluated for \(\mathsf{SEC}\); larger \(n\) should be reachable with more computation \cite[Section 6]{arzaki2026SEC}.

\subsection{Technical Challenges}\label{Section:Technical-Challenges}

At first sight, \(\mathsf{SEC}\) makes native non-interactive identifiable abort immediate: it gives each party one short share, short shares make partial responses directly verifiable, and short reconstruction vectors turn reconstruction into aggregation.
We first recall these three properties and then explain why the move from \(\mathsf{SEC}\) to \(\mathsf{LaTSEC}\) is nonetheless non-trivial.

\paragraph{\textbf{One short share per party}}

\(\mathsf{SEC}\) shares a short secret \(\bm{s}\) through a public matrix \(\mathbf{G}=\begin{bmatrix}\mathbf{g} \ \mathbf{\hat{G}}\end{bmatrix}\in\mathbb{Z}^{n\times k}\), where \(\mathbf{\hat{G}}\in\{-1,0,1\}^{n\times(k-1)}\) is derived from a lossless bipartite expander and \(\mathbf{g}\) is a short secret-bearing column \cite{arzaki2026SEC}.
Party \(i\) receives the \(i\)-th row \(\bm{s}_i\) of \(\mathbf{G}\begin{bmatrix}\bm{s}^\top\\ \mathbf{R}\end{bmatrix}\) for a short Gaussian masking matrix \(\mathbf{R}\), so each share is a short integer combination of \(\bm{s}\) and a few rows of \(\mathbf{R}\).
A set \(T\) is \textit{certified} if it reconstructs with an integer vector \(\bm{\lambda}_T\) satisfying \(\bm{\lambda}_T^\top\mathbf{G}_T=\mathbf{e}_1^\top\).
Signing uses short vectors of this form, and we write \(\mathsf{Rec}_\mathsf{pp}(T,(\bm{x}_i)_{i\in T}):=\sum_{i\in T}\lambda_{T,i}\bm{x}_i\).

\paragraph{\textbf{Direct verification}}

\(\mathsf{LaTSEC}\) follows the Raccoon-style no-abort signature underlying Threshold Raccoon \cite{del2024threshold}, with \(\mathsf{SEC}\) in place of masked Shamir sharing.
Party \(i\) holds the share \(\bm{s}_i\) of the signing key \(\bm{s}\), whose verification key is \(\bm{t}:=\bm{A}\bm{s}\), and is associated with the public abort key \(\bm{t}_i:=\bm{A}\bm{s}_i\).
In three rounds, each active signer commits to \(\bm{w}_i:=\bm{A}\bm{r}_i\) for a fresh short mask \(\bm{r}_i\), opens the commitment, and, given the common challenge \(c\), sends \(\bm{z}_i:=c\bm{s}_i+\bm{r}_i\).
Since \(\bm{s}_i\) is short, \(\bm{z}_i\) is short, and anyone can check \(\bm{A}\bm{z}_i-c\bm{t}_i=\bm{w}_i\) and \(\|\bm{z}_i\|_2\leq B_1\); a failed check identifies the signer.

\paragraph{\textbf{Reconstruction as aggregation}}

The combiner outputs \(\bm{z}:=\mathsf{Rec}_\mathsf{pp}(\mathsf{act},(\bm{z}_i)_{i\in\mathsf{act}})\), with the public vector \(\bm{\lambda}_\mathsf{act}\) as aggregation weights.
Because \(\mathsf{Rec}_\mathsf{pp}\) is integer-linear and commutes with left multiplication by \(\bm{A}\) \cite[Section 4.3]{arzaki2026SEC},
\begin{align}
	\bm{A}\bm{z}-c\bm{t} &= \bm{A}\, \mathsf{Rec}_{\mathsf{pp}}\left(\mathsf{act}, (c\bm{s}_i+\bm{r}_i)_{i\in\mathsf{act}}\right) -c\bm{t} \nonumber\\
	&= c\bm{A}\, \mathsf{Rec}_{\mathsf{pp}} \left(\mathsf{act}, (\bm{s}_i)_{i\in\mathsf{act}} \right) \nonumber \\
	&\quad + \mathsf{Rec}_{\mathsf{pp}}\left(\mathsf{act}, (\bm{A}\bm{r}_i)_{i\in\mathsf{act}}\right) -c\bm{t} \nonumber\\
	&= c\bm{t} + \mathsf{Rec}_{\mathsf{pp}}\left(\mathsf{act}, (\bm{w}_i)_{i\in\mathsf{act}}\right) -c\bm{t} = \bm{w} \label{Eq:Aggregation-Identity},
\end{align}
so \((c,\bm{z})\) satisfies the verification relation of the underlying signature, and \(\|\bm{z}\|_2\leq\|\bm{\lambda}_\mathsf{act}\|_1B_1\) is short because \(\bm{\lambda}_\mathsf{act}\) is.

\paragraph{\textbf{Three challenges}}

Prior work on everywhere-short sharing shows that removing the masks is not automatic: the sDKG framework distills the properties a sharing must satisfy for identifiable abort \cite{del2024shortly}, and Borin et al.\ note that security depends on the specific sharing structure \cite{borin2026hermine}.
For \(\mathsf{SEC}\), this raises three challenges.

\textit{Hiding the shares in unmasked partial responses.}
Direct verification reveals each \(\bm{z}_i=c\bm{s}_i+\bm{r}_i\) without one-time masks, so the Gaussian mask \(\bm{r}_i\) alone must hide \(c\bm{s}_i\), while the privacy of \(\mathsf{SEC}\) is computational.
We rely on the multi-query privacy of \(\mathsf{SEC}\), which keeps \(\bm{t}\) pseudorandom given the corrupted shares and these responses \cite{arzaki2026SEC}, and choose the mask width to meet its conditions (\Cref{Section:LaTSEC-Concrete-Instantiation-Evaluation}).

\textit{Simulating the abort keys.}
\(\mathsf{LaTSEC}\) also publishes the abort keys \(\bm{t}_i=\bm{A}\bm{s}_i\), and the reduction must simulate all of them together with the signing leakage.
Matrix Hint-MLWE cannot model this view directly: it provides one \(\mathsf{MLWE}\) target with hints on its secret, whereas the abort keys are several \(\mathsf{MLWE}\) targets under the same \(\bm{A}\), and the hints act jointly on their correlated secrets.
We therefore introduce \(\mathsf{RA\text{-}MHMLWE}\), prove that it reduces from \(\mathsf{MLWE}\) (\Cref{Section:RA-MHMLWE}), and reduce the abort-key view to it through an algebraic completion of the corrupted keys and a unimodular change of variables that isolates the hidden coordinates (\Cref{Section:LaTSEC-Abort-Key-Simulation}).

\textit{Min-entropy of the aggregate commitment.}
Programming the challenge requires the aggregate commitment \(\bm{w}\) to have high min-entropy given the adversary's view.
In Threshold Raccoon, \(\bm{w}\) is a plain sum of commitments \cite{del2024threshold}.
In \(\mathsf{LaTSEC}\), it is the reconstruction \(\sum_{i\in\mathsf{act}}\lambda_{\mathsf{act},i}\bm{w}_i\), whose entropy the adversary could remove if the honest coefficients vanished, or reduce if they were not invertible modulo \(q\).
We prove a min-entropy bound specific to such reconstructed commitments: since no set of at most \(t_\mathsf{uf}\) parties is certified, some honest coefficient is nonzero, and admissibility bounds its magnitude below \(q\), so it is invertible modulo the prime \(q\).
Thus, \(\bm{w}\) inherits that signer's min-entropy (\Cref{Corollary:Reconstructed-Commitment-MinEntropy-Prime}).

\section{Preliminaries}\label{Section:Preliminaries}

Besides standard background, this section introduces a new problem, \(\mathsf{RA\text{-}MHMLWE}\), and states its reduction from \(\mathsf{MLWE}\) (\Cref{Section:RA-MHMLWE}).

\subsection{Notation and Lattice Background}\label{Section:Notation}

We follow the notation of \(\mathsf{SEC}\) \cite[Section 2 and Appendix A]{arzaki2026SEC}.
Let \([n]:=\{1,\dots,n\}\), let \(\mathcal{R}:=\mathbb{Z}[X]/(X^d+1)\) for a power of two \(d\), and let \(\mathcal{R}_q:=\mathcal{R}/q\mathcal{R}\).
Bold italic letters denote objects over \(\mathcal{R}\) or \(\mathcal{R}_q\), and bold upright letters denote their coefficient embeddings or objects over \(\mathbb{Z}\), \(\mathbb{Z}_q\), or \(\mathbb{R}\).
Thus, \(\bm{a}\in\mathcal{R}^\ell\) is identified with \(\mathbf{a}\in\mathbb{Z}^{d\ell}\), and \(\bm{A}\in\mathcal{R}_q^{p\times\ell}\) with \(\mathbf{A}\in\mathbb{Z}_q^{dp\times d\ell}\).
Throughout, \(D:=d\ell\).
For a matrix \(\mathbf{M}\) and \(T\subseteq[n]\), \(\mathbf{M}_T\) denotes the rows indexed by \(T\).
\(\mathrm{Row}_\mathbb{Z}\), \(\mathrm{Col}_\mathbb{Z}\), and \(\ker\) denote integer row and column modules and kernels, \(\|\mathbf{M}\|_2\) and \(\sigma_{\min}(\mathbf{M})\) the largest and smallest singular values, \(wt_H\) the Hamming weight, \(\mathrm{vec}\) column-wise vectorization, and \(\otimes\) the Kronecker product.
Norms of ring elements are those of their coefficient embeddings, with representatives in \((-q/2,q/2]\) modulo \(q\).
\(H_\infty\) denotes min-entropy.

For a countable \(S\subseteq\mathbb{R}^N\), a center \(\mathbf{c}\), and a positive-definite parameter matrix \(\mathbf{\Sigma}\), \(\mathcal{D}_{S,\mathbf{c},\sqrt{\mathbf{\Sigma}}}\) is the discrete Gaussian with mass proportional to \(\exp(-\frac12(\mathbf{x}-\mathbf{c})^\top\mathbf{\Sigma}^{-1}(\mathbf{x}-\mathbf{c}))\); we write \(\mathcal{D}_\sigma\) when \(S=\mathbb{Z}^N\), \(\mathbf{c}=\mathbf{0}\), and \(\mathbf{\Sigma}=\sigma^2\mathbf{I}\).
The smoothing parameter \(\eta_\epsilon\) is normalized in the same standard-deviation convention, and hence is smaller by a factor \(\sqrt{2\pi}\) than in the \(\exp(-\pi\|\mathbf{x}\|^2/s^2)\) convention \cite[Eq.~(2)]{arzaki2026SEC}.
\(\mathsf{MLWE}_{n,m,q,d,\sigma}\) denotes decisional Module-LWE in the systematic form of \cite[Definition 2]{arzaki2026SEC}: distinguish \((\bm{A},\overline{\bm{A}}\bm{s})\) from \((\bm{A},\bm{u})\), where \(\overline{\bm{A}}:=\begin{bmatrix}\bm{I}_n \ \bm{A}\end{bmatrix}\), \(\bm{A}\randgets\mathcal{R}_q^{n\times m}\), \(\bm{s}\gets\mathcal{D}_\sigma^{n+m}\), and \(\bm{u}\randgets\mathcal{R}_q^n\).
For \(h\in\mathbb{Z}^+\), \(\overline{\bm{A}}^{\langle h\rangle}:=\bm{I}_h\otimes\overline{\bm{A}}\in\mathcal{R}_q^{hn\times h(n+m)}\) repeats \(\overline{\bm{A}}\) on the diagonal.
\(\mathsf{MSIS}\) and the hint-free repeated variant of \(\mathsf{MLWE}\) are recalled in Appendix~\ref{Appendix:Standard-Problems}.

\subsection{Repeated-\texorpdfstring{\(\bm{A}\)}{A} Matrix Hint-MLWE}\label{Section:RA-MHMLWE}

Matrix Hint-MLWE augments an \(\mathsf{MLWE}\) sample with noisy linear images \(\bm{M}\bm{s}+\bm{y}\) of its secret \cite{kim2023toward,espitau2024flood,wang2025post,arzaki2026SEC}.
The abort-key simulation of \Cref{Section:LaTSEC-Abort-Key-Simulation} produces several \(\mathsf{MLWE}\) targets under the same matrix, together with correlated hints on all of their secrets jointly.
We capture this with the following problem.

\begin{definition}[Repeated-\(\bm{A}\) Matrix Hint-MLWE]\label{Definition:RA-MHMLWE}
	Let \(n,m,q,d,h\in\mathbb{Z}^{+}\), \(\nu\in\mathbb{Z}_{\geq0}\), and \((n_j)_{j\in[\nu]}\in(\mathbb{Z}^{+})^\nu\).
	Let \(\chi_{\bm{s}}\) be a distribution over \(\mathcal{R}^{h(n+m)}\) and, for \(j\in[\nu]\), let \(\chi_{\bm{y}_j}\) be a distribution over \(\mathcal{R}^{n_j}\) and \(\mathcal{M}_j\) a distribution over \(\mathcal{R}^{n_j\times h(n+m)}\).
	The problem \(\mathsf{RA\text{-}MHMLWE}[h]\) is to distinguish \(\left( \bm{A}, \overline{\bm{A}}^{\langle h\rangle}\bm{s}, (\bm{M}_j)_{j\in[\nu]}, (\bm{z}_j)_{j\in[\nu]} \right)\) from \(\left( \bm{A}, \bm{u}, (\bm{M}_j)_{j\in[\nu]}, (\bm{z}_j)_{j\in[\nu]} \right)\), where \(\bm{A} \randgets \mathcal{R}_q^{n\times m}\), \(\bm{u} \randgets \mathcal{R}_q^{hn}\), \(\bm{s}\gets\chi_{\bm{s}}\), and, independently for \(j\in[\nu]\), \(\bm{M}_j\gets\mathcal{M}_j\), \(\bm{y}_j\gets\chi_{\bm{y}_j}\), and \(\bm{z}_j := \bm{M}_j\bm{s}+\bm{y}_j\) over \(\mathcal{R}\).
	We write \(\mathsf{Adv}_{\mathcal{A}}^{\mathsf{RA\text{-}MHMLWE}[h]}(\kappa)\) for the distinguishing advantage of \(\mathcal{A}\).
\end{definition}

The hint matrices act on the whole stacked secret and may mix coordinates of different blocks.
For \(h=1\), the problem is Matrix Hint-MLWE with rectangular hints \cite{wang2025post,arzaki2026SEC}.
Without hints, it is the stacked form of multi-secret \(\mathsf{MLWE}\) \cite{brakerski2013classical,boudgoust2021hardness}, which we call \(\mathsf{RA\text{-}MLWE}\) (Appendix~\ref{Appendix:Standard-Problems}).
Other hinted variants of (M)LWE include Leaky LWE \cite{lai2025leaky} and adaptive Hint-MLWE \cite{hwang2026proving}.
\(\mathsf{RA\text{-}MHMLWE}\) is a new problem rather than a new assumption: its hardness follows from standard \(\mathsf{MLWE}\).
Under coefficient embedding, let \(N:=dh(n+m)\) and \(N_j:=dn_j\).

\begin{lemma}[Hardness of \(\mathsf{RA\text{-}MHMLWE}\)]\label{Lemma:Hardness-RA-MHMLWE}
	In \Cref{Definition:RA-MHMLWE}, let \(\mathbf{s}\gets \mathcal{D}_{\mathbb{Z}^{N},\mathbf{c}_{\mathbf{s}}, \sqrt{\mathbf{\Sigma}_{\mathbf{s}}}}\) and \(\mathbf{y}_j\gets \mathcal{D}_{\mathbb{Z}^{N_j},\mathbf{c}_{\mathbf{y}_j}, \sqrt{\mathbf{\Sigma}_{\mathbf{y}_j}}}\) with \(\mathbf{\Sigma}_{\mathbf{s}},\mathbf{\Sigma}_{\mathbf{y}_j}\succ0\).
	For every tuple of hint matrices in the support, with coefficient embeddings \(\mathbf{M}_j\in\mathbb{Z}^{N_j\times N}\), let \(\mathbf{\Xi} := \mathbf{\Sigma}_{\mathbf{s}}^{-1} + \sum_{j=1}^{\nu} \mathbf{M}_j^\top \mathbf{\Sigma}_{\mathbf{y}_j}^{-1} \mathbf{M}_j\).
	Suppose that \(1/\sigma^2 \geq 2\|\mathbf{\Xi}\|_2\) uniformly over all such tuples, and that \(\sigma\geq\sqrt{2}\eta_\epsilon(\mathbb{Z}^{N})\) for some \(0<\epsilon\leq1/\sqrt{2N+4}\).
	Then, for every PPT adversary \(\mathcal{A}\), there exists a PPT adversary \(\mathcal{B}\) such that \(\mathsf{Adv}_{\mathcal{A}}^{ \mathsf{RA\text{-}MHMLWE}[h] }(\kappa) \leq h\cdot \mathsf{Adv}_{ \mathcal{B}}^{\mathsf{MLWE}_{n,m,q,d,\sigma} }(\kappa) + 2\epsilon\).
	The same holds in the presence of public auxiliary information jointly distributed with the secret, provided that this information can be sampled exactly in polynomial time, that, conditioned on it, the secret is an elliptic discrete Gaussian over \(\mathbb{Z}^N\) with publicly computable parameters, and that the precision and smoothing conditions hold uniformly over its outcomes.
\end{lemma}

\begin{proof}[Proof sketch]
    A block-by-block hybrid shows that \(\mathsf{RA\text{-}MLWE}[h]\) is hard under \(\mathsf{MLWE}\), with loss \(h\) (\Cref{Lemma:Hardness-RA-MLWE}).
	The posterior-resampling argument for Matrix Hint-MLWE \cite{kim2023toward,espitau2024flood,wang2025post} then applies jointly to the stacked secret: it creates the correlated elliptic secret required by the hints from a spherical source secret, and the repeated target map, being linear and public, does not otherwise affect it.
    The reduction samples the hint noise as in the problem, and the dummy secret and the correction exactly: for a covariance \(\mathbf{\Sigma}\) with all eigenvalues at least \(\sigma^2\), the exact sampler of \cite[Lemma 2.3]{brakerski2013classical} on the lattice \(\mathbf{\Sigma}^{-1/2}\mathbb{Z}^N\) applies by the smoothing condition, so the bound has no sampling term.
    Details are in Appendix~\ref{Appendix:Hardness-RA-MHMLWE-Sketch}; the full proof is in Appendix~\ref{Appendix:Hardness-RA-MHMLWE}.
\end{proof}

\subsection{Self-Target MSIS}\label{Section:STMSIS}

As in \cite{del2024shortly,del2024threshold}, the final forgery reduction targets Self-Target \(\mathsf{MSIS}\) \cite{ducas2018crystals,kiltz2018concrete}.

\begin{definition}[Self-Target MSIS]\label{Definition:STMSIS}
	Let \(\mathsf{G}: \mathcal{R}_q^n \times \{0,1\}^* \to \mathcal{C}\) be a random oracle with \(\mathcal{C}\subseteq\mathcal{R}_q\).
	Given \(\bm{A} \randgets \mathcal{R}_q^{n\times m}\), the problem \(\mathsf{SelfTargetMSIS}_{ n,m,q,d,\mathcal{C},B_\mathsf{STMSIS} }\) asks for \(\mathsf{msg}\in\{0,1\}^*\) and \(\bm{z} = \begin{bmatrix} \bm{z}'\\ c \end{bmatrix} \in \mathcal{R}_q^{n+m}\) with \(c\in\mathcal{C}\), \(0 < \|\bm{z}\|_2 \leq B_\mathsf{STMSIS}\), and \(\mathsf{G}( \begin{bmatrix} \bm{I}_n \ \bm{A} \end{bmatrix} \bm{z}, \mathsf{msg}) = c\).
	We write \(\mathsf{Adv}_{\mathcal{A}}^\mathsf{STMSIS}(\kappa)\) for the success probability of \(\mathcal{A}\).
\end{definition}

\begin{lemma}[Hardness of \(\mathsf{SelfTargetMSIS}\) {\cite[Lemma B.1]{del2024threshold}}]\label{Lemma:Hardness-STMSIS}
	For every adversary \(\mathcal{A}\) against \(\mathsf{SelfTargetMSIS}_{ n,m,q,d,\mathcal{C},B_\mathsf{STMSIS} }\) making at most \(Q_H\) random-oracle queries, there exists \(\mathcal{B}\) against \(\mathsf{MSIS}_{n,m,q,d,2B_\mathsf{STMSIS}}\) with \(\mathsf{Time}(\mathcal{B}) \approx 2\,\mathsf{Time}(\mathcal{A})\) and \(\mathsf{Adv}_{\mathcal{A}}^\mathsf{STMSIS}(\kappa) \leq \sqrt{ Q_H\, \mathsf{Adv}_{\mathcal{B}}^\mathsf{MSIS}(\kappa) } + Q_H/|\mathcal{C}|\).
\end{lemma}

The proof of \cite[Lemma B.1]{del2024threshold} does not depend on the message length, so the lemma holds for \(\mathsf{msg}\in\{0,1\}^*\), as in \cite{ducas2018crystals}.

\subsection{The \texorpdfstring{\(\bm{\mathsf{SEC}}\)}{SEC} Sharing Scheme}\label{Section:SEC-Summary}

We recall the properties of \(\mathsf{SEC}\) \cite[Definition 11]{arzaki2026SEC} that \(\mathsf{LaTSEC}\) uses.
\(\mathsf{SEC.Setup}(1^\kappa)\) samples a lossless bipartite expander with left degree \(c_\mathrm{exp}\), sets \(\mathbf{\hat{G}}\in\{-1,0,1\}^{n\times(k-1)}\) to its signed bi-adjacency matrix, so that every row of \(\mathbf{\hat{G}}\) has \(c_\mathrm{exp}\) nonzero entries, and appends a secret-bearing column \(\mathbf{g}\in\mathbb{Z}^n\) with \(\|\mathbf{g}\|_\infty\leq\beta_\mathbf{g}\) to form \(\mathbf{G}=\begin{bmatrix}\mathbf{g} \ \mathbf{\hat{G}}\end{bmatrix}\).
It outputs \(\mathsf{pp}\), including \(\mathbf{G}\), a privacy threshold \(k_p\), and an empirical reconstruction threshold \(k_c>k_p\) above which random sets reconstruct except with a small probability \(\epsilon_c\).
\(\mathsf{SEC.Deal}_\mathsf{pp}(\mathbf{s})\), for \(\mathbf{s}\in\mathbb{Z}^D\), samples \(\mathbf{R}\gets\mathcal{D}_{\sigma_\mathbf{R}}^{(k-1)\times D}\) and outputs the rows \(\mathbf{s}_i\) of
\begin{equation}\label{Eq:SEC-Deal}
	\mathbf{S}:=\mathbf{G}\begin{bmatrix}\mathbf{s}^\top\\ \mathbf{R}\end{bmatrix}=\mathbf{g}\mathbf{s}^\top+\mathbf{\hat{G}}\mathbf{R}\in\mathbb{Z}^{n\times D}.
\end{equation}
\(\mathsf{SEC.Rec}_\mathsf{pp}(T,(\mathbf{x}_i)_{i\in T})\) outputs \(\sum_{i\in T}\lambda_{T,i}\mathbf{x}_i\) for a reconstruction vector \(\bm{\lambda}_T\in\mathbb{Z}^{|T|}\) with \(\bm{\lambda}_T^\top\mathbf{G}_T=\mathbf{e}_1^\top\).
We use the following properties.
\begin{enumerate}
	\item\label{Item:SEC-Short} \textit{One short share.} Each party holds one share \(\mathbf{s}_i\in\mathbb{Z}^D\), and if \(\|\mathbf{s}\|_\infty\leq\beta_{\bm{s}}\) and all entries of \(\mathbf{R}\) are bounded by \(\beta_\mathbf{R}\), then \(\|\mathbf{s}_i\|_\infty\leq\beta_\mathbf{g}\beta_{\bm{s}}+c_\mathrm{exp}\beta_\mathbf{R}\) \cite[Lemma 12]{arzaki2026SEC}.
	\item\label{Item:SEC-Rec} \textit{Exact, deterministic reconstruction.} A set \(T\) is \textit{certified} if \(\mathbf{e}_1^\top\in\mathrm{Row}_\mathbb{Z}(\mathbf{G}_T)\), that is, if such a vector exists.
	For certified \(T\), a deterministic algorithm \cite[Algorithm 1]{arzaki2026SEC} computes a short \(\bm{\lambda}_T\) from \(\mathbf{G}_T\) alone; otherwise it returns \(\bot\).
	Reconstruction is exact for every such vector, since \(\bm{\lambda}_T^\top\mathbf{S}_T=\bm{\lambda}_T^\top\mathbf{G}_T\begin{bmatrix}\mathbf{s}\ \mathbf{R}^\top\end{bmatrix}^\top=\mathbf{s}^\top\) \cite[Lemma 5]{arzaki2026SEC}.
	\item\label{Item:SEC-Coeff} \textit{Public coefficient checks.} Given \(T\) and a candidate \(\bm{\lambda}_T\), the identity \(\bm{\lambda}_T^\top\mathbf{G}_T=\mathbf{e}_1^\top\) and the bounds \(\|\bm{\lambda}_T\|_\infty\leq\beta_{\bm{\lambda}}\) and \(wt_H(\bm{\lambda}_T)\leq\omega_{\bm{\lambda}}\) are exact integer checks on public data; for the selected instances, the output of \cite[Algorithm 1]{arzaki2026SEC} meets both bounds on all but a small fraction of certified sets of size at least \(k_c\) (\cite[Section 6.3]{arzaki2026SEC} and \Cref{Section:EmpiricalProfiles-InstantiationMethod}).
	\item\label{Item:SEC-Linear} \textit{Linearity and public-map compatibility.} \(\mathsf{SEC.Rec}_\mathsf{pp}\) is \(\mathbb{Z}\)-linear and commutes with left multiplication by any public matrix modulo \(q\) \cite[Lemmas 17 and 18]{arzaki2026SEC}.
    In particular, \(\mathsf{SEC.Rec}_\mathsf{pp}(T,(\bm{A}\bm{x}_i)_{i\in T})=\bm{A}\,\mathsf{SEC.Rec}_\mathsf{pp}(T,(\bm{x}_i)_{i\in T})\).
	\item\label{Item:SEC-Unimodular} \textit{Non-cancellation and unimodular minors.} For every \(U\subseteq[n]\) with \(1\leq|U|\leq k_p\), the expander yields a unimodular \(|U|\times|U|\) submatrix \(\mathbf{M}_U\) of \(\mathbf{\hat{G}}_U\), computed deterministically by peeling \cite[Lemma 4 and Algorithm 3]{arzaki2026SEC}.
	Hence \(\mathrm{Col}_\mathbb{Z}(\mathbf{\hat{G}}_U)=\mathbb{Z}^{|U|}\), so \((\mathbf{g})_U\in\mathrm{Col}_\mathbb{Z}(\mathbf{\hat{G}}_U)\) and no such \(U\) is certified (\Cref{Lemma:Structural-NonCancellation-NonReconstruction}).
	\item\label{Item:SEC-Privacy} \textit{Multi-query privacy.} For \(\bm{A}=\begin{bmatrix}\bm{I}_p\ \bm{A}'\end{bmatrix}\), the target \(\bm{A}\bm{s}\) is pseudorandom given \(\bm{A}\), the shares of any \(U\) with \(|U|\leq k_p\), and noisy linear leakage \(c_s\bm{s}_i+\bm{r}_{i,s}\) from all shares over \(Q\) queries, under \(\mathsf{MLWE}_{p,\ell-p,q,d,\sigma_\mathrm{MQ}}\) for the width \(\sigma_\mathrm{MQ}\) of \cite[Eq.~(58)]{arzaki2026SEC}, provided its kernel-smoothing and posterior-resampling conditions hold \cite[Theorem 4]{arzaki2026SEC}.
	Formally, privacy is stated through the experiment \(\mathsf{HC\text{-}Leak}\) \cite[Definition 9]{arzaki2026SEC}.
\end{enumerate}

\subsection{Threshold Signatures with Identifiable Abort}\label{Section:TSS-with-IA}

We use the round-based syntax of \cite[Definition 8]{del2024shortly}, which refines that of \cite{del2024threshold} with identifiable abort in the sense of \cite{ishai2014secure}, replacing distributed key generation by a trusted dealer and adding a partial-verification algorithm as in \cite{ruffing2022roast,borin2026hermine}.
To accommodate ramp sharing, we separate the \textit{unforgeability threshold} \(t_\mathsf{uf}\), the maximum number of corrupted parties, from the \textit{signing threshold} \(t_\mathsf{sig}\), the minimum size of a signing set.
An efficiently computable predicate \(\mathcal{P}(\mathsf{tspar},\mathsf{vk},\mathsf{sid},\mathsf{act},\mathsf{msg})\in\{0,1\}\) defines the \textit{admissible} session inputs, each with \(|\mathsf{act}|\geq t_\mathsf{sig}\); a session is admissible if its input is.

\begin{definition}[TSS with IA]\label{Definition:TSS-with-IA}
	Let \(1\leq t_\mathsf{uf}<t_\mathsf{sig}\leq n\).
	An \(\mathsf{rnd}\)-round \((t_\mathsf{uf},t_\mathsf{sig},n)\)-threshold signature scheme with identifiable abort consists of the following algorithms.
	\begin{enumerate}
		\item \(\mathsf{Setup}(1^\kappa,n,t_\mathsf{uf},t_\mathsf{sig})\) outputs public parameters \(\mathsf{tspar}\).
		\item \(\mathsf{Keygen}(\mathsf{tspar})\) outputs a tuple \((\mathsf{vk},\mathsf{ak},(\mathsf{sk}_i)_{i \in [n]})\), i.e., a verification key \(\mathsf{vk}\), public abort information \(\mathsf{ak}=(\mathsf{ak}_i)_{i\in[n]}\), and secret key shares \((\mathsf{sk}_i)_{i\in[n]}\).
		\item For \(k\in[\mathsf{rnd}]\), \(\mathsf{ShareSign}_k(\mathsf{st}_i,\mathsf{sid},\mathsf{act},\cdot)\) outputs an updated state and a message \((\mathsf{st}_i,\mathsf{pm}_i^k)\), or an abort \((\mathsf{abort},S_\bot^i)\) with \(S_\bot^i\subseteq\mathsf{act}\); its last input is \(\mathsf{msg}\) for \(k=1\) and \((\mathsf{pm}_j^{k-1})_{j\in\mathsf{act}}\) otherwise.
		\item \(\mathsf{PartialVerify}(\mathsf{tspar},\allowbreak\mathsf{vk},\allowbreak\mathsf{ak}_i,\allowbreak\mathsf{sid},\allowbreak\mathsf{act},\allowbreak\mathsf{msg},\allowbreak\mathsf{PM})\) outputs a bit that checks party \(i\)'s contribution to the transcript \(\mathsf{PM}:=(\mathsf{pm}_j^k)_{k\in[\mathsf{rnd}],j\in\mathsf{act}}\), or \(\bot\) if another party's message needed for the check is missing or malformed.
		\item \(\mathsf{Combine}(\mathsf{tspar},\allowbreak\mathsf{vk},\allowbreak\mathsf{ak},\allowbreak\mathsf{sid},\allowbreak\mathsf{act},\allowbreak\mathsf{msg},\allowbreak\mathsf{PM})\) outputs a signature \(\mathsf{sig}\) or an abort \((\mathsf{abort},S_\bot)\).
		\item \(\mathsf{Verify}(\mathsf{vk},\mathsf{msg},\mathsf{sig})\) outputs a bit.
	\end{enumerate}
\end{definition}

Signing is synchronous: messages of one round are delivered consistently to all active parties before the next round, for example over an authenticated broadcast channel.
\textit{Unforgeability} (\Cref{Definition:TSS-Unforgeability}) requires that no PPT adversary corrupting at most \(t_\mathsf{uf}\) parties, chosen statically, forges a signature on a message for which no honest party contributed a signing message.
\textit{Identifiable abort} (\Cref{Definition:TSS-IA}) requires that, in every admissible session, either all honest parties abort with the same non-empty blame set contained in the corrupted set, or \(\mathsf{Combine}\) returns such a set, or the output signature verifies.
\textit{Correctness} requires that an honest execution of an admissible session outputs a valid signature; it follows from identifiable abort with \(\mathsf{corrupt}=\emptyset\), since no non-empty blame set is then contained in \(\mathsf{corrupt}\).
The games are given in Appendix~\ref{Appendix:TSS-with-IA}.

\section{The \texorpdfstring{\(\bm{\mathsf{LaTSEC}}\)}{LaTSEC} Scheme}\label{Section:LaTSEC-Main}

\subsection{Overview and Parameters}\label{Section:LaTSEC-Overview}

\(\mathsf{LaTSEC}\) distributes the Raccoon-style no-abort signature underlying Threshold Raccoon \cite{del2024raccoon,del2024threshold}, without compression or rounding (see Appendix~\ref{Appendix:Underlying-No-Abort-Signature}).
Its signing key is \(\bm{s}\gets\mathcal{D}_{\sigma_{\bm{s}}}^\ell\), its verification key is \((\bm{A},\bm{t})\) with \(\bm{A}=\begin{bmatrix}\bm{I}_p\ \bm{A}'\end{bmatrix}\in\mathcal{R}_q^{p\times\ell}\) and \(\bm{t}:=\bm{A}\bm{s}\), and a signature \((c,\bm{z})\) is valid if \(c=H_c(\cdots,\bm{A}\bm{z}-c\bm{t})\) and \(\|\bm{z}\|_2\leq B_2\).
The signing key is shared with \(\mathsf{SEC}\) under the coefficient embedding \(\mathcal{R}^\ell\cong\mathbb{Z}^D\), and we set \(t_\mathsf{uf}:=k_p\) and \(t_\mathsf{sig}:=k_c\).
Party \(i\) holds \(\mathsf{sk}_i:=\bm{s}_i\) and is associated with the public abort key \(\mathsf{ak}_i:=\bm{t}_i:=\bm{A}\bm{s}_i\).
Besides the payload, the message \(\mathsf{msg}\) canonically encodes the session identifier \(\mathsf{sid}\), the active set \(\mathsf{act}\subseteq[n]\), and a reconstruction vector \(\bm{\lambda}_\mathsf{act}\in\mathbb{Z}^{|\mathsf{act}|}\), which \(\mathsf{SEC.Rec}_\mathsf{pp}(\mathsf{act},\cdot)\) uses throughout.
A session input is \textit{admissible}, that is, \(\mathcal{P}(\mathsf{tspar},\mathsf{vk},\mathsf{sid},\mathsf{act},\mathsf{msg})=1\), if \(\mathsf{msg}\) encodes \((\mathsf{sid},\mathsf{act},\bm{\lambda}_\mathsf{act})\), \(|\mathsf{act}|\geq t_\mathsf{sig}\), \(\bm{\lambda}_\mathsf{act}^\top\mathbf{G}_\mathsf{act}=\mathbf{e}_1^\top\), \(\|\bm{\lambda}_\mathsf{act}\|_\infty\leq\beta_{\bm{\lambda}}\), and \(wt_H(\bm{\lambda}_\mathsf{act})\leq\omega_{\bm{\lambda}}\).
These exact integer checks on public data do not depend on how \(\bm{\lambda}_\mathsf{act}\) was computed, so implementations need not share a lattice-reduction routine; the session initiator can take \(\bm{\lambda}_\mathsf{act}\) from \cite[Algorithm 1]{arzaki2026SEC}.
An inadmissible input is rejected before signing without blaming anyone, which affects availability only (\Cref{Section:EmpiricalProfiles-InstantiationMethod}).
\Cref{Table:LaTSEC-Parameters} summarizes the parameters; their conditions are given in \Cref{Section:LaTSEC-Asymptotic-Parameters}.

\begin{table}[!t]
	\caption{Parameters of \(\mathsf{LaTSEC}\)}
	\label{Table:LaTSEC-Parameters}
	\centering
	\footnotesize
	\setlength{\tabcolsep}{3pt}
	\begin{tabular}{@{}l p{0.78\columnwidth}@{}}
		\toprule
		Parameter & Meaning \\
		\midrule
		\(d,q\) & ring degree and prime modulus of \(\mathcal{R}_q\) \\
		\(p,\ell\) & dimensions of \(\bm{A}\in\mathcal{R}_q^{p\times\ell}\), with \(\ell>p\); \(D:=d\ell\) \\
		\(n,k\) & number of parties, columns of \(\mathbf{G}\in\mathbb{Z}^{n\times k}\) \\
		\(t_\mathsf{uf},t_\mathsf{sig}\) & unforgeability and signing thresholds, \(k_p\) and \(k_c\) of \(\mathsf{SEC}\) \\
		\(\mathcal{C},\omega_c\) & challenges \(\{c\in\mathcal{R}:\|c\|_\infty=1,\ \|c\|_1=\omega_c\}\) \\
		\(\sigma_{\bm{s}},\sigma_\mathbf{R},\sigma_{\bm{r}}\) & widths of the signing key, of the entries of \(\mathbf{R}\in\mathbb{Z}^{(k-1)\times D}\), and of the signing masks \\
        \(\beta_{\bm{s}},\beta_\mathbf{R},\beta_{\bm{r}}\) & tail bounds on \(\|\bm{s}\|_2\) (hence on \(\|\bm{s}\|_\infty\)), \(|\mathbf{R}[i',j']|\), and \(\|\bm{r}_i\|_2\) \\
		\(\beta_\mathbf{g},c_\mathrm{exp}\) & bound on \(\|\mathbf{g}\|_\infty\); nonzero entries per row of \(\mathbf{\hat{G}}\) \\
		\(\beta_{\bm{\lambda}},\omega_{\bm{\lambda}}\) & bounds on \(\|\bm{\lambda}_\mathsf{act}\|_\infty\) and \(wt_H(\bm{\lambda}_\mathsf{act})\) in admissible sessions \\
		\(B_1,B_2\) & norm bounds of partial and final verification \\
		\bottomrule
	\end{tabular}
\end{table}

The hash functions \(H_\mathsf{cmt}:\{0,1\}^*\to\{0,1\}^{2\kappa}\) and \(H_c:\{0,1\}^*\to\mathcal{C}\) are modeled as random oracles; \(H_\mathsf{cmt}\) binds each signer to its commitment before the openings are revealed, and \(H_c\) derives the challenge.
For implementations, \(H_c\) can feed the output stream of an extendable-output function on its input directly into a sample-in-ball sampler as in Dilithium \cite{ducas2018crystals}.
This is a random oracle into \(\mathcal{C}\) whenever the extendable-output function is one (\Cref{Section:Implementation-Instantiation}).

\subsection{Protocol}\label{Section:LaTSEC-Construction}

\begin{figure*}[!t]
	\centering
	\footnotesize
	\begin{tcolorbox}[gamebox]
	\begin{minipage}[t]{0.485\textwidth}
		\(\mathsf{Setup}(1^\kappa,n,t_\mathsf{uf},t_\mathsf{sig})\)
		\hrule height 0.4pt
		\begin{algorithmic}[1]
			\State \(\mathsf{pp}\gets\mathsf{SEC.Setup}(1^\kappa)\) for \(n\) parties and secret dimension \(D\), with \((k_p,k_c)=(t_\mathsf{uf},t_\mathsf{sig})\), or return \(\bot\) if no such instance is available
			\State Fix \(\mathcal{C},\sigma_{\bm{s}},\sigma_{\bm{r}},\beta_{\bm{\lambda}},\omega_{\bm{\lambda}},B_1,B_2\) as in \Cref{Section:LaTSEC-Asymptotic-Parameters}.
			\State \Return \(\mathsf{tspar}:=(\mathsf{pp},\mathcal{C},B_1,B_2,\sigma_{\bm{s}},\sigma_{\bm{r}},\beta_{\bm{\lambda}},\omega_{\bm{\lambda}})\)
		\end{algorithmic}
		\smallskip
		\(\mathsf{Keygen}(\mathsf{tspar})\) \hfill (trusted dealer)
		\hrule height 0.4pt
		\begin{algorithmic}[1]
			\State \(\bm{A}'\randgets\mathcal{R}_q^{p\times(\ell-p)}\); \(\bm{A}:=\begin{bmatrix}\bm{I}_p\ \bm{A}'\end{bmatrix}\); \(\bm{s}\gets\mathcal{D}_{\sigma_{\bm{s}}}^\ell\)
			\State \((\bm{s}_i)_{i\in[n]}\gets\mathsf{SEC.Deal}_\mathsf{pp}(\bm{s})\) \Comment{\eqref{Eq:SEC-Deal}}
			\State \(\bm{t}:=\bm{A}\bm{s}\); \(\bm{t}_i:=\bm{A}\bm{s}_i\) for \(i\in[n]\)
			\State \Return \(\mathsf{vk}:=(\bm{A},\bm{t})\), \(\mathsf{ak}:=(\bm{t}_i)_{i\in[n]}\), \((\mathsf{sk}_i:=\bm{s}_i)_{i\in[n]}\)
		\end{algorithmic}
		\smallskip
		\(\mathsf{PartialVerify}(\mathsf{tspar},\mathsf{vk},\mathsf{ak}_i,\mathsf{sid},\mathsf{act},\mathsf{msg},\mathsf{PM})\)
		\hrule height 0.4pt
		\begin{algorithmic}[1]
			\State Parse \(\mathsf{PM}\) for \((\bm{w}_j)_{j\in\mathsf{act}}\) and \(\bm{z}_i\).
			\State \(\bm{w}:=\mathsf{SEC.Rec}_\mathsf{pp}(\mathsf{act},(\bm{w}_j)_{j\in\mathsf{act}})\); \(c:=H_c(\mathsf{sid},\mathsf{vk},\mathsf{act},\mathsf{msg},\bm{w})\)
			\State \Return \(1\) iff \(\bm{A}\bm{z}_i-c\bm{t}_i=\bm{w}_i\) and \(\|\bm{z}_i\|_2\leq B_1\)
		\end{algorithmic}
		\smallskip
		\(\mathsf{Combine}(\mathsf{tspar},\mathsf{vk},\mathsf{ak},\mathsf{sid},\mathsf{act},\mathsf{msg},\mathsf{PM})\)
		\hrule height 0.4pt
		\begin{algorithmic}[1]
			\State Compute \(\bm{w}\) and \(c\) as in \(\mathsf{PartialVerify}\).
			\State \(S_\bot:=\{i\in\mathsf{act}:\mathsf{PartialVerify}(\dots,\mathsf{ak}_i,\dots)=0\}\)
			\State \textbf{if} \(S_\bot\neq\emptyset\) \textbf{then return} \((\mathsf{abort},S_\bot)\)
			\State \(\bm{z}:=\mathsf{SEC.Rec}_\mathsf{pp}(\mathsf{act},(\bm{z}_j)_{j\in\mathsf{act}})\); \Return \(\mathsf{sig}:=(c,\bm{z})\)
		\end{algorithmic}
		\smallskip
		\(\mathsf{Verify}(\mathsf{vk},\mathsf{msg},\mathsf{sig})\)
		\hrule height 0.4pt
		\begin{algorithmic}[1]
			\State Parse \((\mathsf{sid},\mathsf{act})\) from \(\mathsf{msg}\) and \((c,\bm{z}):=\mathsf{sig}\), returning \(0\) on failure; \(\bm{w}:=\bm{A}\bm{z}-c\bm{t}\).
			\State \Return \(1\) iff \(c=H_c(\mathsf{sid},\mathsf{vk},\mathsf{act},\mathsf{msg},\bm{w})\) and \(\|\bm{z}\|_2\leq B_2\)
		\end{algorithmic}
	\end{minipage}\hfill
	\vrule\hfill
	\begin{minipage}[t]{0.485\textwidth}
		\(\mathsf{ShareSign}_1(\mathsf{st}_i,\mathsf{sid},\mathsf{act},\mathsf{msg})\)
		\hrule height 0.4pt
		\begin{algorithmic}[1]
			\State \textbf{assert} \(\mathsf{sid}\) is fresh, \(i\in\mathsf{act}\), and \(\mathcal{P}(\mathsf{tspar},\mathsf{vk},\mathsf{sid},\mathsf{act},\mathsf{msg})=1\)
			\State \(\bm{r}_i\gets\mathcal{D}_{\sigma_{\bm{r}}}^\ell\); \(\bm{w}_i:=\bm{A}\bm{r}_i\)
			\State \(\mathsf{cmt}_i:=H_\mathsf{cmt}(\mathsf{sid},\mathsf{vk},\mathsf{act},i,\bm{w}_i)\)
			\State Store \((\mathsf{act},\mathsf{msg},\bm{r}_i,\bm{w}_i,\mathsf{cmt}_i)\) in \(\mathsf{st}_i.\mathsf{session}[\mathsf{sid}]\).
			\State \Return \(\mathsf{pm}_i^1:=\mathsf{cmt}_i\)
		\end{algorithmic}
		\smallskip
		\(\mathsf{ShareSign}_2(\mathsf{st}_i,\mathsf{sid},\mathsf{act},(\mathsf{cmt}_j)_{j\in\mathsf{act}})\)
		\hrule height 0.4pt
		\begin{algorithmic}[1]
			\State \textbf{assert} the session is in round \(1\) and its own entry equals \(\mathsf{cmt}_i\)
			\State Store \((\mathsf{cmt}_j)_{j\in\mathsf{act}}\).
			\State \Return \(\mathsf{pm}_i^2:=\bm{w}_i\)
		\end{algorithmic}
		\smallskip
		\(\mathsf{ShareSign}_3(\mathsf{st}_i,\mathsf{sid},\mathsf{act},(\bm{w}_j)_{j\in\mathsf{act}})\)
		\hrule height 0.4pt
		\begin{algorithmic}[1]
			\State \textbf{assert} the session is in round \(2\) and its own entry equals \(\bm{w}_i\)
			\For{\(j\in\mathsf{act}\) in canonical order}
				\If{\(\mathsf{cmt}_j\neq H_\mathsf{cmt}(\mathsf{sid},\mathsf{vk},\mathsf{act},j,\bm{w}_j)\)}
					\State \Return \((\mathsf{abort},\{j\})\) \Comment{identify signer \(j\)}
				\EndIf
			\EndFor
			\State \(\bm{w}:=\mathsf{SEC.Rec}_\mathsf{pp}(\mathsf{act},(\bm{w}_j)_{j\in\mathsf{act}})=\sum_{j\in\mathsf{act}}\lambda_{\mathsf{act},j}\bm{w}_j\)
			\State \(c:=H_c(\mathsf{sid},\mathsf{vk},\mathsf{act},\mathsf{msg},\bm{w})\)
			\State \Return \(\mathsf{pm}_i^3:=\bm{z}_i:=c\bm{s}_i+\bm{r}_i\)
		\end{algorithmic}
		\medskip
		\(\mathcal{P}(\mathsf{tspar},\mathsf{vk},\mathsf{sid},\mathsf{act},\mathsf{msg})\)
		\hrule height 0.4pt
		\begin{algorithmic}[1]
			\State Parse \((\mathsf{sid}',\mathsf{act}',\bm{\lambda}_\mathsf{act})\) from \(\mathsf{msg}\); \textbf{if} this fails or \((\mathsf{sid}',\mathsf{act}')\neq(\mathsf{sid},\mathsf{act})\) \textbf{then return} \(0\).
			\State \textbf{if} \(|\mathsf{act}|<t_\mathsf{sig}\) \textbf{then return} \(0\)
			\State \Return \(1\) iff \(\bm{\lambda}_\mathsf{act}^\top\mathbf{G}_\mathsf{act}=\mathbf{e}_1^\top\), \(\|\bm{\lambda}_\mathsf{act}\|_\infty\leq\beta_{\bm{\lambda}}\), and \(wt_H(\bm{\lambda}_\mathsf{act})\leq\omega_{\bm{\lambda}}\)
		\end{algorithmic}
	\end{minipage}
	\end{tcolorbox}
	\caption{The \(\mathsf{LaTSEC}\) protocol.
	\(\mathsf{PM}\) denotes the transcript \((\mathsf{pm}_j^k)_{k\in[3],j\in\mathsf{act}}=((\mathsf{cmt}_j,\bm{w}_j,\bm{z}_j))_{j\in\mathsf{act}}\), and \(\mathsf{SEC.Rec}_\mathsf{pp}(\mathsf{act},\cdot)\) uses the vector \(\bm{\lambda}_\mathsf{act}\) encoded in \(\mathsf{msg}\).
	Each \(\mathsf{st}_i\) holds \(\mathsf{tspar}\), \(\mathsf{vk}\), \(\mathsf{sk}_i\), and a session table, and \(\mathsf{vk}\) implicitly includes \(\mathsf{tspar}\).
    A failed assertion is a malformed invocation and returns \(\bot\) without blame; only the checks in \(\mathsf{ShareSign}_3\) and \(\mathsf{Combine}\) produce identifiable aborts, and a missing or unparsable peer message fails the check that uses it, so its sender is blamed.}
	\label{Figure:LaTSEC-Protocol}
\end{figure*}

\Cref{Figure:LaTSEC-Protocol} gives the protocol.
The complete stateful specification, with session states and all assertions, is in Appendix~\ref{Appendix:LaTSEC-Full-Specification}.
A trusted dealer runs \(\mathsf{Keygen}\), and the active set of an admissible session signs in three rounds.
In the first round, each signer samples a fresh mask \(\bm{r}_i\) and broadcasts a hash commitment to \(\bm{w}_i=\bm{A}\bm{r}_i\).
In the second round, it reveals \(\bm{w}_i\).
In the third round, it checks every opening against its commitment, aborts with the first inconsistent signer in the canonical order of \(\mathsf{act}\), and otherwise reconstructs the aggregate commitment \(\bm{w}\), derives the common challenge \(c\), and sends \(\bm{z}_i=c\bm{s}_i+\bm{r}_i\).
Since all honest parties receive the same transcript, they compute the same blame set.
The combiner runs \(\mathsf{PartialVerify}\) for every active signer, aborts with the set of signers whose responses fail, and otherwise outputs \((c,\bm{z})\) with \(\bm{z}=\sum_{j\in\mathsf{act}}\lambda_{\mathsf{act},j}\bm{z}_j\).
Partial verification needs only the transcript, \(\mathsf{vk}\), and the abort key \(\mathsf{ak}_i\); it requires neither interaction nor zero-knowledge proofs.
By \Cref{Item:SEC-Rec,Item:SEC-Linear} of the \(\mathsf{SEC}\) properties in \Cref{Section:SEC-Summary}, \(\mathsf{SEC.Rec}_\mathsf{pp}(\mathsf{act},(\bm{t}_j)_{j\in\mathsf{act}})=\bm{t}\), and the aggregation identity in \eqref{Eq:Aggregation-Identity} shows that a non-aborting combination of valid partial responses satisfies \(\bm{A}\bm{z}-c\bm{t}=\bm{w}\).

\section{Correctness and Identifiable Abort}\label{Section:LaTSEC-Correctness-Security}

\subsection{Parameter Conditions}\label{Section:LaTSEC-Asymptotic-Parameters}

The following conditions suffice for correctness, identifiable abort, and unforgeability.
All dimensions and the query bounds \(Q_H\) and \(Q_\mathsf{Sign}\) are polynomial in \(\kappa\).
\begin{enumerate}
	\item\label{Item:Cond-Dims} \textit{Dimensions.} \(d\geq\kappa\), \(\ell>p\), and \(q > 2\) is prime.
	\item\label{Item:Cond-Challenge} \textit{Challenges.} \(|\mathcal{C}|=2^{\omega_c}\binom{d}{\omega_c}\geq Q_H2^\kappa\); every \(c\in\mathcal{C}\) satisfies \(\|c\bm{u}\|_2\leq\omega_c\|\bm{u}\|_2\).
	\item\label{Item:Cond-Tail} \textit{Gaussian tails.} The tail factors \(\theta_{\bm{s}},\theta_\mathbf{R},\theta_{\bm{r}}\) satisfy \cite[Eqs.~(64)--(66)]{arzaki2026SEC} with query bound \(Q:=Q_\mathsf{Sign}\), and \(\beta_{\bm{s}}:=\theta_{\bm{s}}\sigma_{\bm{s}}\sqrt{D}\), \(\beta_\mathbf{R}:=\theta_\mathbf{R}\sigma_\mathbf{R}\), \(\beta_{\bm{r}}:=\theta_{\bm{r}}\sigma_{\bm{r}}\sqrt{D}\).
	Let \(\mathsf{Good}_\mathsf{tail}\) be the event that \(\|\bm{s}\|_2\leq\beta_{\bm{s}}\), every entry of \(\mathbf{R}\) is at most \(\beta_\mathbf{R}\) in absolute value, and every signing mask \(\bm{r}_{i,s}\) of an honest signer in any of the \(Q_\mathsf{Sign}\) signing queries satisfies \(\|\bm{r}_{i,s}\|_2\leq\beta_{\bm{r}}\).
	Then \(\Pr[\neg\mathsf{Good}_\mathsf{tail}]\leq3\cdot2^{-\kappa}\) by the tail bound and union bounds of \cite[Lemmas 35 and 36]{arzaki2026SEC},
    and under \(\mathsf{Good}_\mathsf{tail}\), since \(\|\bm{s}\|_\infty\leq\|\bm{s}\|_2\leq\beta_{\bm{s}}\), \(\|\bm{s}_i\|_\infty\leq\beta_\mathsf{sh}:=\beta_\mathbf{g}\beta_{\bm{s}}+c_\mathrm{exp}\beta_\mathbf{R}\) by \Cref{Item:SEC-Short} of the \(\mathsf{SEC}\) properties in \Cref{Section:SEC-Summary}.
	\item\label{Item:Cond-Lambda} \textit{Reconstruction.} Every admissible session input satisfies \(\|\bm{\lambda}_\mathsf{act}\|_\infty\leq\beta_{\bm{\lambda}}<q\) and \(wt_H(\bm{\lambda}_\mathsf{act})\leq\omega_{\bm{\lambda}}\), so \(\|\bm{\lambda}_\mathsf{act}\|_1\leq\omega_{\bm{\lambda}}\beta_{\bm{\lambda}}\).
	\item\label{Item:Cond-Verify} \textit{Verification bounds.} \(B_1\geq\omega_c\sqrt{D}\beta_\mathsf{sh}+\beta_{\bm{r}}\) and \(B_2\geq\omega_{\bm{\lambda}}\beta_{\bm{\lambda}}B_1\).
	\item\label{Item:Cond-Entropy} \textit{Commitment entropy.} \(\sigma_{\bm{r}}>2d\cdot q^{1/\ell+2/(d(\ell-p))}\) (\Cref{Lemma:LaTSEC-Systematic-MLWE-MinEntropy}).
	\item\label{Item:Cond-Privacy} \textit{Key privacy.} The width \(\sigma_\mathrm{MQ}\) of \cite[Eq.~(58)]{arzaki2026SEC} satisfies the kernel-smoothing and posterior-resampling conditions of \cite[Theorem 4]{arzaki2026SEC} with \(Q:=Q_\mathsf{Sign}\), and \(\mathsf{MLWE}_{p,\ell-p,q,d,\sigma_\mathrm{MQ}}\) and \(\mathsf{MLWE}_{p,\ell-p,q,d,\sigma_{\bm{s}}}\) are hard.
    Hence, \(\bm{t}\) is pseudorandom given the corrupted shares and all signing responses (\Cref{Item:SEC-Privacy} of the \(\mathsf{SEC}\) properties in \Cref{Section:SEC-Summary}).
	\item\label{Item:Cond-AK} \textit{Abort keys.} The width \(\sigma_\mathrm{AK}\) of \eqref{Eq:Abort-Key-Gaussian-Parameter} satisfies \(\sigma_\mathrm{AK}\geq\sqrt{2}\eta_\epsilon(\mathbb{Z}^{kD})\) for a negligible \(\epsilon\), and \(\mathsf{MLWE}_{p,\ell-p,q,d,\sigma_\mathrm{AK}}\) is hard.
	\item\label{Item:Cond-MSIS} \textit{Forgery.} \(\mathsf{MSIS}_{p,\ell-p+1,q,d,2B_\mathsf{STMSIS}}\) is hard for \(B_\mathsf{STMSIS}\geq\sqrt{B_2^2+\omega_c}\).
\end{enumerate}
\Cref{Item:Cond-Dims} and \Cref{Item:Cond-Tail,Item:Cond-Lambda,Item:Cond-Verify} above give correctness and identifiable abort.
Unforgeability uses \Cref{Item:Cond-Dims,Item:Cond-Challenge,Item:Cond-Lambda,Item:Cond-Entropy,Item:Cond-AK,Item:Cond-MSIS}.
\Cref{Item:Cond-Privacy} is not a hypothesis of the unforgeability theorem, but guarantees that \(\mathsf{LaTSEC}\) keeps the key privacy of \(\mathsf{SEC}\).

\subsection{Identifiable Abort and Correctness}\label{Section:LaTSEC-IA-Correctness}

Identifiable abort and correctness both follow from the event \(\mathsf{Good}_\mathsf{tail}\); the proofs are in Appendix~\ref{Appendix:LaTSEC-IA-Correctness}.

\begin{lemma}[Honest partial responses]\label{Lemma:Honest-Partial-Response-Bound}
	Under \(\mathsf{Good}_\mathsf{tail}\) and \Cref{Item:Cond-Verify} of the parameter conditions in \Cref{Section:LaTSEC-Asymptotic-Parameters}, every honest partial response \(\bm{z}_i=c\bm{s}_i+\bm{r}_i\) satisfies \(\|\bm{z}_i\|_2\leq\omega_c\sqrt{D}\beta_\mathsf{sh}+\beta_{\bm{r}}\leq B_1\) and \(\bm{A}\bm{z}_i-c\bm{t}_i=\bm{w}_i\).
\end{lemma}

\begin{theorem}[Identifiable abort]\label{Theorem:LaTSEC-TS-IA}
    Assume \Cref{Item:Cond-Dims} and \Cref{Item:Cond-Tail,Item:Cond-Lambda,Item:Cond-Verify} of the parameter conditions in \Cref{Section:LaTSEC-Asymptotic-Parameters}.
	Then, for every adversary \(\mathcal{A}\) in \(\mathsf{Game}_\mathsf{TS\text{-}IA}\), \(\mathsf{Adv}_\mathcal{A}^{\mathsf{TS\text{-}IA}}(\kappa)\leq3\cdot2^{-\kappa}\).
\end{theorem}

\begin{proof}[Proof sketch]
	Condition on \(\mathsf{Good}_\mathsf{tail}\); all honest parties parse the same \(\bm{\lambda}_\mathsf{act}\) from \(\mathsf{msg}\).
	An honest signer reveals the opening it committed to, so every signer blamed in \(\mathsf{ShareSign}_3\) is corrupted, and consistent delivery makes all honest parties blame the same signer.
	By \Cref{Lemma:Honest-Partial-Response-Bound}, every honest signer passes \(\mathsf{PartialVerify}\), so a blame set returned by \(\mathsf{Combine}\) is non-empty and contained in the corrupted set.
	If \(\mathsf{Combine}\) does not abort, every \(i\in\mathsf{act}\) satisfies \(\bm{A}\bm{z}_i-c\bm{t}_i=\bm{w}_i\) and \(\|\bm{z}_i\|_2\leq B_1\).
    By \Cref{Item:SEC-Rec,Item:SEC-Linear} of the \(\mathsf{SEC}\) properties in \Cref{Section:SEC-Summary}, \(\bm{A}\bm{z}-c\bm{t}=\sum_i\lambda_{\mathsf{act},i}(\bm{A}\bm{z}_i-c\bm{t}_i)=\bm{w}\), and \(\|\bm{z}\|_2\leq\|\bm{\lambda}_\mathsf{act}\|_1B_1\leq\omega_{\bm{\lambda}}\beta_{\bm{\lambda}}B_1\leq B_2\).
	Since \(\mathsf{msg}\) determines \(\mathsf{sid}\) and \(\mathsf{act}\), \(\mathsf{Verify}\) recomputes the same challenge and accepts.
	The adversary can therefore win only if \(\mathsf{Good}_\mathsf{tail}\) fails.
	The full proof is in Appendix~\ref{Appendix:LaTSEC-IA-Correctness}.
\end{proof}

\begin{corollary}[Correctness]\label{Corollary:LaTSEC-Correctness}
	Under the assumptions of \Cref{Theorem:LaTSEC-TS-IA}, an honest execution of an admissible session outputs a signature accepted by \(\mathsf{Verify}\), except with probability at most \(3\cdot2^{-\kappa}\).
\end{corollary}

\begin{proof}[Proof sketch]
	Apply \Cref{Theorem:LaTSEC-TS-IA} with \(\mathsf{corrupt}=\emptyset\): no non-empty blame set is contained in \(\emptyset\), so under \(\mathsf{Good}_\mathsf{tail}\) the execution neither aborts nor outputs an invalid signature.
	The full proof is in Appendix~\ref{Appendix:LaTSEC-IA-Correctness}.
\end{proof}

For correctness alone, the honest aggregate \(\bm{z}=c\bm{s}+\sum_i\lambda_{\mathsf{act},i}\bm{r}_i\) satisfies \(\|\bm{z}\|_2\leq\omega_c\beta_{\bm{s}}+\omega_{\bm{\lambda}}\beta_{\bm{\lambda}}\beta_{\bm{r}}\) under \(\mathsf{Good}_\mathsf{tail}\), so \(B_2\geq\omega_c\beta_{\bm{s}}+\omega_{\bm{\lambda}}\beta_{\bm{\lambda}}\beta_{\bm{r}}\) would suffice.
We use \(B_2\geq\omega_{\bm{\lambda}}\beta_{\bm{\lambda}}B_1\), which also covers corrupted responses that pass partial verification.

\section{Simulating the Public Abort Keys}\label{Section:LaTSEC-Abort-Key-Simulation}

The unforgeability proof must eventually replace the verification target \(\bm{t}=\bm{A}\bm{s}\) by a uniform value, so that a forgery can be turned into a \(\mathsf{SelfTargetMSIS}\) solution.
In \(\mathsf{LaTSEC}\), however, the adversary also sees the abort keys \(\bm{t}_i=\bm{A}\bm{s}_i\) of all parties, which must remain consistent with the corrupted shares and must reconstruct to the replaced target.
This section shows how to complete such a tuple from a uniform target (\Cref{Section:AK-Completion}), formalizes the resulting real-or-simulated experiment (\Cref{Section:HC-Leak-AK}), and proves the two views indistinguishable under \(\mathsf{MLWE}\) (\Cref{Section:Hidden-Coordinates,Section:AK-Indistinguishability}).
Throughout, \(\mathbf{A}\in\mathbb{Z}_q^{dp\times D}\) is the coefficient embedding of \(\bm{A}\), and \(\mathbf{X}:=\begin{bmatrix}\mathbf{s}^\top\\ \mathbf{R}\end{bmatrix}\in\mathbb{Z}^{k\times D}\) is the dealing matrix, so that \(\mathbf{S}=\mathbf{G}\mathbf{X}\).

\subsection{Abort-Key Completion and Simulation}\label{Section:AK-Completion}

Let \(\mathbf{t}:=\mathbf{A}\mathbf{s}\), \(\mathbf{t}_i:=\mathbf{A}\mathbf{s}_i\), and let \(\mathbf{T}\in\mathbb{Z}_q^{n\times dp}\) have rows \(\mathbf{t}_i^\top\).
Then
\begin{equation}\label{Eq:Induced-Abort-Key-Sharing}
	\mathbf{T}=\mathbf{S}\mathbf{A}^\top=\mathbf{G}\mathbf{Y}=\left(\mathbf{g}\mathbf{t}^\top+\mathbf{\hat{G}}\mathbf{R}_\mathbf{t}\right)\bmod q,
\end{equation}
where \(\mathbf{Y}:=\mathbf{X}\mathbf{A}^\top=\begin{bmatrix}\mathbf{t}^\top\\ \mathbf{R}_\mathbf{t}\end{bmatrix}\) and \(\mathbf{R}_\mathbf{t}:=\mathbf{R}\mathbf{A}^\top\bmod q\).
Thus the abort keys form an \(\mathsf{SEC}\)-form sharing of \(\mathbf{t}\) over \(\mathbb{Z}_q\), with the \textit{evaluated} matrix \(\mathbf{Y}\) in place of the dealing matrix.

\begin{lemma}[Algebraic completion of abort keys]\label{Lemma:Algebraic-Abort-Key-Completion}
	Let \(q\) be prime, let \(U\subseteq[n]\) with \(|U|\leq k_p\), let \(\mathbf{T}_U\in\mathbb{Z}_q^{|U|\times dp}\), and let \(\mathbf{\breve{t}}\in\mathbb{Z}_q^{dp}\).
	One can efficiently sample \(\mathbf{\breve{R}}_\mathbf{t}\) such that \(\mathbf{\breve{T}}:=(\mathbf{g}\mathbf{\breve{t}}^\top+\mathbf{\hat{G}}\mathbf{\breve{R}}_\mathbf{t})\bmod q\) satisfies \(\mathbf{\breve{T}}_U=\mathbf{T}_U\), and the rows of \(\mathbf{\breve{T}}_T\) reconstruct to \(\mathbf{\breve{t}}\) for every certified set \(T\).
\end{lemma}

\begin{proof}[Proof sketch]
	For \(U\neq\emptyset\), \Cref{Item:SEC-Unimodular} of \Cref{Section:SEC-Summary} gives a unimodular submatrix \(\mathbf{M}_U\) of \(\mathbf{\hat{G}}_U\); placing \(\mathbf{M}_U^{-1}\) in the selected rows yields an integer right inverse \(\mathbf{K}_U\) of \(\mathbf{\hat{G}}_U\).
	Set \(\mathbf{\breve{R}}_\mathbf{t}:=\mathbf{K}_U(\mathbf{T}_U-(\mathbf{g})_U\mathbf{\breve{t}}^\top)+\mathbf{N}\) for any \(\mathbf{N}\) with \(\mathbf{\hat{G}}_U\mathbf{N}=\mathbf{0}\bmod q\); then \(\mathbf{\breve{T}}_U=\mathbf{T}_U\).
	For certified \(T\), \(\bm{\lambda}_T^\top(\mathbf{g})_T=1\) and \(\bm{\lambda}_T^\top\mathbf{\hat{G}}_T=\mathbf{0}^\top\), so \(\bm{\lambda}_T^\top\mathbf{\breve{T}}_T=\mathbf{\breve{t}}^\top\).
	The full proof is in Appendix~\ref{Appendix:LaTSEC-Abort-Key-Simulation}.
\end{proof}

\begin{definition}[Abort-key simulator]\label{Definition:SimAK}
	For prime \(q\), \(U\subseteq[n]\) with \(|U|\leq k_p\), \(\mathbf{\breve{t}}\in\mathbb{Z}_q^{dp}\), and \(\mathbf{T}_U\in\mathbb{Z}_q^{|U|\times dp}\), the simulator \(\mathsf{SimAK}_\mathsf{pp}(U,\mathbf{\breve{t}},\mathbf{T}_U)\) samples \(\mathbf{\breve{R}}_\mathbf{t}\) uniformly from \(\{\mathbf{Z}\in\mathbb{Z}_q^{(k-1)\times dp}:\mathbf{\hat{G}}_U\mathbf{Z}=\mathbf{T}_U-(\mathbf{g})_U\mathbf{\breve{t}}^\top\bmod q\}\) and outputs \(\mathbf{\breve{T}}:=(\mathbf{g}\mathbf{\breve{t}}^\top+\mathbf{\hat{G}}\mathbf{\breve{R}}_\mathbf{t})\bmod q\).
\end{definition}

The particular solution of \Cref{Lemma:Algebraic-Abort-Key-Completion} plus a uniform element of the modular kernel of \(\mathbf{\hat{G}}_U\) gives an efficient uniform sampler.
Algebraic consistency does not imply indistinguishability: the real matrix \(\mathbf{R}_\mathbf{t}=\mathbf{R}\mathbf{A}^\top\) is structured, whereas \(\mathsf{SimAK}\) samples uniformly from an affine space.
The remainder of this section bridges this gap.

\subsection{The \texorpdfstring{\(\bm{\mathsf{HC\text{-}Leak\text{-}AK}}\)}{HC-Leak-AK} Experiment}\label{Section:HC-Leak-AK}

We extend the multi-query privacy experiment of \(\mathsf{SEC}\) \cite[Definition 9]{arzaki2026SEC} by the verification target and the complete abort-key tuple.
A \textit{leakage pattern} \(\mathcal{U}=(U_\mathrm{raw},(V_s)_{s\in[Q_\mathsf{Sign}]})\in\mathcal{U}_{k_p,Q_\mathsf{Sign}}\) \cite[Section 5.5.2]{arzaki2026SEC} consists of a set \(U_\mathrm{raw}\subseteq[n]\) with \(|U_\mathrm{raw}|\leq k_p\), whose raw shares are revealed, and sets \(V_s\subseteq[n]\) of parties whose responses are revealed in query \(s\).

\begin{definition}[\(\mathsf{HC\text{-}Leak\text{-}AK}\) experiment]\label{Definition:HC-Leak-AK}
	Fix a leakage pattern \(\mathcal{U}\) and challenges \(\mathbf{c}=(c_s)_{s\in[Q_\mathsf{Sign}]}\in\mathcal{C}^{Q_\mathsf{Sign}}\), and let \(\mathcal{S}:=\{(s,i):s\in[Q_\mathsf{Sign}],\ i\in V_s\}\).
	The experiment \(\mathsf{HC\text{-}Leak\text{-}AK}^{(\mathsf{pp},\mathcal{U},\mathbf{c})}_{\mathcal{B}}(\kappa)\) samples \(\bm{A}'\randgets\mathcal{R}_q^{p\times(\ell-p)}\), \(\bm{s}\gets\mathcal{D}_{\sigma_{\bm{s}}}^\ell\), \((\bm{s}_i)_{i\in[n]}\gets\mathsf{SEC.Deal}_\mathsf{pp}(\bm{s})\), and, for \((s,i)\in\mathcal{S}\), \(\bm{r}_{i,s}\gets\mathcal{D}_{\sigma_{\bm{r}}}^\ell\) and \(\bm{z}_{i,s}:=c_s\bm{s}_i+\bm{r}_{i,s}\).
	It sets the real view \((\bm{v}_1,\mathsf{ak}_1):=(\bm{t},(\bm{t}_i)_{i\in[n]})\) with \(\bm{t}=\bm{A}\bm{s}\) and \(\bm{t}_i=\bm{A}\bm{s}_i\), and the simulated view \((\bm{v}_0,\mathsf{ak}_0):=(\bm{\breve{t}},(\bm{\breve{t}}_i)_{i\in[n]})\), where \(\bm{\breve{t}}\randgets\mathcal{R}_q^p\) and \((\bm{\breve{t}}_i)_{i\in[n]}\) are the rows of \(\mathsf{SimAK}_\mathsf{pp}(U_\mathrm{raw},\mathbf{\breve{t}},\mathbf{T}_{U_\mathrm{raw}})\).
	It samples \(b\randgets\{0,1\}\) and gives \((\mathsf{pp},\bm{A},\mathcal{U},\mathbf{c},\bm{v}_b,(\bm{s}_i)_{i\in U_\mathrm{raw}},(\bm{z}_{i,s})_{(s,i)\in\mathcal{S}},\mathsf{ak}_b)\) to \(\mathcal{B}\), which outputs \(b'\).
	The advantage is \(\mathsf{Adv}^{(\mathsf{pp},\mathcal{U},\mathbf{c})}_{\mathcal{B},\mathsf{HC\text{-}Leak\text{-}AK}}(\kappa):=2|\Pr[b'=b]-1/2|\).
\end{definition}

Both branches reveal the same raw shares and responses; by \Cref{Lemma:Algebraic-Abort-Key-Completion}, the simulated tuple agrees with the real abort keys on \(U_\mathrm{raw}\) and reconstructs to \(\bm{\breve{t}}\).

\subsection{Hidden Coordinates}\label{Section:Hidden-Coordinates}

Write \(U:=U_\mathrm{raw}\) and \(h_U:=k-|U|\).
To isolate what the raw shares \(\mathbf{S}_U=\mathbf{G}_U\mathbf{X}\) reveal, we change variables in \(\mathbf{X}\) unimodularly.
For \(U\neq\emptyset\), let \(\mathbf{M}_U\) be the peeled unimodular submatrix of \(\mathbf{\hat{G}}_U\) (\Cref{Item:SEC-Unimodular} of \Cref{Section:SEC-Summary}), let \(\mathbf{P}_U\) be the column permutation with \(\mathbf{G}_U\mathbf{P}_U=\begin{bmatrix}\mathbf{M}_U\ \mathbf{N}_U\end{bmatrix}\), and define
\begin{equation}\label{Eq:Hidden-Coordinate-Transformation}
	\mathbf{Q}_U:=\mathbf{P}_U\begin{bmatrix}\mathbf{M}_U^{-1} & -\mathbf{M}_U^{-1}\mathbf{N}_U\\ \mathbf{O} & \mathbf{I}_{h_U}\end{bmatrix}\in\mathsf{GL}_k(\mathbb{Z}),
\end{equation}
so that \(\mathbf{G}_U\mathbf{Q}_U=\begin{bmatrix}\mathbf{I}_{|U|}\ \mathbf{O}\end{bmatrix}\); let \(\mathbf{Q}_\emptyset:=\mathbf{I}_k\).
Then \(\mathbf{X}':=\mathbf{Q}_U^{-1}\mathbf{X}=\begin{bmatrix}\mathbf{X}'_\mathsf{pub}\\ \mathbf{X}'_\mathsf{hid}\end{bmatrix}\) satisfies \(\mathbf{X}'_\mathsf{pub}=\mathbf{S}_U\): the raw shares reveal exactly the public coordinates, and \(\mathbf{X}'_\mathsf{hid}\in\mathbb{Z}^{h_U\times D}\) is hidden.
Consequently \(\mathbf{Y}=\mathbf{Q}_U\begin{bmatrix}\mathbf{S}_U\mathbf{A}^\top\\ \mathbf{X}'_\mathsf{hid}\mathbf{A}^\top\end{bmatrix}\), and the only hidden part of the abort-key view is the evaluated matrix \(\mathbf{X}'_\mathsf{hid}\mathbf{A}^\top\).
Replacing it by a uniform matrix yields a uniform element of \(\{\mathbf{Z}:\mathbf{G}_U\mathbf{Z}=\mathbf{T}_U\}\), which is exactly the distribution of \(\mathsf{SimAK}\).

Let \(\mathbf{x}'_\mathsf{hid}:=\mathrm{vec}({\mathbf{X}'_\mathsf{hid}}^\top)\in\mathbb{Z}^{h_UD}\).
After subtracting the contribution of \(\mathbf{S}_U\), each response becomes a noisy linear image of \(\mathbf{x}'_\mathsf{hid}\), and the stacked centered responses are \(\mathbf{\widetilde{z}}_\mathcal{S}=\mathbf{M}_\mathsf{hid}\mathbf{x}'_\mathsf{hid}+\mathbf{r}_\mathcal{S}\) for a public hint matrix \(\mathbf{M}_\mathsf{hid}\) built from \(\mathbf{G}\mathbf{Q}_U\) and the challenges.
With \(\mathbf{A}^{\langle h_U\rangle}:=\mathbf{I}_{h_U}\otimes\mathbf{A}\), the real and simulated hidden views are
\begin{gather*}
	\left(\mathbf{A},\mathbf{A}^{\langle h_U\rangle}\mathbf{x}'_\mathsf{hid},\mathbf{M}_\mathsf{hid},\mathbf{\widetilde{z}}_\mathcal{S}\right)\ \text{and}\\
	\left(\mathbf{A},\mathbf{u}_\mathsf{hid},\mathbf{M}_\mathsf{hid},\mathbf{\widetilde{z}}_\mathcal{S}\right),\quad
	\mathbf{u}_\mathsf{hid}\randgets\mathbb{Z}_q^{h_Udp},
\end{gather*}
which is an instance of \(\mathsf{RA\text{-}MHMLWE}[h_U]\) with one stacked hint (see Appendix~\ref{Appendix:LaTSEC-Abort-Key-View-Details}).

It remains to describe the secret \(\mathbf{x}'_\mathsf{hid}\) given \(\mathbf{S}_U\).
Each column of \(\mathbf{X}\) is a discrete Gaussian with parameter matrix \(\mathbf{\Sigma}_\mathbf{X}:=\mathrm{diag}(\sigma_{\bm{s}}^2,\sigma_\mathbf{R}^2\mathbf{I}_{k-1})\). After the unimodular change of variables and conditioning on \(\mathbf{X}'_\mathsf{pub}=\mathbf{S}_U\), the hidden coordinates are a discrete Gaussian whose center is determined by \(\mathbf{S}_U\) and whose parameter matrix \(\mathbf{\Sigma}_{U,\mathsf{hid}\mid\mathsf{pub}}\) is a Schur complement.

\begin{lemma}[Conditional precision of the hidden coordinates]\label{Lemma:Conditional-Hidden-Coordinate-Precision}
	Let \(\mathbf{Q}_{U,\mathsf{hid}}\) be the last \(h_U\) columns of \(\mathbf{Q}_U\).
	Then \(\mathbf{\Sigma}_{U,\mathsf{hid}\mid\mathsf{pub}}^{-1}=\mathbf{Q}_{U,\mathsf{hid}}^\top\mathbf{\Sigma}_\mathbf{X}^{-1}\mathbf{Q}_{U,\mathsf{hid}}\), and hence \(\|\mathbf{\Sigma}_{U,\mathsf{hid}\mid\mathsf{pub}}^{-1}\|_2\leq\|\mathbf{\Sigma}_\mathbf{X}^{-1}\|_2\|\mathbf{Q}_U\|_2^2\).
\end{lemma}

The proof identifies the conditional precision with the hidden principal block of the transformed precision \(\mathbf{Q}_U^\top\mathbf{\Sigma}_\mathbf{X}^{-1}\mathbf{Q}_U\) (see Appendix~\ref{Appendix:LaTSEC-Abort-Key-Simulation}).
To make this bound uniform over all coalitions, note that \eqref{Eq:Hidden-Coordinate-Transformation} gives
\begin{equation}\label{Eq:Hidden-Coordinate-Transformation-Pattern-Bound}
	\|\mathbf{Q}_U\|_2\leq1+\frac{\sqrt{1+\|\mathbf{N}_U\|_2^2}}{\sigma_{\min}(\mathbf{M}_U)}.
\end{equation}
Let \(\delta_{k_p}:=\max_{1\leq|U|\leq k_p}\|\mathbf{G}_U\|_2\) and \(\gamma_{k_p}:=\min_{1\leq|U|\leq k_p}\sigma_{\min}(\mathbf{M}_U)>0\).
Since \(\|\mathbf{N}_U\|_2\leq\|\mathbf{G}_U\|_2\), the quantity \(\beta_\mathbf{Q}^*:=\max_{|U|\leq k_p}\|\mathbf{Q}_U\|_2\) satisfies
\begin{equation}\label{Eq:Uniform-Hidden-Coordinate-Transformation-Bound}
	\beta_\mathbf{Q}^*\leq1+\frac{\sqrt{1+\delta_{k_p}^2}}{\gamma_{k_p}}.
\end{equation}
Both \(\delta_{k_p}\) and \(\gamma_{k_p}\) are computed exhaustively for the selected instances (\Cref{Table:LaTSEC-Empirical-Matrix-Parameters-40-60}).
Below, \(\beta_\mathbf{Q}\) denotes any upper bound on \(\beta_\mathbf{Q}^*\).
The concrete instantiations take the right-hand side of \eqref{Eq:Uniform-Hidden-Coordinate-Transformation-Bound}.

\subsection{Abort-Key Indistinguishability}\label{Section:AK-Indistinguishability}

The responses add precision through the hints.
Since the masks are independent with parameter \(\sigma_{\bm{r}}^2\mathbf{I}_D\), every challenge satisfies \(\|\mathbf{L}_s\|_2\leq\omega_c\) for its multiplication matrix \(\mathbf{L}_s\), and the relevant rows of \(\mathbf{G}\mathbf{Q}_U\) have spectral norm at most \(\|\mathbf{G}\|_2\beta_\mathbf{Q}\), a Kronecker-product calculation (see Appendix~\ref{Appendix:LaTSEC-Abort-Key-Simulation}) gives
\begin{equation}\label{Eq:Abort-Key-Hint-Precision-Uniform}
	\left\|\mathbf{M}_\mathsf{hid}^\top\mathbf{\Sigma}_{\mathbf{r}_\mathcal{S}}^{-1}\mathbf{M}_\mathsf{hid}\right\|_2\leq\frac{\omega_c^2Q_\mathsf{Sign}\|\mathbf{G}\|_2^2\beta_\mathbf{Q}^2}{\sigma_{\bm{r}}^2}.
\end{equation}
With \(\|\mathbf{\Sigma}_\mathbf{X}^{-1}\|_2=\max\{\sigma_{\bm{s}}^{-2},\sigma_\mathbf{R}^{-2}\}\) and \Cref{Lemma:Conditional-Hidden-Coordinate-Precision}, the combined precision is at most
\begin{equation}\label{Eq:Abort-Key-Combined-Precision}
	\beta_\mathbf{Q}^2\left(\max\left\{\sigma_{\bm{s}}^{-2},\sigma_\mathbf{R}^{-2}\right\}+\frac{\omega_c^2Q_\mathsf{Sign}\|\mathbf{G}\|_2^2}{\sigma_{\bm{r}}^2}\right),
\end{equation}
uniformly over all leakage patterns and challenge sequences.
We therefore define \(\sigma_\mathrm{AK}>0\) by
\begin{equation}\label{Eq:Abort-Key-Gaussian-Parameter}
	\frac{1}{\sigma_\mathrm{AK}^2}:=2\beta_\mathbf{Q}^2\left(\max\left\{\sigma_{\bm{s}}^{-2},\sigma_\mathbf{R}^{-2}\right\}+\frac{\omega_c^2Q_\mathsf{Sign}\|\mathbf{G}\|_2^2}{\sigma_{\bm{r}}^2}\right),
\end{equation}
so that, for every pattern,
\begin{equation}\label{Eq:Abort-Key-Precision-via-Sigma-AK}
	\left\|\mathbf{\Sigma}_{U,\mathsf{hid}\mid\mathsf{pub}}^{-1}\otimes\mathbf{I}_D+\mathbf{M}_\mathsf{hid}^\top\mathbf{\Sigma}_{\mathbf{r}_\mathcal{S}}^{-1}\mathbf{M}_\mathsf{hid}\right\|_2\leq\frac{1}{2\sigma_\mathrm{AK}^2},
\end{equation}
which is the precision condition of \Cref{Lemma:Hardness-RA-MHMLWE}.

\begin{lemma}[Indistinguishability of the abort-key view]\label{Lemma:Computational-Indistinguishability-Abort-Key-View}
	Fix a leakage pattern \(\mathcal{U}\) with \(h_U:=k-|U_\mathrm{raw}|\) and challenges \(\mathbf{c}\in\mathcal{C}^{Q_\mathsf{Sign}}\).
	If \(\sigma_\mathrm{AK}\geq\sqrt{2}\eta_\epsilon(\mathbb{Z}^{kD})\) for some \(0<\epsilon\leq1/\sqrt{2kD+4}\), then for every PPT \(\mathcal{B}\) there is a PPT \(\mathcal{B}'\) with
	\begin{equation}\label{Eq:Abort-Key-View-Advantage-Bound}
		\mathsf{Adv}^{(\mathsf{pp},\mathcal{U},\mathbf{c})}_{\mathcal{B},\mathsf{HC\text{-}Leak\text{-}AK}}(\kappa)\leq h_U\cdot\mathsf{Adv}_{\mathcal{B}'}^{\mathsf{MLWE}_{p,\ell-p,q,d,\sigma_\mathrm{AK}}}(\kappa)+2\epsilon.
	\end{equation}
\end{lemma}

\begin{proof}[Proof sketch]
	Condition on \(\mathbf{S}_U\).
	By the representation above, the two branches differ only in whether the hidden target is \(\mathbf{A}^{\langle h_U\rangle}\mathbf{x}'_\mathsf{hid}\) or uniform, and the representation is a bijection, so it adds no loss.
	The conditional secret is an elliptic discrete Gaussian whose center is a public function of \(\mathbf{S}_U\), and \eqref{Eq:Abort-Key-Precision-via-Sigma-AK} and \(h_U\leq k\) give the precision and smoothing conditions of \Cref{Lemma:Hardness-RA-MHMLWE} with \(\mathbf{S}_U\) as auxiliary information.
	That lemma bounds the \(\mathsf{RA\text{-}MHMLWE}[h_U]\) advantage by that against \(\mathsf{MLWE}_{p,\ell-p,q,d,\sigma_\mathrm{AK}}\), with loss \(h_U\) and statistical error \(2\epsilon\).
	Since the conditional parameter matrix and the precision bound do not depend on the value of \(\mathbf{S}_U\), the bound holds for the joint distribution.
	The full proof is in Appendix~\ref{Appendix:LaTSEC-Abort-Key-Simulation}.
\end{proof}

The condition \(\sigma_\mathrm{AK}\geq\sqrt{2}\eta_\epsilon(\mathbb{Z}^{kD})\) covers every coalition at once, because \(h_UD\leq kD\).
In the concrete instantiations, this condition and the hardness of \(\mathsf{MLWE}_{\sigma_\mathrm{AK}}\) determine \(\sigma_{\bm{s}}=\sigma_\mathbf{R}\) (\Cref{Section:SecurityEstimates-ParameterSets}).

\section{Unforgeability}\label{Section:LaTSEC-Unforgeability}

We prove unforgeability through a sequence of hybrids \(\mathsf{Hyb}_0,\dots,\mathsf{Hyb}_5\), following the strategy of \cite{del2024shortly}, and then reduce a forgery in \(\mathsf{Hyb}_5\) to \(\mathsf{SelfTargetMSIS}\).
Two steps are specific to \(\mathsf{LaTSEC}\): the challenge is programmed at an aggregate commitment reconstructed through \(\mathsf{SEC.Rec}_\mathsf{pp}\), which requires a new min-entropy argument, and the abort keys are replaced by the simulation of \Cref{Section:LaTSEC-Abort-Key-Simulation}.
We write \(\mathsf{Hyb}_j\) for the \(j\)-th hybrid and \(\Pr[\mathsf{Hyb}_j]\) for the probability that the adversary wins it.
The complete experiments are in Appendix~\ref{Appendix:LaTSEC-Unforgeability-Hybrids}.

\subsection{Hybrid Argument}\label{Section:LaTSEC-Initial-Hybrids}

\paragraph{\textbf{\(\bm{\mathsf{Hyb}_0}\)}}

This is \(\mathsf{Game}_\mathsf{TS\text{-}UF}\) (\Cref{Figure:Game-TS-UF}) for \(\mathsf{LaTSEC}\), so \(\Pr[\mathsf{Hyb}_0]=\mathsf{Adv}_\mathcal{A}^\mathsf{TS\text{-}UF}(\kappa)\).

\paragraph{\textbf{\(\bm{\mathsf{Hyb}_1}\): random honest commitments}}

For every honest active signer, the challenger samples \(\mathsf{cmt}_i\randgets\{0,1\}^{2\kappa}\) in round \(1\) and programs \(H_\mathsf{cmt}(\mathsf{sid},\mathsf{vk},\mathsf{act},i,\bm{w}_i):=\mathsf{cmt}_i\) before revealing \(\bm{w}_i\).
The views differ only if the adversary queried that point earlier.
By \Cref{Lemma:LaTSEC-Systematic-MLWE-MinEntropy}, \(\bm{w}_i=\bm{A}\bm{r}_i\) has min-entropy at least \(d-1\) except with probability \(2^{-d+1}\) over \(\bm{A}'\), so a union bound gives
\begin{equation}\label{Eq:Hybrid0-Hybrid1}
	|\Pr[\mathsf{Hyb}_0]-\Pr[\mathsf{Hyb}_1]|\leq n(Q_H+1)Q_\mathsf{Sign}\cdot2^{-d+1}.
\end{equation}

\paragraph{\textbf{\(\bm{\mathsf{Hyb}_2}\): extracted corrupted openings}}

In round \(2\), the challenger looks up openings \(\widehat{\bm{w}}_j\) of the corrupted commitments in the \(H_\mathsf{cmt}\)-table, reconstructs \(\bm{w}\) from the honest and extracted openings, computes \(c:=H_c(\mathsf{sid},\mathsf{vk},\mathsf{act},\mathsf{msg},\bm{w})\), and precomputes the honest responses, which it returns only in round \(3\).
The views differ only if a corrupted signer later opens to a value that was not extracted, which requires a preimage of a fixed \(2\kappa\)-bit value or a collision:
\begin{equation}\label{Eq:Hybrid1-Hybrid2}
	|\Pr[\mathsf{Hyb}_1]-\Pr[\mathsf{Hyb}_2]|\leq(Q_H^2+t_\mathsf{uf}Q_\mathsf{Sign}Q_H)\cdot2^{-2\kappa}.
\end{equation}

\paragraph{\textbf{\(\bm{\mathsf{Hyb}_3}\): programmed challenges}}

The challenger instead samples \(c\randgets\mathcal{C}\) and programs \(H_c(\mathsf{sid},\mathsf{vk},\mathsf{act},\mathsf{msg},\bm{w}):=c\).
This changes the view only if the adversary queried that point before round \(2\), so \(\bm{w}\) must have high min-entropy given the adversary's view.
Unlike in a single signer, \(\bm{w}=\sum_{i\in\mathsf{act}}\lambda_{\mathsf{act},i}\bm{w}_i\) is a reconstruction in which the corrupted openings are adversarial.
Since no set of at most \(t_\mathsf{uf}=k_p\) parties is certified (\Cref{Item:SEC-Unimodular} of \Cref{Section:SEC-Summary}), some honest \(i^*\) has \(\lambda_{\mathsf{act},i^*}\neq0\) (\Cref{Lemma:Honest-Nonzero-Reconstruction-Coefficient}), and since \(q\) is prime and \(\beta_{\bm{\lambda}}<q\), multiplication by \(\lambda_{\mathsf{act},i^*}\) is a bijection modulo \(q\).
This gives the following corollary of \Cref{Lemma:Reconstructed-Commitment-MinEntropy-General}, proved in Appendix~\ref{Appendix:LaTSEC-Unforgeability-MinEntropy}.

\begin{corollary}[Min-entropy of the aggregate commitment]\label{Corollary:Reconstructed-Commitment-MinEntropy-Prime}
	Let \(q\) be prime, let \(\bm{A}=\begin{bmatrix}\bm{I}_p\ \bm{A}'\end{bmatrix}\) with \(\bm{A}'\randgets\mathcal{R}_q^{p\times(\ell-p)}\), and let \(\sigma_{\bm{r}}>2d\cdot q^{1/\ell+2/(d(\ell-p))}\).
	Let \(\mathsf{act}\) be certified with \(\|\bm{\lambda}_\mathsf{act}\|_\infty\leq\beta_{\bm{\lambda}}<q\), let \(|\mathsf{corrupt}|\leq t_\mathsf{uf}\), let \(\bm{w}_i=\bm{A}\bm{r}_i\) with \(\bm{r}_i\gets\mathcal{D}_{\sigma_{\bm{r}}}^\ell\) for \(i\in\mathsf{act}\setminus\mathsf{corrupt}\), and let the adversary's view be independent of these masks and determine the corrupted openings.
	Then there is \(i^*\in\mathsf{act}\setminus\mathsf{corrupt}\) such that, conditioned on that view and on all randomness except \(\bm{r}_{i^*}\), \(\bm{w}:=\mathsf{SEC.Rec}_\mathsf{pp}(\mathsf{act},(\bm{w}_i)_{i\in\mathsf{act}})\) satisfies \(\Pr_{\bm{A}'}[H_\infty(\bm{w})\geq d-1]\geq1-2^{-d+1}\).
\end{corollary}

A union bound over the random-oracle and signing queries then gives
\begin{equation}\label{Eq:Hybrid2-Hybrid3}
	|\Pr[\mathsf{Hyb}_2]-\Pr[\mathsf{Hyb}_3]|\leq(Q_H+1)Q_\mathsf{Sign}\cdot2^{-d+1}.
\end{equation}

\paragraph{\textbf{\(\bm{\mathsf{Hyb}_4}\): openings from responses}}

The challenger computes each honest opening as \(\bm{w}_i:=\bm{A}\bm{z}_i-c\bm{t}_i\).
Since \(\bm{A}\bm{z}_i-c\bm{t}_i=\bm{A}\bm{r}_i\), nothing changes, and
\begin{equation}\label{Eq:Hybrid3-Hybrid4}
	\Pr[\mathsf{Hyb}_4]=\Pr[\mathsf{Hyb}_3].
\end{equation}
This expresses the honest openings through the responses and the abort keys, which prepares the replacement of the keys in \(\mathsf{Hyb}_5\).

\paragraph{\textbf{\(\bm{\mathsf{Hyb}_5}\): simulated keys}}

The challenger samples \(\bm{\breve{t}}\randgets\mathcal{R}_q^p\), computes \((\bm{\breve{t}}_i)_{i\in[n]}\) with \(\mathsf{SimAK}_\mathsf{pp}(\mathsf{corrupt},\mathbf{\breve{t}},\mathbf{T}_\mathsf{corrupt})\), publishes \((\bm{A},\bm{\breve{t}})\) and \((\bm{\breve{t}}_i)_{i\in[n]}\) instead of the real keys, and sets \(\bm{w}_i:=\bm{A}\bm{z}_i-c\bm{\breve{t}}_i\); the honest responses \(\bm{z}_i=c\bm{s}_i+\bm{r}_i\) are unchanged.
Consider the fixed pattern \(\mathcal{U}_\mathsf{UF}:=(\mathsf{corrupt},(\mathsf{honest})_{s\in[Q_\mathsf{Sign}]})\).
Given an \(\mathsf{HC\text{-}Leak\text{-}AK}\) challenge for \(\mathcal{U}_\mathsf{UF}\), a reduction \(\mathcal{B}_\mathrm{AK}\) keeps, in signing query \(s\), the response blocks of \(\mathsf{honest}\cap\mathsf{act}_s\), discards the others, and simulates everything else as in \(\mathsf{Hyb}_4\); the challenges are sampled inside the experiment and programmed as in \(\mathsf{Hyb}_3\).
It never uses honest shares or masks: it assigns the leakage slots to sessions in the order in which their corrupted openings are extracted, uses \(c_s\) and the leaked responses as the challenge and \(\bm{z}_i\) of slot \(s\), and sets \(\bm{w}_i:=\bm{A}\bm{z}_i-c_s\bm{t}_i\) with the abort keys it receives.
If extraction fails, it opens to \(\bm{A}\bm{r}_i\) for fresh masks \(\bm{r}_i\), and the session then aborts in round \(3\) or on \(\mathsf{Bad}_\mathsf{open}\), so no response is needed.
The real branch yields \(\mathsf{Hyb}_4\) and the simulated branch yields \(\mathsf{Hyb}_5\).
Since \eqref{Eq:Abort-Key-Combined-Precision} is uniform over \(\mathbf{c}\in\mathcal{C}^{Q_\mathsf{Sign}}\) and \(h_\mathsf{corrupt}\leq k\), \Cref{Lemma:Computational-Indistinguishability-Abort-Key-View} gives, under \(\sigma_\mathrm{AK}\geq\sqrt{2}\eta_\epsilon(\mathbb{Z}^{kD})\),
\begin{align}
	&|\Pr[\mathsf{Hyb}_4]-\Pr[\mathsf{Hyb}_5]|\leq\mathsf{Adv}_{\mathcal{B}_\mathrm{AK}}^\mathsf{HC\text{-}Leak\text{-}AK}(\kappa)\label{Eq:Hybrid4-Hybrid5-via-HCLeakAK}\\
	&\qquad\leq k\cdot\mathsf{Adv}_{\mathcal{B}'}^{\mathsf{MLWE}_{\sigma_\mathrm{AK}}}(\kappa)+2\epsilon,\label{Eq:Hybrid4-Hybrid5}
\end{align}
where \(\mathsf{MLWE}_{\sigma_\mathrm{AK}}:=\mathsf{MLWE}_{p,\ell-p,q,d,\sigma_\mathrm{AK}}\).
By \Cref{Lemma:Algebraic-Abort-Key-Completion}, the simulated keys agree with \(\bm{A}\bm{s}_i\) for \(i\in\mathsf{corrupt}\) and reconstruct to \(\bm{\breve{t}}\) from every certified set, while \(\bm{\breve{t}}\) is uniform and independent of the signing key.

\subsection{Reduction and Main Unforgeability Theorem}\label{Section:LaTSEC-STMSIS-Reduction}

In \(\mathsf{Hyb}_5\), \(\bm{A}_\mathsf{ST}:=\begin{bmatrix}\bm{A}'\ -\bm{\breve{t}}\end{bmatrix}\) is uniform.
A valid forgery \((c^*,\bm{z}^*)\) satisfies \(\begin{bmatrix}\bm{I}_p\ \bm{A}_\mathsf{ST}\end{bmatrix}\begin{bmatrix}\bm{z}^*\\ c^*\end{bmatrix}=\bm{A}\bm{z}^*-c^*\bm{\breve{t}}=\bm{w}^*\), and \(c^*=H_c(\mathsf{sid}^*,\mathsf{vk},\mathsf{act}^*,\mathsf{msg}^*,\bm{w}^*)\).

\begin{lemma}[From \(\mathsf{Hyb}_5\) to \(\mathsf{SelfTargetMSIS}\)]\label{Lemma:Hybrid5-to-STMSIS}
	Let \(B_\mathsf{STMSIS}\geq\sqrt{B_2^2+\omega_c}\).
	For every PPT \(\mathcal{A}\), there is a PPT \(\mathcal{B}_\mathsf{STMSIS}\) against \(\mathsf{SelfTargetMSIS}_{p,\ell-p+1,q,d,\mathcal{C},B_\mathsf{STMSIS}}\) with \(\Pr[\mathsf{Hyb}_5]\leq\mathsf{Adv}_{\mathcal{B}_\mathsf{STMSIS}}^\mathsf{STMSIS}(\kappa)\).
\end{lemma}

\begin{proof}[Proof sketch]
	\(\mathcal{B}_\mathsf{STMSIS}\) embeds its challenge as \(\bm{A}_\mathsf{ST}\), samples \(\bm{s}\) and the shares itself, completes the abort keys with \(\mathsf{SimAK}\), and answers every unprogrammed \(H_c\)-query \((\mathsf{sid},\mathsf{vk},\mathsf{act},\mathsf{msg},\bm{w})\) with \(\mathsf{G}(\bm{w},\mathsf{Encode}(\mathsf{sid},\mathsf{vk},\mathsf{act},\mathsf{msg}))\) for a fixed injective encoding \(\mathsf{Encode}\).
	By freshness, the forgery point was not programmed, so \((\mathsf{Encode}(\mathsf{sid}^*,\mathsf{vk},\mathsf{act}^*,\mathsf{msg}^*),\begin{bmatrix}\bm{z}^*\\ c^*\end{bmatrix})\) is a solution with \(0<\|(\bm{z}^*,c^*)\|_2^2\leq B_2^2+\omega_c\).
	The full proof is in Appendix~\ref{Appendix:LaTSEC-Unforgeability-Hybrids}.
\end{proof}

\begin{theorem}[Unforgeability]\label{Theorem:LaTSEC-TS-UF}
	Let \(q\) be prime, \(\beta_{\bm{\lambda}}<q\), and \(\sigma_{\bm{r}}>2d\cdot q^{1/\ell+2/(d(\ell-p))}\).
	Let \(\mathcal{A}\) make at most \(Q_H\) random-oracle and \(Q_\mathsf{Sign}\) signing queries and corrupt at most \(t_\mathsf{uf}=k_p\) parties in \(\mathsf{Game}_\mathsf{TS\text{-}UF}\), where \(Q_H\) also counts the random-oracle evaluations made by the game, such as opening checks and the final verification.
	Then there are PPT adversaries \(\mathcal{B}_\mathrm{AK}\) for \(\mathsf{HC\text{-}Leak\text{-}AK}\) under \(\mathcal{U}_\mathsf{UF}\) and \(\mathcal{B}_\mathsf{STMSIS}\) against \(\mathsf{SelfTargetMSIS}_{p,\ell-p+1,q,d,\mathcal{C},B_\mathsf{STMSIS}}\) with \(B_\mathsf{STMSIS}\geq\sqrt{B_2^2+\omega_c}\), both running in time about \(\mathsf{Time}(\mathcal{A})\), such that
	\begin{equation}\label{Eq:LaTSEC-TS-UF-Modular-Bound}
		\begin{aligned}
			\mathsf{Adv}_\mathcal{A}^\mathsf{TS\text{-}UF}(\kappa)\leq\;&(n+1)(Q_H+1)Q_\mathsf{Sign}\cdot2^{-d+1}\\
			&+(Q_H^2+t_\mathsf{uf}Q_\mathsf{Sign}Q_H)\cdot2^{-2\kappa}\\
			&+\mathsf{Adv}_{\mathcal{B}_\mathrm{AK}}^\mathsf{HC\text{-}Leak\text{-}AK}(\kappa)+\mathsf{Adv}_{\mathcal{B}_\mathsf{STMSIS}}^\mathsf{STMSIS}(\kappa).
		\end{aligned}
	\end{equation}
\end{theorem}

\begin{proof}
	Combine \eqref{Eq:Hybrid0-Hybrid1}--\eqref{Eq:Hybrid4-Hybrid5-via-HCLeakAK} with \Cref{Lemma:Hybrid5-to-STMSIS}.
\end{proof}

\begin{corollary}[Unforgeability under \(\mathsf{MLWE}\) and \(\mathsf{MSIS}\)]\label{Corollary:LaTSEC-TS-UF-MLWE-MSIS}
	In the setting of \Cref{Theorem:LaTSEC-TS-UF}, suppose that \(\sigma_\mathrm{AK}\geq\sqrt{2}\eta_\epsilon(\mathbb{Z}^{kD})\) for some \(0<\epsilon\leq1/\sqrt{2kD+4}\).
	Let \(\mathsf{MSIS}_{2B_\mathsf{STMSIS}}:=\mathsf{MSIS}_{p,\ell-p+1,q,d,2B_\mathsf{STMSIS}}\).
	Then there are PPT adversaries \(\mathcal{B}_\mathsf{MLWE}\) and \(\mathcal{B}_\mathsf{MSIS}\), with \(\mathsf{Time}(\mathcal{B}_\mathsf{MLWE})\approx\mathsf{Time}(\mathcal{A})\) and \(\mathsf{Time}(\mathcal{B}_\mathsf{MSIS})\approx2\,\mathsf{Time}(\mathcal{A})\), such that
	\begin{equation}\label{Eq:LaTSEC-TS-UF-MLWE-MSIS}
		\begin{aligned}
			\mathsf{Adv}_\mathcal{A}^\mathsf{TS\text{-}UF}(\kappa)\leq\;&(n+1)(Q_H+1)Q_\mathsf{Sign}\cdot2^{-d+1}\\
			&+(Q_H^2+t_\mathsf{uf}Q_\mathsf{Sign}Q_H)\cdot2^{-2\kappa}+2\epsilon\\
			&+k\cdot\mathsf{Adv}_{\mathcal{B}_\mathsf{MLWE}}^{\mathsf{MLWE}_{\sigma_\mathrm{AK}}}(\kappa)\\
			&+\sqrt{Q_H\cdot\mathsf{Adv}_{\mathcal{B}_\mathsf{MSIS}}^{\mathsf{MSIS}_{2B_\mathsf{STMSIS}}}(\kappa)}+\frac{Q_H}{|\mathcal{C}|}.
		\end{aligned}
	\end{equation}
	In particular, \(\mathsf{LaTSEC}\) is unforgeable if these \(\mathsf{MLWE}\) and \(\mathsf{MSIS}\) problems are hard and \(\epsilon\) is negligible.
\end{corollary}

\begin{proof}
	Substitute \eqref{Eq:Hybrid4-Hybrid5} and \Cref{Lemma:Hardness-STMSIS} into \eqref{Eq:LaTSEC-TS-UF-Modular-Bound}.
\end{proof}

\section{Concrete Instantiations}\label{Section:LaTSEC-Concrete-Instantiation-Evaluation}

\subsection{Methodology and Requirements}\label{Section:EmpiricalProfiles-InstantiationMethod}

We instantiate \(\mathsf{LaTSEC}\) from the \((n,k)=(40,20)\) and \((60,40)\) \(\mathsf{SEC}\) instances used in \cite[Section 6]{arzaki2026SEC}.
Among the selected matrices with \(n\geq25\), the former has the largest ratio \(k_p/k_c\approx0.19\), while the latter is the largest evaluated configuration.
They yield \((t_\mathsf{uf},t_\mathsf{sig})=(5,26)\) and \((6,46)\), respectively, and fix \(c_\mathrm{exp}=3\), \(\beta_\mathbf{g}=1\), and the reconstruction bounds \(\beta_{\bm{\lambda}}\) and \(\omega_{\bm{\lambda}}\) (\Cref{Table:LaTSEC-Empirical-Matrix-Parameters-40-60}).
These thresholds and their certification probabilities are inherited from \(\mathsf{SEC}\) \cite[Section 6.2.2]{arzaki2026SEC}; instances with larger \(k_p\) or smaller \(k_c\) would correspondingly improve \(t_\mathsf{uf}\) or \(t_\mathsf{sig}\), subject to re-selecting the parameters below.
Moreover, every set of at least \(k_c^\mathsf{det}\) parties is certified, with \(k_c^\mathsf{det}=35\) for \(n=40\) and \(56\) for \(n=60\), by the exhaustive verification of \cite[Section 6.2.2]{arzaki2026SEC}.
Our own exhaustive computation shows that, on every set of exactly \(k_c^\mathsf{det}\) parties, the output of \cite[Algorithm 1]{arzaki2026SEC} meets both bounds; since such a vector extends by zeros to any larger set, every active set of at least \(t_\mathsf{sig}^\mathsf{det}:=k_c^\mathsf{det}\) participants has an admissible reconstruction vector.

The reconstruction bounds lie at or above the maxima observed for the output of \cite[Algorithm 1]{arzaki2026SEC} over the certified sets of size at least \(k_c\) tested in \cite[Section 6.3]{arzaki2026SEC}, with \(3\cdot10^4\) sets per size.
In our expanded tests of \(10^5\) sets per size, \(\|\bm{\lambda}_\mathsf{act}\|_\infty\) reaches \(\beta_{\bm{\lambda}}\) for both instances, and one certified \(27\)-party set for \(n=40\) has weight \(26>\omega_{\bm{\lambda}}=25\) and is rejected.
We retain the bounds of \cite{arzaki2026SEC}, since such rejection affects only availability and the rejected set admits many other valid reconstruction vectors that the initiator may supply (\Cref{Section:Implementation-Correctness-IA}).

For the abort-key analysis, we additionally compute \(\|\mathbf{G}\|_2\), \(\delta_{k_p}\), and \(\gamma_{k_p}\) exhaustively over \(1\leq|U|\leq k_p\), where \(\mathbf{M}_U\) is returned by the deterministic peeling procedure of \cite[Algorithm 3]{arzaki2026SEC}.
Because row restriction cannot increase the spectral norm, \(\delta_{k_p}\) is attained at \(|U|=k_p\).
We set \(\beta_\mathbf{Q}\) to the right-hand side of \eqref{Eq:Uniform-Hidden-Coordinate-Transformation-Bound}.
\Cref{Table:LaTSEC-Empirical-Matrix-Parameters-40-60} reports all resulting values; profiles for all \(17\) instances of \cite{arzaki2026SEC} appear in Appendix~\ref{Appendix:LaTSEC-Detailed-Empirical-Profiles}.

\begin{table}[!t]
	\caption{Matrix and reconstruction quantities of the two \(\mathsf{SEC}\) instances used by \(\mathsf{LaTSEC}\)}
	\label{Table:LaTSEC-Empirical-Matrix-Parameters-40-60}
	\centering
	\footnotesize
	\setlength{\tabcolsep}{3pt}
	\begin{tabular}{@{}ccccccccccc@{}}
		\toprule
        \(n\) & \(k\) & \(t_\mathsf{uf}{=}k_p\) & \(t_\mathsf{sig}{=}k_c\) & \(t_\mathsf{sig}^\mathsf{det}\) & \(\beta_{\bm{\lambda}}\) & \(\omega_{\bm{\lambda}}\) & \(\|\mathbf{G}\|_2\) & \(\delta_{k_p}\) & \(\gamma_{k_p}\) & \(\beta_\mathbf{Q}\) \\
		\midrule
		\(40\) & \(20\) & \(5\) & \(26\) & \(35\) & \(5\) & \(25\) & \(5.861\) & \(3.646\) & \(0.193\) & \(20.635\) \\
		\(60\) & \(40\) & \(6\) & \(46\) & \(56\) & \(9\) & \(45\) & \(7.842\) & \(3.788\) & \(0.152\) & \(26.759\) \\
		\bottomrule
	\end{tabular}
\end{table}

We target \(\kappa\)-bit security for query bounds \(Q_H\) and \(Q_\mathsf{Sign}\), with \(d=512\), a prime \(q\equiv1\pmod{2d}\), and \(\ell=p+\ell'\).
A parameter set must satisfy the following requirements.
\begin{enumerate}
	\item \textit{Challenges.} \(\omega_c:=\min\{\omega:2^\omega\binom{d}{\omega}\geq Q_H2^\kappa\}\), so \(Q_H/|\mathcal{C}|\leq2^{-\kappa}\).
	\item \textit{Query terms.} \((n+1)(Q_H+1)Q_\mathsf{Sign}2^{-d+1}\leq2^{-\kappa}\) and \((Q_H^2+t_\mathsf{uf}Q_\mathsf{Sign}Q_H)2^{-2\kappa}\leq2^{-\kappa}\).
    \item \textit{Abort keys.} With \(\epsilon_\mathrm{AK}:=2^{-(\kappa+2)}\), \(\sigma_\mathrm{AK}\geq\sqrt{2}\eta_{\epsilon_\mathrm{AK}}(\mathbb{Z}^{kD})\), and \(\mathsf{MLWE}_{p,\ell',q,d,\sigma_\mathrm{AK}}\) has at least \(\kappa_\mathsf{MLWE}:=\lceil\kappa+1+\lg k\rceil\) bits of security, so the term \(k\cdot\mathsf{Adv}^\mathsf{MLWE}+2\epsilon_\mathrm{AK}\) is at most \(2^{-\kappa}\).
    \item \textit{Key privacy.} The requirements of \(\mathsf{SEC}\) \cite[Section 6.4.3]{arzaki2026SEC} hold with \(Q=Q_\mathsf{Sign}\) and its smoothing parameters \(\epsilon=\epsilon'=2^{-(\kappa+3)}\).
	In particular, the kernel-smoothing condition \cite[Eq.~(56)]{arzaki2026SEC} holds with margins of at least \(4.66\) for \(n=40\) and \(6.58\) for \(n=60\).
	\item \textit{Norm bounds.} \(B_1:=\omega_c\sqrt{D}(\beta_{\bm{s}}+c_\mathrm{exp}\beta_\mathbf{R})+\beta_{\bm{r}}\), \(B_2:=\omega_{\bm{\lambda}}\beta_{\bm{\lambda}}B_1\), and \(B_\mathsf{STMSIS}:=\sqrt{B_2^2+\omega_c}\), and \(\mathsf{MSIS}_{p,\ell'+1,q,d,2B_\mathsf{STMSIS}}\) has at least \(\kappa\) bits of security.
	\item \textit{Commitment entropy.} \(\sigma_{\bm{r}}>2d\cdot q^{1/\ell+2/(d\ell')}\).
\end{enumerate}

Each statistical term in \Cref{Corollary:LaTSEC-TS-UF-MLWE-MSIS} is then at most \(2^{-\kappa}\), so together they amount to a small constant multiple of \(2^{-\kappa}\).

\subsection{Security Estimates and Parameter Sets}\label{Section:SecurityEstimates-ParameterSets}

We estimate every retained tuple at its exact prime modulus with the full \(\mathsf{LWE}\) attack set of the lattice estimator \cite{albrecht2015concrete}, including primal, BDD, dual, and hybrid attacks, under the reduction-cost model of \cite{alkim2016post}.
For each tuple, \(q\) is the largest prime below \(2^{\lceil\lg q\rceil}\) with \(q\equiv1\pmod{2d}\) and \(q>2B_\mathsf{MSIS}+1\), since the estimator at the pinned commit \texttt{27a581b} treats \(\mathsf{SIS}\) instances with larger bounds as trivial.
Upstream commit \texttt{2f9a6d6} relaxes this check to \(B_\mathsf{MSIS}<q\); a modulus one bit smaller then meets every target, saving one bit per coefficient of \(\mathcal{R}_q\) but raising \(B_\mathsf{MSIS}/q\) from \(0.26\)--\(0.45\) to \(0.52\)--\(0.91\), so we keep the more conservative moduli.
As in prior concrete analyses, we do not charge the square-root loss of the forking reduction \cite{schnorr1989efficient,fersch2016provable,beullens2019csi,espitau2025two,borin2026hermine}.
\Cref{Table:LaTSEC-Selected-Parameters} reports the selected sets.
We choose \((\sigma_{\bm{s}},\sigma_\mathbf{R},\sigma_{\bm{r}})\) to minimize \(B_1\), which determines the signature and communication sizes and the modulus.
In every set, the abort-key requirement is then binding, with \(\sigma_\mathrm{AK}=\sqrt{2}\eta_{\epsilon_\mathrm{AK}}(\mathbb{Z}^{kD})\) up to rounding, and the optimum has \(\sigma_{\bm{s}}=\sigma_\mathbf{R}\) and assigns almost all of its precision budget to the mask term.
Shifting part of this budget to \(\sigma_{\bm{s}}\) and \(\sigma_\mathbf{R}\) would reduce \(|\mathsf{sk}_i|\) further at a small cost in \(B_1\).
We leave this storage-oriented selection to future work.
The parameter-selection notebook, which checks every requirement at \(200\)-bit precision, and its lattice-estimator outputs are in the artifact (\fullversionurl).
The requirements of \(\mathsf{SEC}\) then hold with slack: \(\sigma_\mathrm{MQ}\) exceeds \(\sqrt{2}\eta_{\epsilon}(\mathbb{Z}^D)\) by factors of \(4.4\) (\(n=40\)) and \(8.0\) (\(n=60\)), and \(\sigma_\mathrm{MQ}>\sigma_\mathrm{AK}\), so \(\mathsf{MLWE}_{\sigma_\mathrm{MQ}}\) is estimated to be at least as hard as \(\mathsf{MLWE}_{\sigma_\mathrm{AK}}\).

\begin{table}[!t]
    \caption{Selected \(\mathsf{LaTSEC}\) parameter sets (\(d=512\)), with \((t_\mathsf{uf},t_\mathsf{sig})=(5,26)\) for \(n=40\) and \((6,46)\) for \(n=60\)}
	\label{Table:LaTSEC-Selected-Parameters}
	\centering
	\footnotesize
	\setlength{\tabcolsep}{2.5pt}
	\begin{tabular}{@{}cccccccccc@{}}
		\toprule
		\(n\) & \(\kappa\) & \(\lg Q_H\) & \(\lg Q_\mathsf{Sign}\) & \(\lceil\lg q\rceil\) & \(\ell\) & \(p\) & \(\omega_c\) & \(\lg\sigma_{\bm{s}}{=}\lg\sigma_\mathbf{R}\) & \(\lg\sigma_{\bm{r}}\) \\
		\midrule
		\(40\) & \(128\) & \(63\) & \(62\) & \(61\) & \(12\) & \(6\) & \(30\) & \(15.464\) & \(45.013\) \\
		\(40\) & \(192\) & \(64\) & \(64\) & \(63\) & \(16\) & \(8\) & \(44\) & \(15.982\) & \(46.827\) \\
		\(40\) & \(256\) & \(64\) & \(64\) & \(64\) & \(20\) & \(10\) & \(60\) & \(16.115\) & \(47.465\) \\
		\midrule
		\(60\) & \(128\) & \(63\) & \(62\) & \(64\) & \(12\) & \(6\) & \(30\) & \(15.983\) & \(45.813\) \\
		\(60\) & \(192\) & \(64\) & \(64\) & \(66\) & \(16\) & \(8\) & \(44\) & \(16.499\) & \(47.625\) \\
		\(60\) & \(256\) & \(64\) & \(64\) & \(66\) & \(20\) & \(10\) & \(60\) & \(16.632\) & \(48.263\) \\
		\bottomrule
	\end{tabular}
	\par\smallskip
	\parbox{\columnwidth}{\scriptsize \(\ell'=\ell-p=p\) in every set.
    Logarithms are base two and computed from full-precision values.}
\end{table}

\subsection{Size Estimates}\label{Section:SizeEstimates}

\Cref{Table:LaTSEC-Selected-Sizes} reports the public storage, long-term secret-share storage, signature sizes, and per-party signing communication.
The uniform block \(\bm{A}'\) is represented by a \(256\)-bit seed, and the verification target, the abort keys, and the commitment openings are encoded without bit dropping, at \(\lceil\lg q\rceil\) bits per coefficient: \(|\mathsf{ak}_i|=pd\lceil\lg q\rceil\), \(|\mathsf{vk}|=256+|\mathsf{ak}_i|\), and \(|\mathsf{ak}|=n|\mathsf{ak}_i|\).
As usual, the proofs model \(\bm{A}'\) as uniform, treating its expansion from the public seed as a random oracle.
For the Gaussian components, we report ideal entropy-coded sizes \cite{saarinen2018arithmetic}, using a variance-matched discrete Gaussian \cite[Section 5.1]{espitau2022shorter}: since each share coefficient is \(\mathbf{g}[i]\mathbf{s}[j]\) plus \(c_\mathrm{exp}\) signed entries of \(\mathbf{R}\), \(|\mathsf{sk}_i|\approx D(\lg\sigma_{\bm{s}}+\frac12\lg(1+c_\mathrm{exp})+2.047)\) bits, as in \cite[Appendix I]{arzaki2026SEC}.
The response bounds give \(|\mathsf{sig}_i|\approx b_c+\mathsf{Size}(B_1)\) and \(|\mathsf{sig}|\approx b_c+\mathsf{Size}(B_2)\), where \(\mathsf{Size}(B):=D(\lg(B/\sqrt{D})+2.047)\) and \(b_c:=\lceil\omega_c+\lg\binom{d}{\omega_c}\rceil\) encodes the challenge, and \(\mathsf{Comm}_i\approx2\kappa+pd\lceil\lg q\rceil+\mathsf{Size}(B_1)\) counts the commitment, opening, and response that each signer broadcasts.
The estimates exclude session context, coding, framing, and byte-alignment overheads; their derivation is in Appendix~\ref{Appendix:LaTSEC-Detailed-Size-Estimates}.
Since \(B_2=\omega_{\bm{\lambda}}\beta_{\bm{\lambda}}B_1\) is a worst-case bound, the estimate of \(|\mathsf{sig}|\) charges \(\lg(\omega_{\bm{\lambda}}\beta_{\bm{\lambda}})\) extra bits per coefficient, whereas an honest aggregate response grows only with \(\|\bm{\lambda}_\mathsf{act}\|_2\); an encoding that adapts to the actual width, as in our prototype, recovers much of this gap (\Cref{Section:Implementation-Sizes}).
Bit dropping and a tighter verification bound \(B_2\), which must also cover corrupted responses that pass partial verification, remain open.

\begin{table}[!t]
	\caption{Estimated \(\mathsf{LaTSEC}\) sizes in KiB, without bit dropping}
	\label{Table:LaTSEC-Selected-Sizes}
	\centering
	\footnotesize
	\setlength{\tabcolsep}{2.2pt}
	\begin{tabular}{@{}ccrrrrrrr@{}}
		\toprule
		\(n\) & \(\kappa\) & \(|\mathsf{vk}|\) & \(|\mathsf{ak}_i|\) & \(|\mathsf{ak}|\) & \(|\mathsf{sk}_i|\) & \(|\mathsf{sig}_i|\) & \(|\mathsf{sig}|\) & \(\mathsf{Comm}_i\) \\
		\midrule
		\(40\) & \(128\) & \(22.906\) & \(22.875\) & \(915.0\) & \(13.883\) & \(35.525\) & \(40.750\) & \(58.408\) \\
		\(40\) & \(192\) & \(31.531\) & \(31.500\) & \(1260.0\) & \(19.029\) & \(49.181\) & \(56.147\) & \(80.696\) \\
		\(40\) & \(256\) & \(40.031\) & \(40.000\) & \(1600.0\) & \(23.952\) & \(62.274\) & \(70.981\) & \(102.297\) \\
		\midrule
		\(60\) & \(128\) & \(24.031\) & \(24.000\) & \(1440.0\) & \(14.273\) & \(36.125\) & \(42.622\) & \(60.133\) \\
		\(60\) & \(192\) & \(33.031\) & \(33.000\) & \(1980.0\) & \(19.547\) & \(49.979\) & \(58.641\) & \(82.995\) \\
		\(60\) & \(256\) & \(41.281\) & \(41.250\) & \(2475.0\) & \(24.599\) & \(63.271\) & \(74.098\) & \(104.544\) \\
		\bottomrule
	\end{tabular}
\end{table}

\section{Implementation and Evaluation}\label{Section:Implementation-Evaluation}

We provide two implementations of \(\mathsf{LaTSEC}\): a Rust implementation used for all performance measurements, and a Python implementation that follows \Cref{Figure:LaTSEC-Protocol} line by line and serves as a test-vector generator; both are in the artifact.
Both are still research prototypes: they are not fully constant-time, key generation uses a trusted dealer, and all parties run in one process with consistent message delivery.

\subsection{Implementation Choices}\label{Section:Implementation-Instantiation}

All hashes are SHAKE256 \cite{FIPS202} with domain separation and length-framed inputs; \(H_\mathsf{cmt}\) outputs \(2\kappa\) bits.
For \(H_c\), the SHAKE256 output stream drives an unbiased sample-in-ball shuffle directly, as suggested in \Cref{Section:LaTSEC-Overview}.
The block \(\bm{A}'\) is expanded from a \(256\)-bit seed by rejection sampling, directly in the NTT domain.

The discrete Gaussians are sampled with the algorithm of Canonne et al.~\cite{CaKaSt20} in exact integer arithmetic, with bounded statistical error: proposals of magnitude \(2^{126}\) or more are rejected, which puts the output within statistical distance \(\exp(-2^{48})\) of \(\mathcal{D}_\sigma\) for every width used, and the remaining machine-integer counters overflow with probability below \(\exp(-2^{128})\).

The session message \(\mathsf{msg}\) encodes \((\mathsf{sid},\mathsf{act},\bm{\lambda}_\mathsf{act})\) and the payload; the initiator computes \(\bm{\lambda}_\mathsf{act}\) with \cite[Algorithm 1]{arzaki2026SEC} using an exact integral LLL, and each signer evaluates \(\mathcal{P}\) with exact integer checks.

Residues are encoded at \(\lceil\lg q\rceil\) bits, the challenge as the rank of its support followed by its signs (\(b_c\) bits, \Cref{Section:SizeEstimates}), and all Gaussian components with a Golomb--Rice code \cite{Go66,Ri79}.
For a component of width \(\sigma\), we use the largest \(k\) with \(1.2863\cdot2^k\leq\sigma\).
This is the large-width limit of the length-minimizing parameter.
For every component width in our parameter sets, it gives the lowest computed length in a discrete-Gaussian model (each width is at least \(4\%\) from a switch point), and for these widths (\(\sigma\geq2^{15}\)) the code stays within \(0.17\) bits per coefficient of the entropy.
Since the width of \(\bm{z}\) depends on \(\bm{\lambda}_\mathsf{act}\), a signature carries its own parameter.

\subsection{Correctness and Identifiable Abort}\label{Section:Implementation-Correctness-IA}

For each parameter set, we ran \(100\) all-honest sessions on random sets of \(t_\mathsf{sig}\) parties.
All \(600\) signed and verified, and no honest partial response was rejected; the largest \(\|\bm{z}_i\|_2/B_1\) was \(0.85\) and the largest \(\|\bm{z}\|_2/B_2\) was \(0.037\).
Three of the \(603\) sets drawn were not certified and were refused before the first round; availability rests on \cite[Section 6.2.2]{arzaki2026SEC}.

On average, only \(10.4\) of \(26\) signers (\(n=40\)) and \(23.0\) of \(46\) (\(n=60\)) had a nonzero coefficient in \(\bm{\lambda}_\mathsf{act}\).
The others contribute nothing to \(\bm{z}\), but they must stay online for all three rounds and pass \(\mathsf{PartialVerify}\), and a deviation by any of them aborts the session with that signer blamed.

Tests with tampered commitments, openings, and partial responses confirm that only deviating signers are blamed.

For one certified \(27\)-party set at \(n=40\), the vector returned by \cite[Algorithm 1]{arzaki2026SEC} has weight \(26>\omega_{\bm{\lambda}}\) and fails \(\mathcal{P}\) (\Cref{Section:EmpiricalProfiles-InstantiationMethod}).
A search over small combinations of the reduced kernel basis found \(11{,}051\) vectors for this set that pass \(\mathcal{P}\), the lightest found of weight \(12\), so the initiator can supply one of them and the set can sign.

\subsection{Sizes}\label{Section:Implementation-Sizes}

\Cref{Table:LaTSEC-Measured-Sizes} compares the encoded sizes with \Cref{Table:LaTSEC-Selected-Sizes}.
Shares are within \(0.6\%\) of the entropy estimate, and partial responses and per-party communication are slightly smaller than estimated, since honest responses have \(\|\bm{z}_i\|_2\approx0.83\,B_1\).
Signatures are \(9.6\)--\(11.2\%\) smaller than in \Cref{Table:LaTSEC-Selected-Sizes}, which charges the worst-case bound \(B_2=\omega_{\bm{\lambda}}\beta_{\bm{\lambda}}B_1\): the aggregate response grows with \(\|\bm{\lambda}_\mathsf{act}\|_2\), on average \(3.3\) (\(n=40\)) and \(5.6\) (\(n=60\)), rather than with \(\omega_{\bm{\lambda}}\beta_{\bm{\lambda}}=125\) and \(405\).
This supports the tighter aggregate bound suggested in \Cref{Section:SizeEstimates}; verification still uses \(B_2\).

\begin{table}[!t]
	\caption{Measured \(\mathsf{LaTSEC}\) sizes in KiB, with the estimates of \Cref{Table:LaTSEC-Selected-Sizes} in parentheses}
	\label{Table:LaTSEC-Measured-Sizes}
	\centering
	\footnotesize
	\setlength{\tabcolsep}{2.5pt}
	\begin{tabular}{@{}ccrrrr@{}}
		\toprule
		\(n\) & \(\kappa\) & \(|\mathsf{sk}_i|\) & \(|\bm{z}_i|\) & \(|\mathsf{sig}|\) & \(\mathsf{Comm}_i\) \\
		\midrule
		\(40\) & \(128\) & \(13.93\) (\(13.88\)) & \(35.36\) (\(35.53\)) & \(36.68\) (\(40.75\)) & \(58.27\) (\(58.41\)) \\
		\(40\) & \(192\) & \(19.02\) (\(19.03\)) & \(48.96\) (\(49.18\)) & \(50.71\) (\(56.15\)) & \(80.50\) (\(80.70\)) \\
		\(40\) & \(256\) & \(23.95\) (\(23.95\)) & \(62.06\) (\(62.27\)) & \(64.15\) (\(70.98\)) & \(102.13\) (\(102.30\)) \\
		\midrule
		\(60\) & \(128\) & \(14.32\) (\(14.27\)) & \(35.96\) (\(36.13\)) & \(37.84\) (\(42.62\)) & \(59.99\) (\(60.13\)) \\
		\(60\) & \(192\) & \(19.66\) (\(19.55\)) & \(49.78\) (\(49.98\)) & \(52.24\) (\(58.64\)) & \(82.83\) (\(83.00\)) \\
		\(60\) & \(256\) & \(24.71\) (\(24.60\)) & \(63.05\) (\(63.27\)) & \(66.15\) (\(74.10\)) & \(104.37\) (\(104.54\)) \\
		\bottomrule
	\end{tabular}
	\par\smallskip
	\parbox{\columnwidth}{\scriptsize Means over all parties or \(100\) sessions.
    \(|\bm{z}_i|\) is compared with \(|\mathsf{sig}_i|\), which also counts the challenge.}
\end{table}

\subsection{Running Time}\label{Section:Implementation-Running-Time}

\Cref{Table:LaTSEC-Timings} in Appendix~\ref{Appendix:Running-Times} gives median single-core running times on an AMD Ryzen AI 9 HX 370, at \(|\mathsf{act}|=t_\mathsf{sig}\); the measurement method and further results are in Appendix~\ref{Appendix:Extended-Benchmarks}.
At \(n=60\) and \(\kappa=128\), each signer spends \(7.6\)~ms over the three rounds, \(\mathsf{Combine}\) takes \(33\)~ms, and \(\mathsf{Verify}\) takes \(0.76\)~ms.
The implementation uses portable \(128\)-bit Montgomery arithmetic without vectorization.
The second round only releases the stored opening.
Identifiable abort costs one \(\mathsf{PartialVerify}\) per signer, \(1.0\) to \(2.8\)~ms (a stand-alone call recomputes \(\bm{w}\) and \(c\); \(\mathsf{Combine}\) computes them once), and \(\mathsf{Combine}\), which partially verifies every signer, accounts for most of the combiner's work.

\appendices

\section{Lattice Problems}\label{Appendix:Standard-Problems}

We use \(\mathsf{MLWE}\) in the systematic form of \cite[Definition 2]{arzaki2026SEC} (\Cref{Section:Notation}) and the following two problems.

\begin{definition}[Module-SIS]\label{Definition:MSIS}
	Given \(\bm{A}\randgets\mathcal{R}_q^{n\times m}\), the problem \(\mathsf{MSIS}_{n,m,q,d,B}\) asks for \(\bm{z}\in\mathcal{R}^{n+m}\setminus\{\bm{0}\}\) with \(\begin{bmatrix}\bm{I}_n\ \bm{A}\end{bmatrix}\bm{z}=\bm{0}\) over \(\mathcal{R}_q\) and \(\|\bm{z}\|_2\leq B\).
\end{definition}

\begin{definition}[Repeated-\(\bm{A}\) MLWE]\label{Definition:RA-MLWE}
	For \(h\in\mathbb{Z}^+\) and a distribution \(\psi\) over \(\mathbb{Z}\), the problem \(\mathsf{RA\text{-}MLWE}_{n,m,q,d,\psi}[h]\) is to distinguish \((\bm{A},\overline{\bm{A}}^{\langle h\rangle}\bm{s})\) from \((\bm{A},\bm{u})\), where \(\bm{A}\randgets\mathcal{R}_q^{n\times m}\), the coefficients of \(\bm{s}\in\mathcal{R}^{h(n+m)}\) are i.i.d.\ from \(\psi\), and \(\bm{u}\randgets\mathcal{R}_q^{hn}\).
\end{definition}

Writing \(\bm{s}=(\bm{s}_1,\dots,\bm{s}_h)\), the target is \((\overline{\bm{A}}\bm{s}_1,\dots,\overline{\bm{A}}\bm{s}_h)\) with independent blocks, so \(\mathsf{RA\text{-}MLWE}\) is the stacked presentation of multi-secret \(\mathsf{MLWE}\) under a common matrix \cite{lyubashevsky2012lattice,brakerski2013classical,boudgoust2021hardness}, and \(\mathsf{RA\text{-}MLWE}[1]=\mathsf{MLWE}\).
We use the stacked form because the hints of \(\mathsf{RA\text{-}MHMLWE}\) may couple different blocks.

\begin{lemma}[Hardness of \(\mathsf{RA\text{-}MLWE}\) \cite{brakerski2013classical,boudgoust2021hardness}]\label{Lemma:Hardness-RA-MLWE}
	For every PPT \(\mathcal{A}\) against \(\mathsf{RA\text{-}MLWE}_{n,m,q,d,\psi}[h]\), there is a PPT \(\mathcal{B}\) against \(\mathsf{MLWE}_{n,m,q,d,\psi}\) with \(\mathsf{Adv}_\mathcal{A}^{\mathsf{RA\text{-}MLWE}[h]}(\kappa)\leq h\cdot\mathsf{Adv}_\mathcal{B}^\mathsf{MLWE}(\kappa)\).
\end{lemma}

\begin{proof}
	Replace the \(h\) target blocks by independent uniform values one at a time; each step embeds one \(\mathsf{MLWE}\) challenge and samples the other blocks itself, as in \cite[Lemma 4.8]{brakerski2013classical} and \cite[Lemma 17]{boudgoust2021hardness}.
\end{proof}

\section{Hardness of \texorpdfstring{\(\bm{\mathsf{RA\text{-}MHMLWE}}\)}{RA-MHMLWE}: Proof Sketch}\label{Appendix:Hardness-RA-MHMLWE-Sketch}

We sketch the proof of \Cref{Lemma:Hardness-RA-MHMLWE}; the full proof is in Appendix~\ref{Appendix:Hardness-RA-MHMLWE}.
Let \(\mathcal{B}_0\) receive an \(\mathsf{RA\text{-}MLWE}_{n,m,q,d,\sigma}[h]\) challenge \((\bm{A},\bm{b})\), where \(\bm{b}=\overline{\bm{A}}^{\langle h\rangle}\bm{w}\) for \(\mathbf{w}\gets\mathcal{D}_{\mathbb{Z}^N,\sigma}\) or \(\bm{b}\) is uniform.
It samples a dummy secret \(\mathbf{s}'\gets\mathcal{D}_{\mathbb{Z}^N,\mathbf{c}_\mathbf{s},\sqrt{\mathbf{\Sigma}_\mathbf{s}}}\), hint matrices \(\bm{M}_j\gets\mathcal{M}_j\), and perturbations \(\mathbf{y}_j\), and sets \(\mathbf{z}_j:=\mathbf{M}_j\mathbf{s}'+\mathbf{y}_j\).
Given the hints, the secret has the posterior distribution \(\mathcal{D}_{\mathbb{Z}^N,\mathbf{c}_0,\sqrt{\mathbf{\Sigma}_0}}\) with \(\mathbf{\Sigma}_0:=\mathbf{\Xi}^{-1}\) and a center \(\mathbf{c}_0\) computable from the hints \cite[Lemma 7]{kim2023toward}, \cite[Lemma 1]{espitau2024flood}, \cite[Lemma G.2]{wang2025post}.
Since \(\|\mathbf{\Xi}\|_2\leq1/(2\sigma^2)\), we have \(\mathbf{\Sigma}_0-\sigma^2\mathbf{I}_N\succeq\sigma^2\mathbf{I}_N\), so \(\mathcal{B}_0\) can sample, exactly as \(\mathbf{s}'\), \(\mathbf{t}\gets\mathcal{D}_{\mathbb{Z}^N,\mathbf{c}_0,\sqrt{\mathbf{\Sigma}_0-\sigma^2\mathbf{I}_N}}\) and give \((\bm{A},\bm{b}+\overline{\bm{A}}^{\langle h\rangle}\bm{t},(\bm{M}_j,\bm{z}_j)_j)\) to \(\mathcal{A}\).
In the real case, \(\bm{b}+\overline{\bm{A}}^{\langle h\rangle}\bm{t}=\overline{\bm{A}}^{\langle h\rangle}(\bm{w}+\bm{t})\) by linearity, and the convolution lemma \cite[Lemmas 3 and 5]{kim2023toward}, whose hypothesis is \(\sigma\geq\sqrt{2}\eta_\epsilon(\mathbb{Z}^N)\) in our normalization \cite[Appendix B]{arzaki2026SEC}, puts \(\mathbf{w}+\mathbf{t}\) within statistical distance \(2\epsilon\) of the posterior; in the uniform case, the target stays uniform.
Hence \(\mathsf{Adv}_\mathcal{A}^{\mathsf{RA\text{-}MHMLWE}[h]}\leq\mathsf{Adv}_{\mathcal{B}_0}^{\mathsf{RA\text{-}MLWE}[h]}+2\epsilon\), and \Cref{Lemma:Hardness-RA-MLWE} completes the proof.
With auxiliary information, the reduction first samples it and then runs conditionally on the outcome, with the conditional center and parameter matrix.
The uniform precision and smoothing conditions keep every sample exact and make the bound hold after averaging.

\section{Security Games}\label{Appendix:TSS-with-IA}

\Cref{Figure:Game-TS-UF} gives the games for a TSS with IA, adapted from \cite{del2024shortly}.
\(\mathcal{A}\) statically chooses \(\mathsf{corrupt}\) and receives the corrupted shares.
\(L_\mathsf{sid}\) stores each session's status, round, active set, message, transcript table \(\mathsf{PM}\) (\(\mathsf{PM}[k,i]=\mathsf{pm}_i^k\)), and blame table \(\mathsf{BL}\) (\(\mathsf{BL}[k,i]=S_\bot^i\)); \(L_\mathsf{sig}\) records the messages for which honest signing messages were returned; and \(H(\mathsf{str},\cdot)\) is a domain-separated random oracle.
\(\mathsf{OSign}_k\) completes the round-\((k-1)\) messages of the honest parties stored in \(\mathsf{PM}\) with those supplied by \(\mathcal{A}\) for \(\mathsf{corrupt}\cap\mathsf{act}\), where a missing message is \(\bot\); a failed \textbf{assert} returns \(\bot\).

\begin{figure*}[!t]
	\centering
	\footnotesize
	\begin{tcolorbox}[gamebox]
	\begin{minipage}[t]{0.485\textwidth}
		\(\mathsf{Game}_\mathsf{TS\text{-}UF}\) and \(\mathsf{Game}_\mathsf{TS\text{-}IA}\)
		\hrule height 0.4pt
		\begin{algorithmic}[1]
			\State \(L_\mathsf{sid},L_H,L_\mathsf{sig}:=\emptyset\); \(\mathsf{tspar}\gets\mathsf{Setup}(1^\kappa,n,t_\mathsf{uf},t_\mathsf{sig})\)
			\State \(\mathsf{corrupt}\gets\mathcal{A}(1^\kappa,\mathsf{tspar})\); \textbf{assert} \(\mathsf{corrupt}\subseteq[n]\), \(|\mathsf{corrupt}|\leq t_\mathsf{uf}\)
			\State \(\mathsf{honest}:=[n]\setminus\mathsf{corrupt}\); \((\mathsf{vk},\mathsf{ak},(\mathsf{sk}_i)_{i\in[n]})\gets\mathsf{Keygen}(\mathsf{tspar})\)
			\State Give \((\mathsf{sk}_i)_{i\in\mathsf{corrupt}}\) to \(\mathcal{A}\); initialize \(\mathsf{st}_i\) for \(i\in\mathsf{honest}\)
			\LineComment{\(\mathsf{Game}_\mathsf{TS\text{-}UF}\):}
			\State \((\mathsf{msg}^*,\mathsf{sig}^*)\gets\mathcal{A}^{H,(\mathsf{OSign}_k)_{k\in[\mathsf{rnd}]}}(\mathsf{tspar},\mathsf{vk},\mathsf{ak})\)
			\State \Return \(\mathsf{msg}^*\notin L_\mathsf{sig}\land\mathsf{Verify}(\mathsf{vk},\mathsf{msg}^*,\mathsf{sig}^*)=1\)
			\LineComment{\(\mathsf{Game}_\mathsf{TS\text{-}IA}\):}
            \State \((\mathsf{sid}^*,(\mathsf{pm}_j^*)_{j\in\mathsf{corrupt}})\gets\mathcal{A}^{H,(\mathsf{OSign}_k)_{k\in[\mathsf{rnd}]}}(\mathsf{tspar},\mathsf{vk},\mathsf{ak})\)
			\State Fetch \((\mathsf{status},r,\mathsf{act}^*,\mathsf{msg}^*,\mathsf{PM}^*,\mathsf{BL}^*):=L_\mathsf{sid}[\mathsf{sid}^*]\)
			\State \textbf{assert} \(\mathcal{P}(\mathsf{tspar},\mathsf{vk},\mathsf{sid}^*,\mathsf{act}^*,\mathsf{msg}^*)=1\) and (\(\mathsf{status}=\mathsf{abort}\) or \(r=\mathsf{rnd}\))
			\If{\(\mathsf{status}=\mathsf{abort}\)}
				\State Let \(S_\bot^i:=\mathsf{BL}^*[r,i]\) for \(i\in\mathsf{honest}\cap\mathsf{act}^*\)
				\State \Return \(1\) iff some \(S_\bot^i\neq S_\bot^j\), \(S_\bot^i=\emptyset\), or \(S_\bot^i\not\subseteq\mathsf{corrupt}\)
			\EndIf
            \State Record \((\mathsf{pm}_j^*)_{j\in\mathsf{corrupt}\cap\mathsf{act}^*}\) in \(\mathsf{PM}^*\) as the corrupted round-\(\mathsf{rnd}\) messages.
			\State \(\mathsf{out}\gets\mathsf{Combine}(\mathsf{tspar},\mathsf{vk},\mathsf{ak},\mathsf{sid}^*,\mathsf{act}^*,\mathsf{msg}^*,\mathsf{PM}^*)\)
			\If{\(\mathsf{out}=(\mathsf{abort},S_\bot)\)}
				\State \Return \(S_\bot=\emptyset\lor S_\bot\not\subseteq\mathsf{corrupt}\)
			\EndIf
			\State \Return \(\mathsf{Verify}(\mathsf{vk},\mathsf{msg}^*,\mathsf{out})=0\)
		\end{algorithmic}
	\end{minipage}\hfill
	\vrule\hfill
	\begin{minipage}[t]{0.485\textwidth}
		\(\mathsf{OSign}_1(\mathsf{sid},\mathsf{act},\mathsf{msg})\)
		\hrule height 0.4pt
		\begin{algorithmic}[1]
			\State \textbf{assert} \(\mathsf{sid}\notin L_\mathsf{sid}\), \(\mathsf{act}\subseteq[n]\), \(\mathcal{P}(\mathsf{tspar},\mathsf{vk},\mathsf{sid},\mathsf{act},\mathsf{msg})=1\), \(\mathsf{honest}\cap\mathsf{act}\neq\emptyset\)
			\State \((\mathsf{st}_i,\mathsf{pm}_i^1)\gets\mathsf{ShareSign}_1(\mathsf{st}_i,\mathsf{sid},\mathsf{act},\mathsf{msg})\) for \(i\in\mathsf{honest}\cap\mathsf{act}\)
			\State Record the honest messages and blame sets in \(\mathsf{PM}\) and \(\mathsf{BL}\).
			\If{some honest \(\mathsf{pm}_i^1=(\mathsf{abort},S_\bot^i)\)}
				\State \(L_\mathsf{sid}[\mathsf{sid}]:=(\mathsf{abort},1,\mathsf{act},\mathsf{msg},\mathsf{PM},\mathsf{BL})\); \Return \(\bot\)
			\EndIf
			\State \(L_\mathsf{sig}:=L_\mathsf{sig}\cup\{\mathsf{msg}\}\); \(L_\mathsf{sid}[\mathsf{sid}]:=(\mathsf{running},1,\mathsf{act},\mathsf{msg},\mathsf{PM},\mathsf{BL})\)
			\State \Return \((\mathsf{pm}_i^1)_{i\in\mathsf{honest}\cap\mathsf{act}}\)
		\end{algorithmic}
		\smallskip
		\(\mathsf{OSign}_k(\mathsf{sid},(\mathsf{pm}_j^{k-1})_{j\in\mathsf{corrupt}\cap\mathsf{act}})\), \(k\in[2,\mathsf{rnd}]\)
		\hrule height 0.4pt
		\begin{algorithmic}[1]
			\State \textbf{assert} \(\mathsf{sid}\in L_\mathsf{sid}\); fetch \((\mathsf{status},r,\mathsf{act},\mathsf{msg},\mathsf{PM},\mathsf{BL})\)
			\State \textbf{assert} \(\mathsf{status}=\mathsf{running}\) and \(r=k-1\)
			\State Record the corrupted messages in \(\mathsf{PM}\), completing \((\mathsf{pm}_u^{k-1})_{u\in\mathsf{act}}\).
			\State \((\mathsf{st}_i,\mathsf{pm}_i^k)\gets\mathsf{ShareSign}_k(\mathsf{st}_i,\mathsf{sid},\mathsf{act},(\mathsf{pm}_u^{k-1})_{u\in\mathsf{act}})\) for \(i\in\mathsf{honest}\cap\mathsf{act}\)
			\State Record the honest messages and blame sets in \(\mathsf{PM}\) and \(\mathsf{BL}\).
			\If{some honest \(\mathsf{pm}_i^k=(\mathsf{abort},S_\bot^i)\)}
				\State \(L_\mathsf{sid}[\mathsf{sid}]:=(\mathsf{abort},k,\dots)\); \Return \(\bot\)
			\EndIf
			\State \(L_\mathsf{sid}[\mathsf{sid}]:=(\mathsf{running},k,\dots)\); \Return \((\mathsf{pm}_i^k)_{i\in\mathsf{honest}\cap\mathsf{act}}\)
		\end{algorithmic}
		\smallskip
		\(H(\mathsf{str},\mathsf{digest})\)
		\hrule height 0.4pt
		\begin{algorithmic}[1]
			\State \textbf{if} \((\mathsf{str},\mathsf{digest})\notin L_H\) \textbf{then} sample \(L_H[(\mathsf{str},\mathsf{digest})]\) uniformly from the range
			\State \Return \(L_H[(\mathsf{str},\mathsf{digest})]\)
		\end{algorithmic}
	\end{minipage}
	\end{tcolorbox}
	\caption{Unforgeability and identifiable-abort games for a TSS with IA, and their oracles.}
	\label{Figure:Game-TS-UF}
	\label{Figure:Game-TS-IA}
	\label{Figure:Game-TS-UF-Oracle}
\end{figure*}

\begin{definition}[Unforgeability]\label{Definition:TSS-Unforgeability}
	A TSS with IA is \emph{unforgeable} if \(\mathsf{Adv}_\mathcal{A}^\mathsf{TS\text{-}UF}(\kappa):=\Pr[\mathsf{Game}_\mathsf{TS\text{-}UF}=1]\) is negligible for every PPT \(\mathcal{A}\), where \(\mathsf{Game}_\mathsf{TS\text{-}UF}\) is described in \Cref{Figure:Game-TS-UF}.
\end{definition}

\begin{definition}[Identifiable abort]\label{Definition:TSS-IA}
	A TSS with IA \emph{identifies aborts} for the admissible session inputs of \(\mathcal{P}\) if \(\mathsf{Adv}_\mathcal{A}^\mathsf{TS\text{-}IA}(\kappa):=\Pr[\mathsf{Game}_\mathsf{TS\text{-}IA}=1]\) is negligible for every PPT \(\mathcal{A}\), where \(\mathsf{Game}_\mathsf{TS\text{-}IA}\) is described in \Cref{Figure:Game-TS-IA}.
\end{definition}

Thus, in every admissible session, either all honest parties abort with the same non-empty blame set contained in \(\mathsf{corrupt}\), or \(\mathsf{Combine}\) returns such a set, or the output signature verifies.
Correctness of admissible sessions follows by running \(\mathsf{Game}_\mathsf{TS\text{-}IA}\) with \(\mathsf{corrupt}=\emptyset\) and an honest execution.
Inadmissible inputs are rejected before signing; for a random active set of size at least \(t_\mathsf{sig}\) with the vector of \cite[Algorithm 1]{arzaki2026SEC}, this happens if the set is not certified, with probability about the coverage error \(\epsilon_c\) of \(\mathsf{SEC}\) \cite[Definition 11]{arzaki2026SEC}, or, rarely, if the vector exceeds the bounds (\Cref{Section:EmpiricalProfiles-InstantiationMethod}). The correctness game is in Appendix~\ref{Appendix:LaTSEC-Full-Specification}.

\section{The Abort-Key View as an \texorpdfstring{\(\bm{\mathsf{RA\text{-}MHMLWE}}\)}{RA-MHMLWE} Instance}\label{Appendix:LaTSEC-Abort-Key-View-Details}

We detail the representation of \Cref{Section:Hidden-Coordinates}.
Write \(U:=U_\mathrm{raw}\) and \(h_U:=k-|U|\).
Products involving \(\mathbf{A}\) are modulo \(q\).

\subsection{Evaluated Matrices and Affine Completions}

With \(\mathbf{X}=\begin{bmatrix}\mathbf{s}^\top\\ \mathbf{R}\end{bmatrix}\) and \(\mathbf{Y}=\mathbf{X}\mathbf{A}^\top=\begin{bmatrix}\mathbf{t}^\top\\ \mathbf{R}_\mathbf{t}\end{bmatrix}\), the abort keys are \(\mathbf{T}=\mathbf{G}\mathbf{Y}=\mathbf{S}\mathbf{A}^\top\) and \(\mathbf{T}_U=\mathbf{G}_U\mathbf{Y}=\mathbf{S}_U\mathbf{A}^\top\).
Let \(\mathcal{Y}_U(\mathbf{T}_U):=\{\mathbf{Z}\in\mathbb{Z}_q^{k\times dp}:\mathbf{G}_U\mathbf{Z}=\mathbf{T}_U\}\).
For \(\mathbf{\breve{Y}}=\begin{bmatrix}\mathbf{\breve{t}}^\top\\ \mathbf{\breve{R}}_\mathbf{t}\end{bmatrix}\), the constraint \(\mathbf{G}_U\mathbf{\breve{Y}}=\mathbf{T}_U\) is exactly the completion equation \((\mathbf{g})_U\mathbf{\breve{t}}^\top+\mathbf{\hat{G}}_U\mathbf{\breve{R}}_\mathbf{t}=\mathbf{T}_U\) of \Cref{Definition:SimAK}.
Since \(\mathbf{\breve{t}}\) is uniform and \(\mathsf{SimAK}\) samples \(\mathbf{\breve{R}}_\mathbf{t}\) uniformly from the corresponding affine space, the simulated \(\mathbf{\breve{Y}}\) is uniform over \(\mathcal{Y}_U(\mathbf{T}_U)\), and both views are deterministic functions of \(\mathbf{Y}\) or \(\mathbf{\breve{Y}}\): the target is the first row, and the abort keys are \(\mathbf{G}\mathbf{Y}\) or \(\mathbf{G}\mathbf{\breve{Y}}\).

\subsection{Separation of Exposed and Hidden Coordinates}

With \(\mathbf{Q}_U\) from \eqref{Eq:Hidden-Coordinate-Transformation} and \(\mathbf{X}'=\mathbf{Q}_U^{-1}\mathbf{X}\), we get \(\mathbf{S}_U=\mathbf{G}_U\mathbf{Q}_U\mathbf{X}'=\mathbf{X}'_\mathsf{pub}\) and \(\mathbf{Y}=\mathbf{Q}_U\begin{bmatrix}\mathbf{S}_U\mathbf{A}^\top\\ \mathbf{X}'_\mathsf{hid}\mathbf{A}^\top\end{bmatrix}\).
For \(\mathbf{U}_\mathsf{hid}\randgets\mathbb{Z}_q^{h_U\times dp}\), the matrix \(\mathbf{\breve{Y}}:=\mathbf{Q}_U\begin{bmatrix}\mathbf{S}_U\mathbf{A}^\top\\ \mathbf{U}_\mathsf{hid}\end{bmatrix}\) satisfies \(\mathbf{G}_U\mathbf{\breve{Y}}=\mathbf{S}_U\mathbf{A}^\top\), and since \(\mathbf{Q}_U\) is invertible modulo \(q\), the map \(\mathbf{U}_\mathsf{hid}\mapsto\mathbf{\breve{Y}}\) is a bijection onto \(\mathcal{Y}_U(\mathbf{S}_U\mathbf{A}^\top)\).
Hence a uniform hidden target induces exactly the distribution of \(\mathsf{SimAK}\).

\subsection{Centered Response-Leakage Representation}

Write \(\mathbf{G}\mathbf{Q}_U=\begin{bmatrix}\mathbf{G}'_\mathsf{pub}\ \mathbf{G}'_\mathsf{hid}\end{bmatrix}\), so that \(\mathbf{S}=\mathbf{G}'_\mathsf{pub}\mathbf{S}_U+\mathbf{G}'_\mathsf{hid}\mathbf{X}'_\mathsf{hid}\).
Let \(\mathbf{L}_s\in\mathbb{Z}^{D\times D}\) represent multiplication by \(c_s\), so that \(\mathbf{z}_{i,s}=\mathbf{L}_s\mathbf{s}_i+\mathbf{r}_{i,s}\).
The centered response \(\mathbf{\widetilde{z}}_{i,s}:=\mathbf{z}_{i,s}-\mathbf{L}_s(\mathbf{G}'_\mathsf{pub}[i,\cdot]\mathbf{S}_U)^\top\) equals \(\mathbf{M}_{i,s}\mathbf{x}'_\mathsf{hid}+\mathbf{r}_{i,s}\) with \(\mathbf{M}_{i,s}:=\mathbf{L}_s(\mathbf{G}'_\mathsf{hid}[i,\cdot]\otimes\mathbf{I}_D)\) and \(\mathbf{x}'_\mathsf{hid}=\mathrm{vec}({\mathbf{X}'_\mathsf{hid}}^\top)\).
Stacking over \(\mathcal{S}\) gives \(\mathbf{\widetilde{z}}_\mathcal{S}=\mathbf{M}_\mathsf{hid}\mathbf{x}'_\mathsf{hid}+\mathbf{r}_\mathcal{S}\).
The centering depends only on \(\mathbf{S}_U\) and public data.

\subsection{Identification with \texorpdfstring{\(\bm{\mathsf{RA\text{-}MHMLWE}}\)}{RA-MHMLWE}}

With \(\mathbf{A}^{\langle h_U\rangle}=\mathbf{I}_{h_U}\otimes\mathbf{A}\), we have \(\mathrm{vec}((\mathbf{X}'_\mathsf{hid}\mathbf{A}^\top)^\top)=\mathbf{A}^{\langle h_U\rangle}\mathbf{x}'_\mathsf{hid}\), and \(\mathbf{u}_\mathsf{hid}:=\mathrm{vec}(\mathbf{U}_\mathsf{hid}^\top)\) is uniform.
Conditioned on \(\mathbf{X}'_\mathsf{pub}=\mathbf{S}_U\), the real and simulated views are therefore \((\mathbf{A},\mathbf{A}^{\langle h_U\rangle}\mathbf{x}'_\mathsf{hid},\mathbf{M}_\mathsf{hid},\mathbf{M}_\mathsf{hid}\mathbf{x}'_\mathsf{hid}+\mathbf{r}_\mathcal{S})\) and \((\mathbf{A},\mathbf{u}_\mathsf{hid},\mathbf{M}_\mathsf{hid},\mathbf{M}_\mathsf{hid}\mathbf{x}'_\mathsf{hid}+\mathbf{r}_\mathcal{S})\): an \(\mathsf{RA\text{-}MHMLWE}[h_U]\) instance with repeated matrix \(\mathbf{A}^{\langle h_U\rangle}\), joint secret \(\mathbf{x}'_\mathsf{hid}\), one stacked hint \(\mathbf{M}_\mathsf{hid}\), and perturbation \(\mathbf{r}_\mathcal{S}\) of parameter \(\sigma_{\bm{r}}^2\mathbf{I}\).
Since \(\bm{A}=\begin{bmatrix}\bm{I}_p\ \bm{A}'\end{bmatrix}\), each block is a systematic \(\mathsf{MLWE}\) target as in \Cref{Definition:RA-MHMLWE} with \((n,m)=(p,\ell-p)\).
The conditional secret has parameter matrix \(\mathbf{\Sigma}_{U,\mathsf{hid}\mid\mathsf{pub}}\otimes\mathbf{I}_D\) (\Cref{Lemma:Conditional-Hidden-Coordinate-Precision}) and a center determined by \(\mathbf{S}_U\), which we treat as public auxiliary information.

\section{Min-Entropy of Reconstructed Commitments}\label{Appendix:LaTSEC-Unforgeability-MinEntropy}

\begin{definition}[Systematic \(\mathsf{MLWE}\) samples {\cite[Definition 3.7]{del2024threshold}}]\label{Definition:LaTSEC-Systematic-MLWE-Sample}
	For \(\bm{A}'\in\mathcal{R}_q^{p\times(\ell-p)}\), let \(\mathcal{D}^\mathsf{MLWE}_{p,\ell-p,q,d,\sigma}(\bm{A}')\) be the distribution of \(\begin{bmatrix}\bm{I}_p\ \bm{A}'\end{bmatrix}\bm{r}\bmod q\) for \(\bm{r}\gets\mathcal{D}_\sigma^\ell\).
\end{definition}

\begin{lemma}[Min-entropy of systematic samples {\cite[Lemma 3.8]{del2024threshold}}]\label{Lemma:LaTSEC-Systematic-MLWE-MinEntropy}
	If \(\sigma>2d\cdot q^{1/\ell+2/(d(\ell-p))}\), then \(\Pr_{\bm{A}'}[H_\infty(\mathcal{D}^\mathsf{MLWE}_{p,\ell-p,q,d,\sigma}(\bm{A}'))\geq d-1]\geq1-2^{-d+1}\).
\end{lemma}

This is the no-bit-dropping case of \cite[Lemma 3.8]{del2024threshold}.

\begin{lemma}[Structural non-cancellation]\label{Lemma:Structural-NonCancellation-NonReconstruction}
	Let \(U\subseteq[n]\).
	If \((\mathbf{g})_U\in\mathrm{Col}_\mathbb{Z}(\mathbf{\hat{G}}_U)\), then \(U\) is not certified.
\end{lemma}

\begin{proof}
	Suppose, for contradiction, that \(U\) is certified, that is, \(\bm{\lambda}_U^\top\mathbf{G}_U=\mathbf{e}_1^\top\) for some \(\bm{\lambda}_U\in\mathbb{Z}^{|U|}\).
	Since \(\mathbf{G}_U=\begin{bmatrix}(\mathbf{g})_U\ \mathbf{\hat{G}}_U\end{bmatrix}\), this means \(\bm{\lambda}_U^\top(\mathbf{g})_U=1\) and \(\bm{\lambda}_U^\top\mathbf{\hat{G}}_U=\mathbf{0}^\top\).
	By hypothesis, \((\mathbf{g})_U=\mathbf{\hat{G}}_U\mathbf{a}\) for some \(\mathbf{a}\in\mathbb{Z}^{k-1}\).
	Then \(1=\bm{\lambda}_U^\top(\mathbf{g})_U=(\bm{\lambda}_U^\top\mathbf{\hat{G}}_U)\mathbf{a}=0\), a contradiction.
\end{proof}

\begin{lemma}[Honest nonzero coefficient]\label{Lemma:Honest-Nonzero-Reconstruction-Coefficient}
	Suppose that no set of size at most \(t_\mathsf{uf}\) is certified, let \(\mathsf{act}\) be certified with \(\bm{\lambda}_\mathsf{act}^\top\mathbf{G}_\mathsf{act}=\mathbf{e}_1^\top\), and let \(|\mathsf{corrupt}|\leq t_\mathsf{uf}\).
	Then \(\lambda_{\mathsf{act},i^*}\neq0\) for some \(i^*\in\mathsf{act}\setminus\mathsf{corrupt}\).
\end{lemma}

\begin{proof}
	Suppose, for contradiction, that \(\lambda_{\mathsf{act},i}=0\) for every \(i\in\mathsf{act}\setminus\mathsf{corrupt}\), and let \(U:=\mathsf{act}\cap\mathsf{corrupt}\), so that \(|U|\leq t_\mathsf{uf}\).
	Let \(\bm{\lambda}_U\) be the restriction of \(\bm{\lambda}_\mathsf{act}\) to \(U\).
	Since the omitted coefficients vanish, \(\bm{\lambda}_U^\top\mathbf{G}_U=\bm{\lambda}_\mathsf{act}^\top\mathbf{G}_\mathsf{act}=\mathbf{e}_1^\top\), so \(U\) is certified.
	This contradicts the hypothesis; for \(U=\emptyset\), the left-hand side is the zero vector, which is also impossible.
\end{proof}

\begin{lemma}[Entropy loss of a scalar multiple]\label{Lemma:Entropy-Loss-NonUnit-Coefficient}
	Let \(\bm{X}\) be a random variable over \(\mathcal{R}_q^p\), let \(\lambda\in\mathbb{Z}\), and let \(\bm{\delta}\in\mathcal{R}_q^p\) be fixed.
	Then \(H_\infty(\lambda\bm{X}+\bm{\delta})\geq H_\infty(\bm{X})-dp\log_2\gcd(\lambda,q)\).
\end{lemma}

\begin{proof}
	Let \(g:=\gcd(\lambda,q)\) and \(\phi(\bm{x}):=\lambda\bm{x}+\bm{\delta}\).
	The map \(\phi\) acts independently on the \(dp\) coefficients in \(\mathbb{Z}_q\), and on each coefficient \(y\mapsto\lambda y\bmod q\) is a group homomorphism whose kernel \(\{y\in\mathbb{Z}_q:(q/g)\mid y\}\) has \(g\) elements.
	Hence every fiber of \(\phi\) has at most \(g^{dp}\) elements, and for every \(\bm{v}\),
	\(\Pr[\phi(\bm{X})=\bm{v}]=\sum_{\bm{x}\in\phi^{-1}(\bm{v})}\Pr[\bm{X}=\bm{x}]\leq g^{dp}\,2^{-H_\infty(\bm{X})}\).
	Taking the maximum over \(\bm{v}\) and the negative logarithm gives the claim.
\end{proof}

\begin{lemma}[Min-entropy of the aggregate commitment]\label{Lemma:Reconstructed-Commitment-MinEntropy-General}
	Assume the setting of \Cref{Corollary:Reconstructed-Commitment-MinEntropy-Prime}, but with an arbitrary modulus \(q\) and without the bound \(\beta_{\bm{\lambda}}<q\), and suppose that no set of size at most \(t_\mathsf{uf}\) is certified.
	Let \(i^*\in\mathsf{act}\setminus\mathsf{corrupt}\) minimize \(g:=\gcd(\lambda_{\mathsf{act},i^*},q)\) among the honest active signers with \(\lambda_{\mathsf{act},i^*}\neq0\).
	Then, except with probability \(2^{-d+1}\) over \(\bm{A}'\), conditioned on the adversary's view and on all randomness except \(\bm{r}_{i^*}\), \(H_\infty(\bm{w})\geq d-1-dp\log_2g\).
\end{lemma}

\begin{proof}
	By \Cref{Lemma:Honest-Nonzero-Reconstruction-Coefficient}, some honest active signer has a nonzero coefficient, so \(i^*\) is well defined.
	Fix the adversary's view and all randomness except \(\bm{r}_{i^*}\).
	The corrupted openings are then determined by the view, and the honest openings \(\bm{w}_i=\bm{A}\bm{r}_i\) for \(i\neq i^*\) by the fixed randomness, so
	\(\bm{w}=\sum_{i\in\mathsf{act}}\lambda_{\mathsf{act},i}\bm{w}_i=\lambda_{\mathsf{act},i^*}\bm{A}\bm{r}_{i^*}+\bm{\delta}\) for a fixed \(\bm{\delta}\).
	Since the view is independent of \(\bm{r}_{i^*}\), the conditional distribution of \(\bm{A}\bm{r}_{i^*}\) is still \(\mathcal{D}^\mathsf{MLWE}_{p,\ell-p,q,d,\sigma_{\bm{r}}}(\bm{A}')\).
	By \Cref{Lemma:LaTSEC-Systematic-MLWE-MinEntropy}, except with probability \(2^{-d+1}\) over \(\bm{A}'\), its min-entropy is at least \(d-1\).
	\Cref{Lemma:Entropy-Loss-NonUnit-Coefficient}, applied with \(\bm{X}:=\bm{A}\bm{r}_{i^*}\), \(\lambda:=\lambda_{\mathsf{act},i^*}\), and \(\bm{\delta}\), gives \(H_\infty(\bm{w})\geq d-1-dp\log_2g\).
\end{proof}

\begin{proof}[Proof of \Cref{Corollary:Reconstructed-Commitment-MinEntropy-Prime}]
	For every \(U\) with \(|U|\leq t_\mathsf{uf}=k_p\), \Cref{Item:SEC-Unimodular} of \Cref{Section:SEC-Summary} gives \((\mathbf{g})_U\in\mathrm{Col}_\mathbb{Z}(\mathbf{\hat{G}}_U)\), so no such \(U\) is certified by \Cref{Lemma:Structural-NonCancellation-NonReconstruction}, and \Cref{Lemma:Reconstructed-Commitment-MinEntropy-General} applies.
	Let \(i^*\) be the signer it provides.
	Since \(0<|\lambda_{\mathsf{act},i^*}|\leq\beta_{\bm{\lambda}}<q\) and \(q\) is prime, \(q\nmid\lambda_{\mathsf{act},i^*}\), so \(g=\gcd(\lambda_{\mathsf{act},i^*},q)=1\), and the bound becomes \(H_\infty(\bm{w})\geq d-1\), except with probability \(2^{-d+1}\) over \(\bm{A}'\).
\end{proof}

\section{Running Times}\label{Appendix:Running-Times}

\Cref{Table:LaTSEC-Timings} gives the median running times discussed in \Cref{Section:Implementation-Running-Time}.

\begin{table}[!htb]
	\caption{Median running times of \(\mathsf{LaTSEC}\) in ms on one core, at \(|\mathsf{act}|=t_\mathsf{sig}\)}
	\label{Table:LaTSEC-Timings}
	\centering
	\footnotesize
	\setlength{\tabcolsep}{3pt}
	\begin{tabular}{@{}ccrrrrrrr@{}}
		\toprule
		\(n\) & \(\kappa\) & \(\mathsf{Keygen}\) & \(\mathsf{SS}_1\) & \(\mathsf{SS}_2\) & \(\mathsf{SS}_3\) & \(\mathsf{PV}\) & \(\mathsf{Comb}\) & \(\mathsf{Vfy}\) \\
		\midrule
		\(40\) & \(128\) & \(63\) & \(4.0\) & \({<}0.01\) & \(2.3\) & \(1.0\) & \(18\) & \(0.76\) \\
		\(40\) & \(192\) & \(82\) & \(4.3\) & \({<}0.01\) & \(3.2\) & \(1.5\) & \(29\) & \(1.2\) \\
		\(40\) & \(256\) & \(145\) & \(5.5\) & \({<}0.01\) & \(4.1\) & \(2.2\) & \(44\) & \(1.7\) \\
		\midrule
		\(60\) & \(128\) & \(109\) & \(3.3\) & \({<}0.01\) & \(4.3\) & \(1.3\) & \(33\) & \(0.76\) \\
		\(60\) & \(192\) & \(167\) & \(4.3\) & \({<}0.01\) & \(6.0\) & \(2.0\) & \(52\) & \(1.2\) \\
		\(60\) & \(256\) & \(220\) & \(7.1\) & \({<}0.01\) & \(7.6\) & \(2.8\) & \(76\) & \(1.8\) \\
		\bottomrule
	\end{tabular}
	\par\smallskip
	\parbox{\columnwidth}{\scriptsize \(|\mathsf{act}|=26\) for \(n=40\) and \(46\) for \(n=60\). \(\mathsf{SS}_k\), \(\mathsf{PV}\), \(\mathsf{Comb}\), and \(\mathsf{Vfy}\) denote \(\mathsf{ShareSign}_k\), \(\mathsf{PartialVerify}\), \(\mathsf{Combine}\), and \(\mathsf{Verify}\); signing rounds are per signer, and \(\mathsf{Keygen}\) covers all \(n\) parties.}
\end{table}

\section{Hardness of \texorpdfstring{\(\bm{\mathsf{RA\text{-}MHMLWE}}\)}{RA-MHMLWE}}\label{Appendix:Hardness-RA-MHMLWE}

\subsection{Posterior-Resampling Reduction}\label{Appendix:Posterior-Resampling-Reduction}

We prove the base case of \Cref{Lemma:Hardness-RA-MHMLWE}, without public auxiliary information, by adapting the Matrix Hint-MLWE posterior-resampling argument to the repeated target map.
The source Repeated-\(\bm{A}\) MLWE secret is spherical with independent blocks, whereas the potentially correlated elliptic secret required by the hinted problem is created jointly by posterior resampling.

\begin{proof}[Proof of \Cref{Lemma:Hardness-RA-MHMLWE} without public auxiliary information]
	Let \(N:=dh(n+m)\) and \(N_j:=dn_j\) for \(j\in[\nu]\).
	We first construct an adversary \(\mathcal{B}_0\) against \(\mathsf{RA\text{-}MLWE}_{n,m,q,d,\sigma}[h]\).
	The adversary receives a challenge \((\bm{A},\bm{b})\), where either \(\bm{b}=\overline{\bm{A}}^{\langle h\rangle}\bm{w}\) for a spherical secret whose coefficient embedding satisfies \(\mathbf{w}\gets\mathcal{D}_{\mathbb{Z}^{N},\sigma}\), or \(\bm{b}\stackrel{\$}{\gets}\mathcal{R}_q^{hn}\).

	The reduction samples a dummy secret \(\mathbf{s}'\gets\mathcal{D}_{\mathbb{Z}^{N},\mathbf{c}_{\mathbf{s}},\sqrt{\mathbf{\Sigma}_{\mathbf{s}}}}\).
	For every \(j\in[\nu]\), it independently samples \(\bm{M}_j\gets\mathcal{M}_j\) and \(\mathbf{y}_j\gets\mathcal{D}_{\mathbb{Z}^{N_j},\mathbf{c}_{\mathbf{y}_j},\sqrt{\mathbf{\Sigma}_{\mathbf{y}_j}}}\).
	Let \(\mathbf{M}_j\in\mathbb{Z}^{N_j\times N}\) denote the coefficient embedding of \(\bm{M}_j\), and set \(\mathbf{z}_j:=\mathbf{M}_j\mathbf{s}'+\mathbf{y}_j\).
	For the sampled tuple, define \(\mathbf{\Xi}:=\mathbf{\Sigma}_{\mathbf{s}}^{-1}+\sum_{j=1}^{\nu}\mathbf{M}_j^\top\mathbf{\Sigma}_{\mathbf{y}_j}^{-1}\mathbf{M}_j\).
	By the uniform precision assumption, \(\|\mathbf{\Xi}\|_2\leq1/(2\sigma^2)\) for every tuple that may be sampled.

	For the sampled hint matrices and values, define \(\mathbf{\Sigma}_0:=\mathbf{\Xi}^{-1}\) and \(\mathbf{c}_0 := \mathbf{\Sigma}_0 \big( \mathbf{\Sigma}_{\mathbf{s}}^{-1}\mathbf{c}_{\mathbf{s}} + \sum_{j=1}^{\nu} \mathbf{M}_j^\top \mathbf{\Sigma}_{\mathbf{y}_j}^{-1} (\mathbf{z}_j-\mathbf{c}_{\mathbf{y}_j}) \big)\).
	By the posterior-resampling identity of \cite[Lemma 7]{kim2023toward}, \cite[Lemma 1]{espitau2024flood}, and \cite[Lemma G.2]{wang2025post}, conditioned on the hint matrices and values, the desired secret has distribution \(\mathcal{D}_{\mathbb{Z}^{N},\mathbf{c}_0,\sqrt{\mathbf{\Sigma}_0}}\).

	Since \(\|\mathbf{\Xi}\|_2\leq1/(2\sigma^2)\), we have \(\lambda_{\min}(\mathbf{\Sigma}_0)\geq2\sigma^2\), so \(\mathbf{\Sigma}_0-\sigma^2\mathbf{I}_N\succeq\sigma^2\mathbf{I}_N\succ0\).
	The reduction samples \(\mathbf{t}\gets\mathcal{D}_{\mathbb{Z}^{N},\mathbf{c}_0,\sqrt{\mathbf{\Sigma}_0-\sigma^2\mathbf{I}_N}}\).
	Let \(\bm{t}\in\mathcal{R}^{h(n+m)}\) denote the corresponding ring vector, reduced coefficient-wise modulo \(q\) when used in the principal target.
	The reduction sets \(\bm{b}':=\bm{b}+\overline{\bm{A}}^{\langle h\rangle}\bm{t}\) and gives \(\left(\bm{A},\bm{b}',(\bm{M}_j)_{j\in[\nu]},(\bm{z}_j)_{j\in[\nu]}\right)\) to \(\mathcal{A}\).

	All samples above are exact.
	The hint noise is sampled as in the problem.
	For the dummy secret and the correction, let \(\mathbf{c}\) be a center and \(\mathbf{\Sigma}\succeq\sigma^2\mathbf{I}_N\) a parameter matrix.
	The exact sampler of \cite[Lemma 2.3]{brakerski2013classical}, run on the coset \(\mathbf{\Sigma}^{-1/2}(\mathbb{Z}^N-\mathbf{c})\) of the lattice \(\mathbf{\Sigma}^{-1/2}\mathbb{Z}^N\) with parameter \(\sqrt{2\pi}\) in its normalization, outputs \(\mathbf{y}\) such that \(\mathbf{\Sigma}^{1/2}\mathbf{y}+\mathbf{c}\) follows \(\mathcal{D}_{\mathbb{Z}^N,\mathbf{c},\sqrt{\mathbf{\Sigma}}}\) exactly.
	The basis \(\mathbf{\Sigma}^{-1/2}\) has Gram--Schmidt norms at most \(\|\mathbf{\Sigma}^{-1/2}\|_2\leq1/\sigma\), so the width condition of that lemma holds when \(\sigma^2\geq\ln(2N+4)/(2\pi^2)\).
	In our normalization, \(\eta_\epsilon(\mathbb{Z}^N)\geq\eta_\epsilon(\mathbb{Z})\geq\sqrt{\ln(2/\epsilon)}/(\sqrt{2}\pi)\), so \(\sigma\geq\sqrt{2}\eta_\epsilon(\mathbb{Z}^N)\) gives \(\sigma^2\geq\ln(2/\epsilon)/\pi^2\), which suffices because \(\epsilon\leq1/\sqrt{2N+4}\).
	Both \(\mathbf{\Sigma}_\mathbf{s}\succeq\mathbf{\Xi}^{-1}\succeq2\sigma^2\mathbf{I}_N\) and \(\mathbf{\Sigma}_0-\sigma^2\mathbf{I}_N\succeq\sigma^2\mathbf{I}_N\) satisfy \(\mathbf{\Sigma}\succeq\sigma^2\mathbf{I}_N\).

	Suppose first that the source challenge is real.
	By linearity of the repeated target map, \(\bm{b}'=\overline{\bm{A}}^{\langle h\rangle}(\bm{w}+\bm{t})\).
	Set \(\mathbf{\Sigma}_1:=\sigma^2\mathbf{I}_N\), \(\mathbf{\Sigma}_2:=\mathbf{\Sigma}_0-\sigma^2\mathbf{I}_N\), and \(\mathbf{\Sigma}_3^{-1}:=\mathbf{\Sigma}_1^{-1}+\mathbf{\Sigma}_2^{-1}\).
	Since \(\mathbf{\Sigma}_2\succeq\sigma^2\mathbf{I}_N\), we have \(\|\mathbf{\Sigma}_3^{-1}\|_2\leq2/\sigma^2\leq1/\eta_\epsilon(\mathbb{Z}^{N})^2\).
	As in the proof of \cite[Corollary~1]{arzaki2026SEC}, this is the hypothesis of the convolution lemma \cite[Lemmas 3 and 5]{kim2023toward} in the normalization \cite[Eq.~(2)]{arzaki2026SEC}, which therefore shows that \(\mathbf{w}+\mathbf{t}\) is within statistical distance at most \(2\epsilon\) of \(\mathcal{D}_{\mathbb{Z}^{N},\mathbf{c}_0,\sqrt{\mathbf{\Sigma}_0}}\).
	Together with the posterior-resampling identity, this implies that the reduction's output is within statistical distance at most \(2\epsilon\) of the real Repeated-\(\bm{A}\) Matrix Hint-MLWE distribution.

	Suppose instead that \(\bm{b}\) is independently uniform over \(\mathcal{R}_q^{hn}\).
	Then \(\bm{b}+\overline{\bm{A}}^{\langle h\rangle}\bm{t}\) remains uniformly distributed over \(\mathcal{R}_q^{hn}\), so the reduction produces the random Repeated-\(\bm{A}\) Matrix Hint-MLWE distribution exactly.
	Hence, \(\mathsf{Adv}_{\mathcal{A}}^{\mathsf{RA\text{-}MHMLWE}[h]}(\kappa) \leq \mathsf{Adv}_{\mathcal{B}_0}^{\mathsf{RA\text{-}MLWE}[h]}(\kappa) + 2\epsilon\).
	Applying \Cref{Lemma:Hardness-RA-MLWE} yields a PPT adversary \(\mathcal{B}\) against \(\mathsf{MLWE}_{n,m,q,d,\sigma}\) such that \(\mathsf{Adv}_{\mathcal{B}_0}^{\mathsf{RA\text{-}MLWE}[h]}(\kappa) \leq h\,\mathsf{Adv}_{\mathcal{B}}^{\mathsf{MLWE}}(\kappa)\). Combining the two inequalities proves \(\mathsf{Adv}_{\mathcal{A}}^{\mathsf{RA\text{-}MHMLWE}[h]}(\kappa) \leq h\,\mathsf{Adv}_{\mathcal{B}}^{\mathsf{MLWE}}(\kappa)+2\epsilon\).
\end{proof}

\subsection{Extension to Public Auxiliary Information}\label{Appendix:Extension-Public-Aux}

We now prove the public-auxiliary-information extension in the final claim of \Cref{Lemma:Hardness-RA-MHMLWE}.
Let \(\mathsf{aux}\) denote the public auxiliary information jointly distributed with the hinted secret, and generate their joint distribution by first sampling \(\mathsf{aux}\) from its marginal distribution and then sampling the secret from its conditional distribution given \(\mathsf{aux}\).
The reduction first samples \(\mathsf{aux}\), exactly by assumption, and then runs conditionally on its outcome.

Fix an outcome \(a\) of \(\mathsf{aux}\).
By assumption, conditioned on \(\mathsf{aux}=a\), the secret is an elliptic discrete Gaussian over \(\mathbb{Z}^N\) with publicly computable center \(\mathbf{c}_{\mathbf{s}\mid a}\) and parameter matrix \(\mathbf{\Sigma}_{\mathbf{s}\mid a}\).
The reduction in Appendix~\ref{Appendix:Posterior-Resampling-Reduction} therefore applies with these parameters, and its conditional dummy secret and correction are sampled exactly as there, since the uniform precision condition gives \(\mathbf{\Sigma}_{\mathbf{s}\mid a}\succeq2\sigma^2\mathbf{I}_N\) and \(\mathbf{\Sigma}_0-\sigma^2\mathbf{I}_N\succeq\sigma^2\mathbf{I}_N\) for every outcome \(a\).
Since the precision and smoothing conditions hold uniformly over all auxiliary-information outcomes and supported tuples of hint matrices, the conditional real branch is simulated within statistical distance at most \(2\epsilon\), whereas the conditional random branch is simulated exactly.
Averaging over the marginal distribution of \(\mathsf{aux}\) preserves the same bound, so the reduction and factor-\(h\) advantage loss established above continue to hold.
\section{The Underlying No-Abort Signature}\label{Appendix:Underlying-No-Abort-Signature}

We recall the non-threshold signature scheme underlying \(\mathsf{LaTSEC}\).
The construction follows the Raccoon-style variant of Lyubashevsky's Fiat-Shamir signature scheme \cite{lyubashevsky2009fiat,lyubashevsky2012lattice,del2024raccoon,del2024threshold}, adapted to the notation of this work and without compression or rounding.

The scheme avoids the rejection-sampling step of the original Fiat-Shamir construction.
Such rejection sampling is undesirable in a threshold setting because a rejection by any active signer may require the entire signing session to restart in order to preserve the privacy of the shared signing key.
Instead, following the noise-flooding approach of \cite{goldwasser2010robustness}, the signing mask is sampled from a sufficiently wide distribution so that the response hides the signing key without rejection sampling.
The same general approach has also been used in lattice-based threshold signature constructions \cite{agrawal2022round,gur2024two,del2024threshold}. Here, ``no-abort'' refers specifically to the absence of rejection-sampling aborts during signature generation.

Let \(\bm{A} := \begin{bmatrix} \bm{I}_p \ \bm{A}' \end{bmatrix} \in \mathcal{R}_q^{p\times\ell}\), where \(\bm{A}' \stackrel{\$}{\gets} \mathcal{R}_q^{p\times(\ell-p)}\) and \(\ell>p\).
The signing key is sampled as \(\bm{s} = \begin{bmatrix} \bm{e}\\ \bm{s}' \end{bmatrix} \gets \mathcal{D}_{\sigma_{\bm{s}}}^\ell\), where \(\bm{e}\in\mathcal{R}^{p}\) and \(\bm{s}'\in\mathcal{R}^{\ell-p}\).
The verification target is \(\bm{t} := \bm{A}\bm{s} = \bm{A}'\bm{s}' + \bm{e} \pmod q\).
Consequently, \((\bm{A}',\bm{t})\) has the usual systematic \(\mathsf{MLWE}\) form with secret \(\bm{s}'\) and error \(\bm{e}\).
The complete construction is given in \Cref{Figure:Underlying-Signature}.

\begin{figure}[!htp]
	\centering
	\begin{tcolorbox}[gamebox]
		\((\mathsf{sk},\mathsf{vk}) \gets \mathsf{Keygen}(1^\kappa)\)
		\hrule height 0.4pt
		\begin{algorithmic}[1]
			\State \(\bm{A}' \stackrel{\$}{\gets} \mathcal{R}_q^{p\times(\ell-p)}\) and set \(\bm{A} := \begin{bmatrix} \bm{I}_p \ \bm{A}' \end{bmatrix}\).
			\State \(\bm{s} \gets \mathcal{D}_{\sigma_{\bm{s}}}^\ell\) \Comment{ \(\bm{s} = \begin{bmatrix} \bm{e}\\ \bm{s}' \end{bmatrix} \in \mathcal{R}^\ell\) }
			\State \(\bm{t}:=\bm{A}\bm{s}\) \Comment{\(\bm{t}\in\mathcal{R}_q^p\)}
			\State \Return \(\mathsf{sk}:=\bm{s}\); \(\mathsf{vk}:=(\bm{A},\bm{t})\)
		\end{algorithmic}

		\smallskip
		\(\mathsf{sig} \gets \mathsf{Sign} (\mathsf{vk},\mathsf{sk},\mathsf{msg})\)
		\hrule height 0.4pt
		\begin{algorithmic}[1]
			\State \(\bm{r} \gets \mathcal{D}_{\sigma_{\bm{r}}}^\ell\) \Comment{signing mask in \(\mathcal{R}^\ell\)}
			\State \(\bm{w}:=\bm{A}\bm{r}\) \Comment{commitment in \(\mathcal{R}_q^p\)}
			\State \(c := H_c(\mathsf{vk},\mathsf{msg},\bm{w})\) \Comment{challenge in \(\mathcal{C}\)}
			\State \(\bm{z}:=c\bm{s}+\bm{r}\) \Comment{response in \(\mathcal{R}^\ell\)}
			\State \Return \(\mathsf{sig}:=(c,\bm{z})\)
		\end{algorithmic}

		\smallskip
		\(\{0,1\} \gets \mathsf{Verify} (\mathsf{vk},\mathsf{msg},\mathsf{sig})\)
		\hrule height 0.4pt
		\begin{algorithmic}[1]
			\State \((c,\bm{z}) := \mathrm{Parse}(\mathsf{sig})\); \textbf{if} this fails, \textbf{return} \(0\)
			\State \(\bm{w}' := \bm{A}\bm{z} - c\bm{t}\)
			\State \(c' := H_c(\mathsf{vk},\mathsf{msg},\bm{w}')\)
			\If{ \(c=c'\) \textbf{and} \(\|\bm{z}\|_2\leq B_2\) }
				\State \Return \(1\) \Comment{\(\mathsf{sig}\) is a valid signature}
			\Else
				\State \Return \(0\) \Comment{\(\mathsf{sig}\) is an invalid signature}
			\EndIf
		\end{algorithmic}
	\end{tcolorbox}
	\caption{The underlying Raccoon-style no-abort signature scheme, adapted from \cite{del2024threshold} without compression or rounding.}
	\label{Figure:Underlying-Signature}
\end{figure}

The commitment and verification relation are interpreted over \(\mathcal{R}_q\), whereas the response \(\bm{z}\) is viewed as an element of \(\mathcal{R}^\ell\) and must satisfy the prescribed norm bound.
For an honestly generated signature, \(\bm{A}\bm{z} - c\bm{t} = \bm{A}(c\bm{s}+\bm{r}) - c\bm{A}\bm{s} = \bm{A}\bm{r} = \bm{w}\).
Hence, verification reconstructs the original commitment and recomputes the same challenge.
The threshold construction in \Cref{Section:LaTSEC-Construction} preserves the signature format \((c,\bm{z})\) and this verification relation.
\section{Full Specification of \texorpdfstring{\(\bm{\mathsf{LaTSEC}}\)}{LaTSEC}}\label{Appendix:LaTSEC-Full-Specification}

\subsection{Syntax and Session States}\label{Section:LaTSEC-FormalSyntax}

We model \(\mathsf{LaTSEC}\) as a synchronous three-round threshold signature scheme with identifiable abort.
Each signing execution is associated with a unique public session identifier \(\mathsf{sid}\), a message \(\mathsf{msg}\), and an active set \(\mathsf{act}\subseteq[n]\).
Messages sent in one round are assumed to be delivered consistently to all active parties before the next round begins, for example, through an authenticated broadcast channel or a public bulletin board.
The scheme is defined by the following algorithm interfaces, where \(\mathsf{PM}:=(\mathsf{pm}_j^k)_{k\in[3],j\in\mathsf{act}}\):
\begingroup\raggedright
    \begin{enumerate}
    	\item \(\mathsf{tspar}\gets\mathsf{LaTSEC.Setup}(1^\kappa,n,t_\mathsf{uf},t_\mathsf{sig})\);
    	\item \((\mathsf{vk},\mathsf{ak},(\mathsf{sk}_i)_{i\in[n]})\gets\mathsf{LaTSEC.Keygen}(\mathsf{tspar})\);
    	\item \((\mathsf{st}_i,\mathsf{pm}_i^1)\gets\mathsf{LaTSEC.ShareSign}_1(\mathsf{st}_i,\allowbreak\mathsf{sid},\allowbreak\mathsf{act},\allowbreak\mathsf{msg})\);
    	\item \((\mathsf{st}_i,\mathsf{pm}_i^2)\gets\mathsf{LaTSEC.ShareSign}_2(\mathsf{st}_i,\allowbreak\mathsf{sid},\allowbreak\mathsf{act},\allowbreak(\mathsf{pm}_j^1)_{j\in\mathsf{act}})\);
    	\item \((\mathsf{st}_i,\mathsf{pm}_i^3)\mid(\mathsf{abort},S_\bot^i)\gets\mathsf{LaTSEC.ShareSign}_3(\mathsf{st}_i,\allowbreak\mathsf{sid},\allowbreak\mathsf{act},\allowbreak(\mathsf{pm}_j^2)_{j\in\mathsf{act}})\);
    	\item \(\mathsf{sig}\mid(\mathsf{abort},S_\bot)\gets\mathsf{LaTSEC.Combine}(\mathsf{tspar},\allowbreak\mathsf{vk},\allowbreak\mathsf{ak},\allowbreak\mathsf{sid},\allowbreak\mathsf{act},\allowbreak\mathsf{msg},\allowbreak\mathsf{PM})\);
    	\item \(\{0,1,\bot\}\ni b\gets\mathsf{LaTSEC.PartialVerify}(\mathsf{tspar},\allowbreak\mathsf{vk},\allowbreak\mathsf{ak}_i,\allowbreak\mathsf{sid},\allowbreak\mathsf{act},\allowbreak\mathsf{msg},\allowbreak\mathsf{PM})\);
    	\item \(\{0,1\}\ni b\gets\mathsf{LaTSEC.Verify}(\mathsf{vk},\allowbreak\mathsf{msg},\allowbreak\mathsf{sig})\).
    \end{enumerate}
\endgroup
Here, \(\mathsf{ak}:=(\mathsf{ak}_i)_{i\in[n]}\) denotes the public abort information, \(\mathsf{pm}_i^k\) denotes the message broadcast by party \(i\) in round \(k\), and \(\mathsf{pm}^k := (\mathsf{pm}_i^k)_{i\in\mathsf{act}}\) denotes the corresponding round transcript.
The key-generation algorithm initializes the local state \(\mathsf{st}_i:=(\mathsf{tspar},\mathsf{vk},\mathsf{sk}_i)\) of each party \(i\in[n]\), with an empty session table, and \(\mathsf{vk}\) implicitly includes \(\mathsf{tspar}\).
Whenever an algorithm returns an identifiable abort, \(S_\bot^i\) or \(S_\bot\) contains the parties identified as having supplied invalid protocol messages.
The parameters and hash functions are those of \Cref{Section:LaTSEC-Overview}.

Signer \(i\) maintains a long-lived user state containing its index \(i\), its secret signing share \(\mathsf{sk}_i\), its public abort key \(\mathsf{ak}_i\), the global verification key \(\mathsf{vk}\), and the public parameters \(\mathsf{tspar}\).
To avoid ambiguity, we reserve the notation \(\mathsf{pp}\) for the public parameters of the underlying secret-sharing scheme.
In our construction, \(\mathsf{tspar}\) includes \(\mathsf{pp}\) together with the other public parameters of the threshold signature scheme.
Each execution of the signing protocol is indexed by a unique public session identifier \(\mathsf{sid}\).
Signer \(i\)'s per-session state is stored as \(\mathsf{st}_i.\mathsf{session}[\mathsf{sid}] = \left(\mathsf{status},\mathsf{rnd},\mathsf{act},\mathsf{msg},\mathsf{open},\mathsf{aux}\right)\), where \(\mathsf{status}\) records whether the session is running, completed, or aborted; \(\mathsf{rnd}\) records the current signing round; \(\mathsf{act}\) and \(\mathsf{msg}\) record the public signing context (the active signer set and the message, respectively); \(\mathsf{open}\) stores the signer's local ephemeral opening information, including the ephemeral randomness, the revealed opening, and the corresponding hash commitment; and \(\mathsf{aux}\) stores any earlier-round transcript data or other auxiliary values needed in subsequent rounds.
Once a session is completed or aborted, its masks and other ephemeral data are erased, but its session identifier stays recorded in \(\mathsf{st}_i\), so the same \(\mathsf{sid}\) is never accepted as fresh again.

\subsection{Algorithms}\label{Section:LaTSEC-Setup-KeyGeneration}

The setup algorithm initializes the underlying \(\mathsf{SEC}\) instance and collects the public parameters required by the threshold signature scheme.
We set \(t_\mathsf{uf}:=k_p\) and \(t_\mathsf{sig}:=k_c\), where \(k_p\) and \(k_c\) are the privacy and empirical reconstruction thresholds of \(\mathsf{SEC}\), respectively.
The secret dimension of the \(\mathsf{SEC}\) instance is set to \(d\ell\), corresponding to the coefficient embedding of vectors in \(\mathcal{R}^\ell\).
For notational simplicity, the globally fixed ring and module parameters are left implicit in \(\mathsf{tspar}\).
The setup algorithm is given in \Cref{Algorithm:LaTSEC-Setup}.

\begin{algorithm*}[!t]
	\caption{ \(\mathsf{tspar} \gets \mathsf{LaTSEC.Setup} (1^\kappa,n,t_\mathsf{uf},t_\mathsf{sig})\) }
	\label{Algorithm:LaTSEC-Setup}
	\small
	\begin{algorithmic}[1]
		\small
		\Require Security parameter \(1^\kappa\), number of parties \(n\), unforgeability threshold \(t_\mathsf{uf}\), and signing threshold \(t_\mathsf{sig}\).
		\Ensure Public parameters \(\mathsf{tspar} = (\mathsf{pp},\mathcal{C},B_1,B_2, \sigma_{\bm{s}},\sigma_{\bm{r}},\beta_{\bm{\lambda}},\omega_{\bm{\lambda}})\).
		\State Instantiate \(\mathsf{SEC}\) for \(n\) parties, privacy threshold \(k_p=t_\mathsf{uf}\), empirical reconstruction threshold \(k_c=t_\mathsf{sig}\), and secret dimension \(d\ell\).
        If no such instance is available, return \(\bot\).
		\State \(\mathsf{pp} \gets \mathsf{SEC.Setup}(1^\kappa)\)
		\State Fix the challenge set \(\mathcal{C} := \left\{ c\in\mathcal{R}: \|c\|_\infty=1 \land \|c\|_1=\omega_c \right\}\).
		\State Fix the Gaussian widths \(\sigma_{\bm{s}}\) and \(\sigma_{\bm{r}}\).
		\State Fix the reconstruction bounds \((\beta_{\bm{\lambda}},\omega_{\bm{\lambda}})\), which, with \(\mathsf{pp}\), determine the admissible session inputs.
		\State Fix the verification bounds \(B_1\) and \(B_2\) subject to the parameter conditions in \Cref{Section:LaTSEC-Asymptotic-Parameters}.
		\State \(\mathsf{tspar} := (\mathsf{pp},\mathcal{C},B_1,B_2, \sigma_{\bm{s}},\sigma_{\bm{r}},\beta_{\bm{\lambda}},\omega_{\bm{\lambda}})\)
		\State \Return \(\mathsf{tspar}\)
	\end{algorithmic}
\end{algorithm*}

The trusted key-generation algorithm samples the public matrix and the short signing key of the underlying no-abort signature scheme.
In particular, a short secret \(\bm{s}\gets\mathcal{D}_{\sigma_{\bm{s}}}^\ell\) is sampled and then shared among the \(n\) parties using the dealing algorithm \(\mathsf{SEC.Deal}_\mathsf{pp}\) from \cite[Definition~11]{arzaki2026SEC}, as defined in \cite[Eq.~(3)]{arzaki2026SEC}.
Under the coefficient embedding, \(\bm{s}\in\mathcal{R}^\ell\) is identified with an integer vector \(\mathbf{s}\in\mathbb{Z}^{d\ell}\). The resulting integer shares are identified with ring-vector shares \((\bm{s}_i)_{i\in[n]}\) in \(\mathcal{R}^\ell\).
For each party \(i\in[n]\), the secret signing-key share and public abort key are defined as \(\mathsf{sk}_i:=\bm{s}_i\) and \(\mathsf{ak}_i:=\bm{t}_i:=\bm{A}\bm{s}_i\), respectively.
The key-generation algorithm is given in \Cref{Algorithm:LaTSEC-Keygen}.

\begin{algorithm*}[!t]
	\caption{ \((\mathsf{vk},\mathsf{ak},(\mathsf{sk}_i)_{i\in[n]}) \gets \mathsf{LaTSEC.Keygen}(\mathsf{tspar})\) }
	\label{Algorithm:LaTSEC-Keygen}
	\small
	\begin{algorithmic}[1]
		\small
		\Require Public parameters \(\mathsf{tspar} = (\mathsf{pp},\mathcal{C},B_1,B_2, \sigma_{\bm{s}},\sigma_{\bm{r}},\beta_{\bm{\lambda}},\omega_{\bm{\lambda}})\).
		\Ensure Verification key \(\mathsf{vk}\), public abort information \(\mathsf{ak}\), and secret key shares \((\mathsf{sk}_i)_{i\in[n]}\).
		\State Parse \(\mathsf{tspar} = (\mathsf{pp},\mathcal{C},B_1,B_2, \sigma_{\bm{s}},\sigma_{\bm{r}},\beta_{\bm{\lambda}},\omega_{\bm{\lambda}})\).
		\State \(\bm{A}' \stackrel{\$}{\gets} \mathcal{R}_q^{p\times(\ell-p)}\) and set \(\bm{A} := \begin{bmatrix} \bm{I}_p \ \bm{A}' \end{bmatrix}\).
		\State \(\bm{s} \gets \mathcal{D}_{\sigma_{\bm{s}}}^\ell\) \Comment{ \(\bm{s}\in\mathcal{R}^\ell\), with coefficient embedding \(\mathbf{s}\in\mathbb{Z}^{d\ell}\) }
		\State \((\bm{s}_i)_{i\in[n]} \gets \mathsf{SEC.Deal}_{\mathsf{pp}}(\bm{s})\) \Comment{Share \(\bm{s}\) using \cite[Eq.~(3)]{arzaki2026SEC}}
		\State \(\bm{t}:=\bm{A}\bm{s}\) \Comment{\(\bm{t}\in\mathcal{R}_q^p\)}
		\State \(\bm{t}_i:=\bm{A}\bm{s}_i\) for every \(i\in[n]\) \Comment{\(\bm{t}_i\in\mathcal{R}_q^p\)}
		\State \(\mathsf{vk}:=(\bm{A},\bm{t})\)
		\State \(\mathsf{ak}:=(\mathsf{ak}_i)_{i\in[n]} :=(\bm{t}_i)_{i\in[n]}\)
		\State \(\mathsf{sk}_i:=\bm{s}_i\) for every \(i\in[n]\)
		\State Initialize \(\mathsf{st}_i:=(\mathsf{tspar},\mathsf{vk},\mathsf{sk}_i)\) with an empty session table, for every \(i\in[n]\).
		\State \Return \((\mathsf{vk},\mathsf{ak},(\mathsf{sk}_i)_{i\in[n]})\)
	\end{algorithmic}
\end{algorithm*}

\subsubsection{Three-Round Signing Protocol}\label{Section:LaTSEC-Signing}

Let \(\mathsf{sid}\) identify a fresh signing execution with an admissible session input \((\mathsf{sid},\mathsf{act},\mathsf{msg})\), where \(\mathsf{act}\subseteq[n]\) and \(\mathsf{msg}\) encodes \((\mathsf{sid},\mathsf{act},\bm{\lambda}_\mathsf{act})\) as in \Cref{Section:LaTSEC-Overview}.
Throughout, \(\mathsf{SEC.Rec}_\mathsf{pp}(\mathsf{act},\cdot)\) uses this vector \(\bm{\lambda}_\mathsf{act}\), which every signer parses from the stored \(\mathsf{msg}\).
The three round messages are given by \(\mathsf{pm}_i^1:=\mathsf{cmt}_i\), \(\mathsf{pm}_i^2:=\bm{w}_i\), and \(\mathsf{pm}_i^3:=\bm{z}_i\).
The local session state \(\mathsf{st}_i.\mathsf{session}[\mathsf{sid}]\) stores signer \(i\)'s ephemeral opening information and the transcript data required in subsequent rounds.
Failed assertion checks in the algorithms below represent malformed or inconsistent invocations and are distinct from identifiable aborts caused by invalid protocol messages.

In the first round, each signer \(i\in\mathsf{act}\) samples a fresh signing mask \(\bm{r}_i\), computes its local commitment \(\bm{w}_i:=\bm{A}\bm{r}_i\), and broadcasts the hash commitment \(\mathsf{pm}_i^1 := \mathsf{cmt}_i := H_\mathsf{cmt} (\mathsf{sid},\mathsf{vk},\mathsf{act},i,\bm{w}_i)\).
The field \(\mathsf{open}\) stores signer \(i\)'s local opening information \((\bm{r}_i,\bm{w}_i,\mathsf{cmt}_i)\) for the session, while \(\mathsf{aux}\) is reserved for auxiliary per-session data required in later rounds.
The first round is given in \Cref{Algorithm:LaTSEC-ShareSign1}.

\begin{algorithm*}[!t]
	\caption{ \((\mathsf{st}_i,\mathsf{pm}_i^1) \gets \mathsf{LaTSEC.ShareSign}_1 (\mathsf{st}_i,\mathsf{sid},\mathsf{act},\mathsf{msg})\) }
	\label{Algorithm:LaTSEC-ShareSign1}
	\small
	\begin{algorithmic}[1]
		\Require Local state \(\mathsf{st}_i\), fresh session identifier \(\mathsf{sid}\), active set \(\mathsf{act}\), and message \(\mathsf{msg}\), where \(\mathsf{st}_i\) contains \(\mathsf{sk}_i=\bm{s}_i\), \(\mathsf{vk}=(\bm{A},\bm{t})\), and \(\mathsf{tspar}\).
		\Ensure Updated state \(\mathsf{st}_i\) and first-round message \(\mathsf{pm}_i^1=\mathsf{cmt}_i\).
		\State \textbf{assert} \(\{ \mathsf{sid} \notin \mathsf{st}_i.\mathsf{session} \}\) \Comment{session consistency check}
		\State \textbf{assert} \(\{ i\in\mathsf{act} \}\)
		\State \textbf{assert} \(\{ \mathcal{P}(\mathsf{tspar},\mathsf{vk},\mathsf{sid},\mathsf{act},\mathsf{msg}) =1 \}\) \Comment{admissibility check}
		\State \(\bm{r}_i \gets \mathcal{D}_{\sigma_{\bm{r}}}^\ell\) \Comment{\(\bm{r}_i\in\mathcal{R}^\ell\)}
		\State \(\bm{w}_i:=\bm{A}\bm{r}_i\) \Comment{\(\bm{w}_i\in\mathcal{R}_q^p\)}
		\State \(\mathsf{cmt}_i := H_\mathsf{cmt} (\mathsf{sid},\mathsf{vk},\mathsf{act},i,\bm{w}_i)\) \Comment{\(\mathsf{cmt}_i\in\{0,1\}^{2\kappa}\)}
		\State \(\mathsf{st}_i.\mathsf{session}[\mathsf{sid}] := ( \mathsf{status}=\mathsf{running}, \mathsf{rnd}=1, \mathsf{act}, \mathsf{msg}, \mathsf{open} = (\bm{r}_i,\bm{w}_i,\mathsf{cmt}_i), \mathsf{aux}=\emptyset )\)
		\State \(\mathsf{pm}_i^1:=\mathsf{cmt}_i\)
		\State \Return \((\mathsf{st}_i,\mathsf{pm}_i^1)\)
	\end{algorithmic}
\end{algorithm*}

In the second round, each signer receives the first-round transcript \((\mathsf{pm}_j^1)_{j\in\mathsf{act}}\), stores it in the auxiliary part of its session state, and reveals its local commitment by broadcasting \(\mathsf{pm}_i^2:=\bm{w}_i\).
Although the generic TSS syntax permits every round algorithm to output \((\mathsf{abort},S_\bot^i)\), the concrete algorithm \(\mathsf{LaTSEC.ShareSign}_2\) does not abort on well-formed inputs.
Verification of the commitment openings is deferred to the third round, where an inconsistent opening can be attributed to the corresponding signer.
The second round is given in \Cref{Algorithm:LaTSEC-ShareSign2}.

\begin{algorithm*}[!t]
	\caption{ \((\mathsf{st}_i,\mathsf{pm}_i^2) \gets \mathsf{LaTSEC.ShareSign}_2 (\mathsf{st}_i,\mathsf{sid},\mathsf{act}, (\mathsf{pm}_j^1)_{j\in\mathsf{act}})\) }
	\label{Algorithm:LaTSEC-ShareSign2}
	\small
	\begin{algorithmic}[1]
		\Require Local state \(\mathsf{st}_i\), session identifier \(\mathsf{sid}\), active set \(\mathsf{act}\), and first-round transcript \((\mathsf{pm}_j^1)_{j\in\mathsf{act}}\).
		\Ensure Updated state \(\mathsf{st}_i\) and second-round message \(\mathsf{pm}_i^2=\bm{w}_i\).
		\State \textbf{assert} \(\{ \mathsf{sid} \in \mathsf{st}_i.\mathsf{session} \}\) \Comment{session consistency check}
		\State Fetch \(\mathsf{st}_i.\mathsf{session}[\mathsf{sid}] = ( \mathsf{status}, \mathsf{rnd}, \mathsf{act}', \mathsf{msg}, \mathsf{open}, \mathsf{aux} )\).
		\State \textbf{assert} \(\{ \mathsf{status}=\mathsf{running} \land \mathsf{rnd}=1 \land \mathsf{act}'=\mathsf{act} \}\)
		\State \textbf{assert} \(\{ i\in\mathsf{act} \}\)
		\State Parse \(\mathsf{open} = (\bm{r}_i,\bm{w}_i,\mathsf{cmt}_i)\).
		\State \textbf{assert} \(\{ \mathsf{pm}_i^1=\mathsf{cmt}_i \}\)
		\State \(\mathsf{st}_i.\mathsf{session}[\mathsf{sid}] := ( \mathsf{status}=\mathsf{running}, \mathsf{rnd}=2, \mathsf{act}, \mathsf{msg}, \mathsf{open} = (\bm{r}_i,\bm{w}_i,\mathsf{cmt}_i), \mathsf{aux} = (\mathsf{pm}_j^1)_{j\in\mathsf{act}} )\)
		\State \(\mathsf{pm}_i^2:=\bm{w}_i\)
		\State \Return \((\mathsf{st}_i,\mathsf{pm}_i^2)\)
	\end{algorithmic}
\end{algorithm*}

In the third round, each signer receives the commitment openings \((\mathsf{pm}_j^2)_{j\in\mathsf{act}} = (\bm{w}_j)_{j\in\mathsf{act}}\) and retrieves the first-round commitments \((\mathsf{cmt}_j)_{j\in\mathsf{act}}\) from its session state.
For every \(j\in\mathsf{act}\), the signer checks whether \(\mathsf{cmt}_j = H_\mathsf{cmt} (\mathsf{sid},\mathsf{vk},\mathsf{act},j,\bm{w}_j)\).
If the check fails for some \(j\), signer \(i\) outputs \((\mathsf{abort},\{j\})\), thereby identifying signer \(j\) as having supplied an invalid commitment opening.
If all openings are valid, the aggregate commitment is reconstructed using the public reconstruction map of \(\mathsf{SEC}\): \(\bm{w} := \mathsf{SEC.Rec}_\mathsf{pp} \left( \mathsf{act}, (\bm{w}_j)_{j\in\mathsf{act}} \right) \in \mathcal{R}_q^p\).
Equivalently, if \(\bm{W}[j,\cdot]:=\bm{w}_j^\top\) and \(\bm{\lambda}_\mathsf{act}\) is the reconstruction vector associated with \(\mathsf{act}\), then \( \bm{w} = \bm{W}^\top\bm{\lambda}_\mathsf{act}\).
The common challenge and partial response are then computed as \(c := H_c (\mathsf{sid},\mathsf{vk},\mathsf{act},\mathsf{msg},\bm{w})\) and  \(\bm{z}_i := c\bm{s}_i+\bm{r}_i\), respectively.
The third round is given in \Cref{Algorithm:LaTSEC-ShareSign3}.

\begin{algorithm*}[!t]
	\caption{ \((\mathsf{st}_i,\mathsf{pm}_i^3) \mid (\mathsf{abort},S_\bot^i) \gets \mathsf{LaTSEC.ShareSign}_3 (\mathsf{st}_i,\mathsf{sid},\mathsf{act}, (\mathsf{pm}_j^2)_{j\in\mathsf{act}})\) }
	\label{Algorithm:LaTSEC-ShareSign3}
	\small
	\begin{algorithmic}[1]
		\Require Local state \(\mathsf{st}_i\), session identifier \(\mathsf{sid}\), active set \(\mathsf{act}\), and second-round transcript \((\mathsf{pm}_j^2)_{j\in\mathsf{act}}\).
		\Ensure Updated state and third-round message \(\mathsf{pm}_i^3=\bm{z}_i\), or an identifiable abort \((\mathsf{abort},S_\bot^i)\).
		\State \textbf{assert} \(\{ \mathsf{sid} \in \mathsf{st}_i.\mathsf{session} \}\) \Comment{session consistency check}
		\State Fetch \(\mathsf{st}_i.\mathsf{session}[\mathsf{sid}] = ( \mathsf{status}, \mathsf{rnd}, \mathsf{act}', \mathsf{msg}, \mathsf{open}, \mathsf{aux} )\).
		\State \textbf{assert} \(\{ \mathsf{status}=\mathsf{running} \land \mathsf{rnd}=2 \land \mathsf{act}'=\mathsf{act} \}\)
		\State \textbf{assert} \(\{ i\in\mathsf{act} \}\)
		\State Parse \(\mathsf{open} = (\bm{r}_i,\bm{w}_i,\mathsf{cmt}_i)\).
		\State Parse \(\mathsf{aux} = (\mathsf{pm}_j^1)_{j\in\mathsf{act}} = (\mathsf{cmt}_j)_{j\in\mathsf{act}}\).
		\State Parse \((\mathsf{pm}_j^2)_{j\in\mathsf{act}} = (\bm{w}_j)_{j\in\mathsf{act}}\).
        \State \textbf{assert} \(\{ \mathsf{pm}_i^2=\bm{w}_i \}\) \Comment{own opening delivered unchanged}
		\For{\(j\in\mathsf{act}\)}
			\If{ \(\mathsf{cmt}_j \neq H_\mathsf{cmt} (\mathsf{sid},\mathsf{vk}, \mathsf{act},j,\bm{w}_j)\) }
                \State Set \(\mathsf{status}:=\mathsf{aborted}\) in \(\mathsf{st}_i.\mathsf{session}[\mathsf{sid}]\) and erase \(\bm{r}_i\).
				\State \Return \((\mathsf{abort},\{j\})\) \Comment{identify signer \(j\)}
			\EndIf
		\EndFor
		\State \(\bm{w} := \mathsf{SEC.Rec}_\mathsf{pp} \left( \mathsf{act}, (\bm{w}_j)_{j\in\mathsf{act}} \right)\) \Comment{\(\bm{w}\in\mathcal{R}_q^p\)}
		\State \(c := H_c (\mathsf{sid},\mathsf{vk}, \mathsf{act},\mathsf{msg},\bm{w})\) \Comment{\(c\in\mathcal{C}\subseteq\mathcal{R}\)}
		\State \(\bm{z}_i := c\bm{s}_i+\bm{r}_i\) \Comment{\(\bm{z}_i\in\mathcal{R}^\ell\)}
		\State \(\mathsf{st}_i.\mathsf{session}[\mathsf{sid}] := ( \mathsf{status}=\mathsf{completed}, \mathsf{rnd}=3, \mathsf{act}, \mathsf{msg}, \mathsf{open} = (\bm{w}_i,\mathsf{cmt}_i), \mathsf{aux} = ( (\mathsf{pm}_j^1)_{j\in\mathsf{act}}, (\mathsf{pm}_j^2)_{j\in\mathsf{act}} ) )\)
		\State \(\mathsf{pm}_i^3:=\bm{z}_i\)
		\State \Return \((\mathsf{st}_i,\mathsf{pm}_i^3)\)
	\end{algorithmic}
\end{algorithm*}

\subsubsection{Combination and Verification}\label{Section:LaTSEC-Combination-Verification}

We first define the public partial-verification procedure used to identify invalid partial responses.
The combination algorithm applies this procedure to every active signer and either returns an identifiable abort or reconstructs the final response.
Final verification then follows the verification relation of the underlying no-abort signature scheme in Appendix~\ref{Appendix:Underlying-No-Abort-Signature}.

\paragraph{\textbf{Partial verification}}

The deterministic algorithm \(\mathsf{LaTSEC.PartialVerify}\) publicly checks the partial response associated with the signer-specific abort key \(\mathsf{ak}_i=\bm{t}_i\).
Given the round-\(2\) openings \((\bm{w}_j)_{j\in\mathsf{act}}\), the algorithm reconstructs the aggregate commitment \( \bm{w} := \mathsf{SEC.Rec}_\mathsf{pp} \left( \mathsf{act}, (\bm{w}_j)_{j\in\mathsf{act}} \right) \) and recomputes the common challenge \( c := H_c \left( \mathsf{sid}, \mathsf{vk}, \mathsf{act}, \mathsf{msg}, \bm{w} \right)\).
It accepts the partial response \(\bm{z}_i\) if and only if \(\bm{A}\bm{z}_i-c\bm{t}_i = \bm{w}_i\) and \(\|\bm{z}_i\|_2 \leq B_1\).
Thus, partial verification checks both the algebraic consistency and the shortness of the partial response using only public information.
The procedure is given in \Cref{Algorithm:LaTSEC-PartialVerify}.

\begin{algorithm*}[!t]
	\caption{ \(\{0,1,\bot\} \gets \mathsf{LaTSEC.PartialVerify} (\mathsf{tspar},\mathsf{vk},\mathsf{ak}_i, \mathsf{sid},\mathsf{act},\mathsf{msg}, (\mathsf{pm}_j^k)_{k\in[3],\,j\in\mathsf{act}})\) }
	\label{Algorithm:LaTSEC-PartialVerify}
	\small
	\begin{algorithmic}[1]
		\Require Public parameters \(\mathsf{tspar} = (\mathsf{pp},\mathcal{C},B_1,B_2, \sigma_{\bm{s}},\sigma_{\bm{r}},\beta_{\bm{\lambda}},\omega_{\bm{\lambda}})\), verification key \(\mathsf{vk}=(\bm{A},\bm{t})\), abort key \(\mathsf{ak}_i=\bm{t}_i\), session identifier \(\mathsf{sid}\), active set \(\mathsf{act}\), message \(\mathsf{msg}\), and signing transcript \((\mathsf{pm}_j^k)_{k\in[3],\,j\in\mathsf{act}}\).
		\Ensure \(1\) if the partial response associated with \(\mathsf{ak}_i\) is valid, and \(0\) otherwise.
		\State Parse \(\mathsf{pm}_i^2=\bm{w}_i\) and \(\mathsf{pm}_i^3=\bm{z}_i\).
		\State Parse \((\mathsf{pm}_j^2)_{j\in\mathsf{act}} = (\bm{w}_j)_{j\in\mathsf{act}}\).
		\State \(\bm{w} := \mathsf{SEC.Rec}_\mathsf{pp} \left( \mathsf{act}, (\bm{w}_j)_{j\in\mathsf{act}} \right)\) \Comment{aggregate commitment}
		\State \(c := H_c (\mathsf{sid},\mathsf{vk}, \mathsf{act},\mathsf{msg},\bm{w})\) \Comment{common challenge}
		\If{ \(\bm{A}\bm{z}_i-c\bm{t}_i\neq\bm{w}_i\) \textbf{or} \(\|\bm{z}_i\|_2>B_1\) }
			\State \Return \(0\)
		\Else
			\State \Return \(1\)
		\EndIf
	\end{algorithmic}
\end{algorithm*}

\paragraph{\textbf{Combination}}

The aggregator applies \(\mathsf{LaTSEC.PartialVerify}\) to every signer in the active set and defines the blame set \(S_\bot\) containing all indices \(i \in \mathsf{act}\) such that \(\mathsf{LaTSEC.PartialVerify}\) for an input associated with \(\mathsf{ak}_i\) returns \(0\).
If \(S_\bot\neq\emptyset\), then the algorithm returns \((\mathsf{abort},S_\bot)\), identifying every signer whose partial response fails verification.
Otherwise, the aggregator reconstructs the final response as \(\bm{z} := \mathsf{SEC.Rec}_\mathsf{pp} \left( \mathsf{act}, (\bm{z}_j)_{j\in\mathsf{act}} \right)\).
Equivalently, if \(\bm{Z}[j,\cdot]:=\bm{z}_j^\top\), then \(\bm{z} = \bm{Z}^\top\bm{\lambda}_{\mathsf{act}}\).
The algorithm returns the final signature \(\mathsf{sig}:=(c,\bm{z})\).
The combination procedure is given in \Cref{Algorithm:LaTSEC-Combine}.

\begin{algorithm*}[!t]
	\caption{ \(\mathsf{sig} \mid (\mathsf{abort},S_\bot) \gets \mathsf{LaTSEC.Combine} (\mathsf{tspar},\mathsf{vk},\mathsf{ak}, \mathsf{sid},\mathsf{act},\mathsf{msg}, (\mathsf{pm}_j^k)_{k\in[3],\,j\in\mathsf{act}})\) }
	\label{Algorithm:LaTSEC-Combine}
	\small
	\begin{algorithmic}[1]
		\Require Public parameters \(\mathsf{tspar}\), verification key \(\mathsf{vk}=(\bm{A},\bm{t})\), public abort information \(\mathsf{ak} = (\mathsf{ak}_i)_{i\in[n]} = (\bm{t}_i)_{i\in[n]}\), session identifier \(\mathsf{sid}\), active set \(\mathsf{act}\), message \(\mathsf{msg}\), and signing transcript \((\mathsf{pm}_j^k)_{k\in[3],\,j\in\mathsf{act}}\).
		\Ensure A signature \(\mathsf{sig}=(c,\bm{z})\) or an identifiable abort \((\mathsf{abort},S_\bot)\).
		\State Parse \((\mathsf{pm}_j^2)_{j\in\mathsf{act}} = (\bm{w}_j)_{j\in\mathsf{act}}\).
		\State Parse \((\mathsf{pm}_j^3)_{j\in\mathsf{act}} = (\bm{z}_j)_{j\in\mathsf{act}}\).
		\State \(\bm{w} := \mathsf{SEC.Rec}_\mathsf{pp} \left( \mathsf{act}, (\bm{w}_j)_{j\in\mathsf{act}} \right)\) \Comment{aggregate commitment}
		\State \(c := H_c (\mathsf{sid},\mathsf{vk}, \mathsf{act},\mathsf{msg},\bm{w})\) \Comment{common challenge}
		\State \(S_\bot := \left\{ i\in\mathsf{act}: \mathsf{LaTSEC.PartialVerify} \left( \mathsf{tspar}, \mathsf{vk}, \mathsf{ak}_i, \mathsf{sid}, \mathsf{act}, \mathsf{msg}, (\mathsf{pm}_j^k)_ {k\in[3],\,j\in\mathsf{act}} \right) =0 \right\}\)
		\If{\(S_\bot\neq\emptyset\)}
			\State \Return \((\mathsf{abort},S_\bot)\)
		\EndIf
		\State \(\bm{z} := \mathsf{SEC.Rec}_\mathsf{pp} \left( \mathsf{act}, (\bm{z}_j)_{j\in\mathsf{act}} \right)\) \Comment{aggregate response}
		\State \(\mathsf{sig}:=(c,\bm{z})\)
		\State \Return \(\mathsf{sig}\)
	\end{algorithmic}
\end{algorithm*}

\paragraph{\textbf{Missing and malformed messages}}

The algorithms above are stated for transcripts in which every message is present and parses.
As stated in \Cref{Figure:LaTSEC-Protocol}, a missing or unparsable peer message fails the check that uses it; we make the cases explicit.
In \(\mathsf{LaTSEC.ShareSign}_3\), a signer \(j\) whose commitment or opening is missing or unparsable fails the commitment check, and the first such signer in the canonical order of \(\mathsf{act}\) is returned.
\(\mathsf{LaTSEC.PartialVerify}\) for the signer \(i\) associated with \(\mathsf{ak}_i\) first applies the checks that need no challenge: it returns \(0\) if \(\bm{w}_i\) is missing or unparsable, or if \(\bm{z}_i\) is missing, unparsable, or satisfies \(\|\bm{z}_i\|_2>B_1\).
Otherwise, if the opening of another signer is missing or unparsable, \(\bm{w}\) and \(c\) are undefined; the call is then a malformed invocation and returns \(\bot\), which is not a verdict on \(i\).
\(\mathsf{LaTSEC.Combine}\) first collects the signers whose openings are missing or unparsable.
If there are any, it returns \((\mathsf{abort},S_\bot)\), where \(S_\bot\) consists of these signers and of the signers \(j\) whose responses are missing, unparsable, or satisfy \(\|\bm{z}_j\|_2>B_1\); otherwise it proceeds as in \Cref{Algorithm:LaTSEC-Combine}.
In each case, a blamed signer sent a message that is missing or unparsable or that fails a public check, and by \Cref{Lemma:Honest-Partial-Response-Bound} no honest signer does so under \(\mathsf{Good}_\mathsf{tail}\).
If signer \(i\) finds its own commitment or opening missing from, or altered in, the messages delivered to it, the assertions of \(\mathsf{LaTSEC.ShareSign}_2\) and \(\mathsf{LaTSEC.ShareSign}_3\) fail, so the call is a malformed invocation and blames nobody.
Both implementations follow these rules.

\paragraph{\textbf{Verification}}

Final verification follows the underlying no-abort signature scheme from Appendix~\ref{Appendix:Underlying-No-Abort-Signature}.
The public message context canonically determines the session identifier \(\mathsf{sid}\) and active set \(\mathsf{act}\).
Equivalently, if \(\mathsf{m}\) denotes the external payload, then \(\mathsf{msg} := \mathrm{Encode}(\mathsf{sid},\mathsf{act},\bm{\lambda}_\mathsf{act},\mathsf{m})\), as in \Cref{Section:LaTSEC-Overview}.
We retain the expanded challenge-hash notation \(H_c (\mathsf{sid},\mathsf{vk}, \mathsf{act},\mathsf{msg},\bm{w})\) throughout the construction, where \(\mathsf{sid}\) and \(\mathsf{act}\) are recovered from the public message context.
Given \(\mathsf{sig}=(c,\bm{z})\), the verifier computes \(\bm{w} := \bm{A}\bm{z}-c\bm{t} \in \mathcal{R}_q^p\) and accepts if and only if \( c = H_c (\mathsf{sid},\mathsf{vk}, \mathsf{act},\mathsf{msg},\bm{w})\) and \(\|\bm{z}\|_2 \leq B_2\).
The verification procedure is given in \Cref{Algorithm:LaTSEC-Verify}.

\begin{algorithm*}[!t]
	\caption{ \(\{0,1\} \gets \mathsf{LaTSEC.Verify} (\mathsf{vk},\mathsf{msg},\mathsf{sig})\) }
	\label{Algorithm:LaTSEC-Verify}
	\small
	\begin{algorithmic}[1]
		\Require Verification key \(\mathsf{vk}=(\bm{A},\bm{t})\), context-bound message \(\mathsf{msg}\), and signature \(\mathsf{sig}=(c,\bm{z})\).
		\Ensure \(1\) if the signature is valid and \(0\) otherwise.
        \State Parse \((\mathsf{sid},\mathsf{act})\) from \(\mathsf{msg}\) and \((c,\bm{z}):=\mathsf{sig}\); \textbf{if} either fails, \textbf{return} \(0\).
		\State \(\bm{w} := \bm{A}\bm{z}-c\bm{t}\) \Comment{\(\bm{w}\in\mathcal{R}_q^p\)}
		\If{ \(c = H_c (\mathsf{sid},\mathsf{vk}, \mathsf{act},\mathsf{msg},\bm{w})\) \textbf{and} \(\|\bm{z}\|_2\leq B_2\) }
			\State \Return \(1\) \Comment{\(\mathsf{sig}\) is valid}
		\Else
			\State \Return \(0\) \Comment{\(\mathsf{sig}\) is invalid}
		\EndIf
	\end{algorithmic}
\end{algorithm*}

\subsection{Correctness Game}\label{Section:TSS-Correctness}

Correctness is required only for admissible session inputs.
More precisely, for every session input \((\mathsf{sid},\mathsf{act},\mathsf{msg})\) with \(\mathsf{act}\subseteq[n]\), \(|\mathsf{act}|\geq t_\mathsf{sig}\), and \(\mathcal{P}(\mathsf{tspar},\mathsf{vk},\mathsf{sid},\mathsf{act},\mathsf{msg})=1\), an honest execution of the signing protocol should output a signature \(\mathsf{sig}\) such that \(\mathsf{Verify}(\mathsf{vk},\mathsf{msg},\mathsf{sig})=1\).
We formalize this through the experiment \(\mathsf{Game}_\mathsf{correct}\) shown in \Cref{Figure:Game-Correct}.

\begin{figure*}[!t]
	\centering
	\begin{tcolorbox}[gamebox]
		\small
		\(\mathsf{Game}_\mathsf{correct}\)
		\begin{algorithmic}[1]
			\State \(\mathsf{tspar}\gets\mathsf{TS.Setup}(1^\kappa,n,t_\mathsf{uf},t_\mathsf{sig})\)
			\State \((\mathsf{vk},\mathsf{ak},(\mathsf{sk}_i)_{i\in[n]})\gets\mathsf{Keygen}(\mathsf{tspar})\)
			\State \textbf{assert}\{\(\mathsf{act}\subseteq[n]\ \land\ |\mathsf{act}|\geq t_\mathsf{sig} \land \mathcal{P}(\mathsf{tspar},\mathsf{vk},\mathsf{sid},\mathsf{act},\mathsf{msg})=1 \)\}
			\State Set \(\mathsf{st}_i.\mathsf{sk}:=\mathsf{sk}_i\), \(\mathsf{st}_i.\mathsf{ak}:=\mathsf{ak}_i\), \(\mathsf{st}_i.\mathsf{vk}:=\mathsf{vk}\), and \(\mathsf{st}_i.\mathsf{tspar}:=\mathsf{tspar}\) for all \(i\in\mathsf{act}\).
			\For{\(i\in\mathsf{act}\)}
				\State \((\mathsf{st}_i,\mathsf{pm}_i^1)\gets \mathsf{ShareSign}_1(\mathsf{st}_i,\mathsf{sid},\mathsf{act},\mathsf{msg})\)
				\If{\(\mathsf{pm}_i^1=(\mathsf{abort},\cdot)\)}
					\State \Return \(0\) \Comment{early termination}
				\EndIf
			\EndFor
			\For{\(k=2\) \textbf{to} \(\mathsf{rnd}\)}
				\For{\(i\in\mathsf{act}\)}
					\State \((\mathsf{st}_i,\mathsf{pm}_i^k)\gets \mathsf{ShareSign}_k(\mathsf{st}_i,\mathsf{sid},\mathsf{act},(\mathsf{pm}_j^{k-1})_{j\in\mathsf{act}})\)
					\If{\(\mathsf{pm}_i^k=(\mathsf{abort},\cdot)\)}
						\State \Return \(0\) \Comment{early termination}
					\EndIf
				\EndFor
			\EndFor
			\State \(\mathsf{out}\gets \mathsf{Combine}(\mathsf{tspar},\mathsf{vk},\mathsf{ak},\mathsf{sid},\mathsf{act},\mathsf{msg},(\mathsf{pm}_i^k)_{k\in[\mathsf{rnd}],\,i\in\mathsf{act}})\)
			\If{\(\mathsf{out}=(\mathsf{abort},\cdot)\)}
				\State \Return \(0\)
			\EndIf
			\State Parse \(\mathsf{out}=\mathsf{sig}\)
			\State \Return \(\mathsf{Verify}(\mathsf{vk},\mathsf{msg},\mathsf{sig})\)
		\end{algorithmic}
	\end{tcolorbox}
	\caption{Correctness game for a threshold signature with IA; the scheme is correct if \(\mathsf{Game}_\mathsf{correct}\) returns \(1\) with high probability.}
	\label{Figure:Game-Correct}
\end{figure*}

In the correctness experiment, the parameters \((n,t_\mathsf{uf},t_\mathsf{sig})\) are fixed as part of the scheme instance, while \(\mathsf{act}\), \(\mathsf{sid}\), and \(\mathsf{msg}\) describe the honest signing session under consideration.
If \(\mathcal{P}(\mathsf{tspar},\mathsf{vk},\mathsf{sid},\mathsf{act},\mathsf{msg})=0\), then the session input is malformed or not supported by the underlying sharing structure, and any resulting failure is attributed to the entity or mechanism responsible for selecting the session input, rather than to the honest signing protocol itself.

In our concrete instantiation, the admissibility predicate is the one described in \Cref{Section:LaTSEC-Overview}: \(\mathcal{P}(\mathsf{tspar},\mathsf{vk},\mathsf{sid},\mathsf{act},\mathsf{msg})=1\) if and only if \(\mathsf{msg}\) canonically encodes \((\mathsf{sid},\mathsf{act},\bm{\lambda}_\mathsf{act})\) for some \(\bm{\lambda}_\mathsf{act}\in\mathbb{Z}^{|\mathsf{act}|}\), \(|\mathsf{act}|\geq t_\mathsf{sig}\), \(\bm{\lambda}_\mathsf{act}^\top\mathbf{G}_\mathsf{act}=\mathbf{e}_1^\top\), \(\|\bm{\lambda}_\mathsf{act}\|_\infty\leq\beta_{\bm{\lambda}}\), and \(wt_H(\bm{\lambda}_\mathsf{act})\leq\omega_{\bm{\lambda}}\).
This predicate is deterministic and consists of exact integer checks on public data, so it does not depend on how \(\bm{\lambda}_\mathsf{act}\) was computed.
Every admissible input satisfies \(\mathbf{e}_1^\top\in\mathrm{Row}_\mathbb{Z}(\mathbf{G}_\mathsf{act})\), so reconstruction is exact.
If the initiator takes \(\bm{\lambda}_\mathsf{act}\) from \cite[Algorithm~1]{arzaki2026SEC}, admissibility nearly coincides with certification, which holds with estimated probability at least \(0.99\) for every tested size \(t\geq k_c\); the bounds reject only a small fraction of certified sets (\Cref{Section:EmpiricalProfiles-InstantiationMethod}).

\begin{definition}[Operational correctness of ramp-TSS]\label{Definition:TSS-Operational-Correctness}
	Fix \(n\in\mathbb{N}\), a signing threshold \(t_\mathsf{sig}\in[n]\), and an admissibility predicate \(\mathcal{P}(\mathsf{tspar},\mathsf{vk},\cdot,\cdot,\cdot)\).
	Let \(\epsilon_\mathrm{cor}\in[0,1]\).
	We say that an \((n,t_\mathsf{uf},t_\mathsf{sig})\)-TSS with IA is \emph{\(\epsilon_\mathrm{cor}\)-operationally correct with respect to \(\mathcal{P}\)} if, for every session input \((\mathsf{sid},\mathsf{act},\mathsf{msg})\) with \(\mathsf{act}\subseteq[n]\), \(|\mathsf{act}| \geq t_\mathsf{sig}\), and \(\mathcal{P}(\mathsf{tspar},\mathsf{vk},\mathsf{sid},\mathsf{act},\mathsf{msg}) = 1\), \(\Pr[ \mathsf{Game}_\mathsf{correct}=1 ] \geq 1-\epsilon_\mathrm{cor}\), where \(\mathsf{Game}_\mathsf{correct}\) is defined in \Cref{Figure:Game-Correct}.
	The probability is taken over the randomness of \(\mathsf{TS.Setup}\), \(\mathsf{Keygen}\), the honest signing algorithms \(\mathsf{ShareSign}_k\), and \(\mathsf{Combine}\), if randomized.
	If no additional set-dependent restriction is required, we may set \(\mathcal{P}\equiv1\), while retaining the separate requirement \(|\mathsf{act}|\geq t_\mathsf{sig}\).
\end{definition}

Finally, correctness on admissible session inputs follows from identifiable aborts in the all-honest setting.
Consider an adversary \(\mathcal{B}\) in \(\mathsf{Game}_\mathsf{TS\text{-}IA}\) that chooses \(\mathsf{corrupt} = \emptyset\) and executes the signing protocol honestly on an admissible session input \((\mathsf{sid},\mathsf{act},\mathsf{msg})\).
Since admissibility includes \(\bm{\lambda}_\mathsf{act}^\top\mathbf{G}_\mathsf{act}=\mathbf{e}_1^\top\), reconstruction is exact, and any abort by an honest party violates identifiable abort, because a valid blamed abort would require \(\emptyset \neq S_\bot \subseteq \mathsf{corrupt} = \emptyset\), which is impossible.
The same reasoning applies if \(\mathsf{Combine}\) returns an abort.
If instead \(\mathsf{Combine}\) outputs a signature rejected by \(\mathsf{Verify}\), then the non-aborting branch of \(\mathsf{Game}_\mathsf{TS\text{-}IA}\) returns \(1\).
Consequently, \(\Pr\left[\mathsf{Game}_\mathsf{correct}=0\right]\leq\mathsf{Adv}_\mathcal{B}^\mathsf{TS\text{-}IA}(\kappa)\), so the operational correctness error satisfies \(\epsilon_\mathrm{cor}\leq\mathsf{Adv}_\mathcal{B}^\mathsf{TS\text{-}IA}(\kappa)\).
The coverage error \(\epsilon_c\) of \(\mathsf{SEC}\) \cite[Definition~11]{arzaki2026SEC} affects availability rather than correctness: it bounds the probability that a random active set of size at least \(t_\mathsf{sig}\) is not certified and hence has no admissible input.

\section{Proofs for \texorpdfstring{\Cref{Section:LaTSEC-Correctness-Security}}{Section \getrefnumber{Section:LaTSEC-Correctness-Security}}}\label{Appendix:LaTSEC-IA-Correctness}

The tail bounds behind \(\mathsf{Good}_\mathsf{tail}\) are proved in \cite[Section 6.4.1 and Appendix I]{arzaki2026SEC}; we prove the remaining statements of \Cref{Section:LaTSEC-IA-Correctness}.

\subsection{Honest Partial-Response Bounds}

\begin{proof}[Proof of \Cref{Lemma:Honest-Partial-Response-Bound}]
	Assume that \(\mathsf{Good}_\mathsf{tail}\) holds and let \(i\) be an honest signer participating in any signing query.
	By \cite[Lemma~12]{arzaki2026SEC}, the bounds \(\|\bm{s}\|_\infty\leq\|\bm{s}\|_2 \leq \beta_{\bm{s}}\) and \(\lvert\mathbf{R}[i',j']\rvert \leq \beta_\mathbf{R} \) imply \(\|\bm{s}_i\|_\infty \leq \beta_\mathbf{g}\beta_{\bm{s}} + c_\mathrm{exp}\beta_\mathbf{R}\).
	Since \(\bm{s}_i\in\mathcal{R}^\ell\) has coefficient dimension \(D=d\ell\), \(\|\bm{s}_i\|_2 \leq \sqrt{D} \left( \beta_\mathbf{g}\beta_{\bm{s}} + c_\mathrm{exp}\beta_\mathbf{R} \right)\).
	The honest partial response is \(\bm{z}_i = c\bm{s}_i+\bm{r}_i\).
	Since \(c\in\mathcal{C}\) satisfies \(\|c\|_1=\omega_c\), multiplication by \(c\) satisfies \(\|c\bm{s}_i\|_2 \leq \omega_c\|\bm{s}_i\|_2\).
	Moreover, \(\mathsf{Good}_\mathsf{tail}\) implies \(\|\bm{r}_i\|_2\leq\beta_{\bm{r}}\).
	Consequently,
	\begin{align*}
		\|\bm{z}_i\|_2 &\leq \|c\bm{s}_i\|_2 + \|\bm{r}_i\|_2 \\
		&\leq \omega_c\sqrt{D} \left( \beta_\mathbf{g}\beta_{\bm{s}} + c_\mathrm{exp}\beta_\mathbf{R} \right) + \beta_{\bm{r}} \leq B_1.
	\end{align*}
	It remains to verify the algebraic relation checked by \(\mathsf{LaTSEC.PartialVerify}\).
	For an honest signer, \(\bm{t}_i = \bm{A}\bm{s}_i\) and \(\bm{w}_i = \bm{A}\bm{r}_i\).
	Therefore,
	\begin{align*}
		\bm{A}\bm{z}_i-c\bm{t}_i = \bm{A} \left( c\bm{s}_i+\bm{r}_i \right) - c\bm{A}\bm{s}_i = \bm{A}\bm{r}_i = \bm{w}_i.
	\end{align*}
	Thus, an honest partial response passes both checks of \(\mathsf{LaTSEC.PartialVerify}\), proving \Cref{Lemma:Honest-Partial-Response-Bound}.
\end{proof}

\subsection{Proof of Identifiable-Abort Security}

\begin{proof}[Proof of \Cref{Theorem:LaTSEC-TS-IA}]
	By \Cref{Item:Cond-Tail} of the parameter conditions in \Cref{Section:LaTSEC-Asymptotic-Parameters}, the event \(\mathsf{Good}_\mathsf{tail}\) fails with probability at most \(3\cdot 2^{-\kappa}\).
	It therefore suffices to show that, conditioned on \(\mathsf{Good}_\mathsf{tail}\), no adversary can make \(\mathsf{Game}_\mathsf{TS\text{-}IA}\) return \(1\).
	Consider a session \(\mathsf{sid}^*\) selected by the adversary, with active set \(\mathsf{act}^*\) and context-bound message \(\mathsf{msg}^*\).
	The game proceeds only if \( \mathcal{P} \left( \mathsf{tspar}, \mathsf{vk}, \mathsf{sid}^*, \mathsf{act}^*, \mathsf{msg}^* \right) = 1\); in particular, all honest signers parse the same \(\bm{\lambda}_{\mathsf{act}^*}\) from \(\mathsf{msg}^*\).

	\textbf{\textit{(Consistency of honest commitment openings.)}}
	In \(\mathsf{LaTSEC.ShareSign}_1\), an honest signer \(i\in\mathsf{act}^*\) samples \(\bm{r}_i\), computes \(\bm{w}_i = \bm{A}\bm{r}_i\), and broadcasts \(\mathsf{cmt}_i = H_\mathsf{cmt} \left( \mathsf{sid}^*, \mathsf{vk}, \mathsf{act}^*, i, \bm{w}_i \right)\).
	In \(\mathsf{LaTSEC.ShareSign}_2\), the signer reveals the same stored value \(\bm{w}_i\).
	Hence, every honest signer supplies a valid opening of its first-round commitment.
	The assertion checks in \(\mathsf{LaTSEC.ShareSign}_1\) and \(\mathsf{LaTSEC.ShareSign}_2\) enforce session consistency and well-formed invocation conditions.
	They do not produce blamed aborts in \(\mathsf{Game}_\mathsf{TS\text{-}IA}\).

	\textbf{\textit{(Abort during \(\mathsf{LaTSEC.ShareSign}_3\).)}}
	An abort during \(\mathsf{LaTSEC.ShareSign}_3\) occurs only if some \(j\in\mathsf{act}^*\) fails \(\mathsf{cmt}_j = H_\mathsf{cmt} \left( \mathsf{sid}^*, \mathsf{vk}, \mathsf{act}^*, j, \bm{w}_j \right)\).
	No honest signer can fail this check because an honest signer reveals the same value to which it committed.
	Thus, every signer blamed in this round is corrupted.
	By the consistent-delivery assumption, all honest active parties receive the same first- and second-round transcripts.
	They therefore perform the same deterministic opening checks in the canonical order of \(\mathsf{act}^*\) and return the same singleton blame set \(\{j\}\).
	Consequently, the common blame set is non-empty and contained in \(\mathsf{corrupt}\), so the abort branch of \(\mathsf{Game}_\mathsf{TS\text{-}IA}\) returns \(0\).

	\textbf{\textit{(Abort returned by \(\mathsf{LaTSEC.Combine}\).)}}
	Suppose that all active parties complete the third signing round and that \(\mathsf{LaTSEC.Combine} \left( \mathsf{tspar}, \mathsf{vk}, \mathsf{ak}, \mathsf{sid}^*, \mathsf{act}^*, \mathsf{msg}^*, \mathsf{PM}^* \right) \) returns \((\mathsf{abort},S_\bot)\).
    If an opening in \(\mathsf{PM}^*\) is missing or unparsable, \(\mathsf{LaTSEC.Combine}\) blames the senders of such openings and of the responses that are missing, unparsable, or longer than \(B_1\) (\Cref{Appendix:LaTSEC-Full-Specification}).
    Under consistent delivery, every honest signer's opening and response are present and parse, and by \Cref{Lemma:Honest-Partial-Response-Bound} its response has norm at most \(B_1\), so this blame set is non-empty and contained in \(\mathsf{corrupt}\).
	Otherwise, by \Cref{Algorithm:LaTSEC-Combine}, \(S_\bot = \left\{ i\in\mathsf{act}^*: \mathsf{LaTSEC.PartialVerify}(\cdots)=0 \right\}\), and the algorithm returns an abort only if \(S_\bot\neq\emptyset\).
	Let \(i\in\mathsf{honest}\cap\mathsf{act}^*\).
	By \Cref{Lemma:Honest-Partial-Response-Bound}, the honest response satisfies \(\bm{A}\bm{z}_i-c\bm{t}_i = \bm{w}_i\) and \(\|\bm{z}_i\|_2 \leq B_1\).
	Hence, \(\mathsf{LaTSEC.PartialVerify}\) returns \(1\) for every honest signer, and therefore \(S_\bot \subseteq \mathsf{corrupt}\).
	Since \(S_\bot\neq\emptyset\), the output is a valid identifiable abort, and the game returns \(0\).

	\textbf{\textit{(Signature returned by \(\mathsf{LaTSEC.Combine}\).)}}
	It remains to consider the case in which \(\mathsf{LaTSEC.Combine}\) returns \(\mathsf{sig} = (c,\bm{z})\).
	Since the algorithm did not abort, every active signer passed \(\mathsf{LaTSEC.PartialVerify}\).
	Therefore, for every \(i\in\mathsf{act}^*\), \(\bm{A}\bm{z}_i-c\bm{t}_i = \bm{w}_i\) and \(\|\bm{z}_i\|_2 \leq B_1\).
	Let \(\bm{\lambda}_{\mathsf{act}^*} = (\lambda_i)_{i\in\mathsf{act}^*}\) be the reconstruction vector encoded in \(\mathsf{msg}^*\) and used by \(\mathsf{SEC.Rec}_\mathsf{pp}\).
	Since the session input is admissible, \(\|\bm{\lambda}_{\mathsf{act}^*}\|_\infty \leq \beta_{\bm{\lambda}}\) and \(wt_H(\bm{\lambda}_{\mathsf{act}^*})\leq\omega_{\bm{\lambda}}\), so \(\|\bm{\lambda}_{\mathsf{act}^*}\|_1\leq\omega_{\bm{\lambda}}\beta_{\bm{\lambda}}\).
	By the linearity and matrix-compatibility properties of \(\mathsf{SEC.Rec}_\mathsf{pp}\) from \cite[Lemma~17 and Lemma~18]{arzaki2026SEC}, \(\bm{z} = \sum_{i\in\mathsf{act}^*} \lambda_i\bm{z}_i\), \(\bm{w} = \sum_{i\in\mathsf{act}^*} \lambda_i\bm{w}_i\), and \(\bm{t} = \sum_{i\in\mathsf{act}^*} \lambda_i\bm{t}_i\).
	It follows that
	\begin{align*}
		\bm{A}\bm{z}-c\bm{t} &= \sum_{i\in\mathsf{act}^*} \lambda_i \left( \bm{A}\bm{z}_i-c\bm{t}_i \right) \\
		&= \sum_{i\in\mathsf{act}^*} \lambda_i\bm{w}_i = \bm{w}.
	\end{align*}
	Moreover,
	\begin{align*}
		\|\bm{z}\|_2 &= \left\| \sum_{i\in\mathsf{act}^*} \lambda_i\bm{z}_i \right\|_2 \leq \sum_{i\in\mathsf{act}^*} |\lambda_i| \|\bm{z}_i\|_2 \\
		&\leq \|\bm{\lambda}_{\mathsf{act}^*}\|_1 B_1 \leq \omega_{\bm{\lambda}}\beta_{\bm{\lambda}}B_1 \leq B_2.
	\end{align*}
	The challenge used by \(\mathsf{LaTSEC.Combine}\) is \(c = H_c \left( \mathsf{sid}^*, \mathsf{vk}, \mathsf{act}^*, \mathsf{msg}^*, \bm{w} \right)\).
	Since the context-bound message \(\mathsf{msg}^*\) determines \(\mathsf{sid}^*\) and \(\mathsf{act}^*\), the verification algorithm recomputes the same challenge.
	Together with \(\bm{A}\bm{z}-c\bm{t} = \bm{w}\) and \(\|\bm{z}\|_2 \leq B_2\), this gives \(\mathsf{LaTSEC.Verify} \left( \mathsf{vk}, \mathsf{msg}^*, (c,\bm{z}) \right) = 1\).
	Therefore, the game also returns \(0\) in the non-aborting branch.

	Conditioned on \(\mathsf{Good}_\mathsf{tail}\), the adversary cannot make \(\mathsf{Game}_\mathsf{TS\text{-}IA}\) return \(1\). Consequently, \(\mathsf{Adv}_\mathcal{A}^{\mathsf{TS\text{-}IA}}(\kappa) \leq \Pr \left[ \neg\mathsf{Good}_\mathsf{tail} \right] \leq 3\cdot2^{-\kappa}\), where the final inequality follows from \Cref{Item:Cond-Tail} of the parameter conditions in \Cref{Section:LaTSEC-Asymptotic-Parameters}.
	This proves \Cref{Theorem:LaTSEC-TS-IA}.
\end{proof}

\subsection{Proof of Correctness}

\Cref{Corollary:LaTSEC-Correctness} follows from the non-aborting branch of \Cref{Theorem:LaTSEC-TS-IA}.
Conditioned on \(\mathsf{Good}_\mathsf{tail}\), that branch shows that every signature output by \(\mathsf{LaTSEC.Combine}\) satisfies both checks of \(\mathsf{LaTSEC.Verify}\).
The remaining failure probability is at most \(3\cdot2^{-\kappa}\) by \Cref{Item:Cond-Tail} of the parameter conditions in \Cref{Section:LaTSEC-Asymptotic-Parameters}.

\section{Constructions and Proofs for \texorpdfstring{\Cref{Section:LaTSEC-Abort-Key-Simulation}}{Section \getrefnumber{Section:LaTSEC-Abort-Key-Simulation}}}\label{Appendix:LaTSEC-Abort-Key-Simulation}

\subsection{Algebraic Completion and Uniform Sampling}

We give the explicit affine completion underlying \(\mathsf{SimAK}\).
Recall from \eqref{Eq:Induced-Abort-Key-Sharing} that the coefficient-embedded public abort-key tuple satisfies \(\mathbf{T} = \left( \mathbf{g}\mathbf{t}^\top + \mathbf{\hat{G}}\mathbf{R}_\mathbf{t} \right) \bmod q\), where \(\mathbf{R}_\mathbf{t} = \mathbf{R}\mathbf{A}^\top\).
Since \(\mathbf{G} = \begin{bmatrix} \mathbf{g} \ \mathbf{\hat{G}} \end{bmatrix} \) and \(\mathbf{t}\) are public, the abort-key tuple reveals \(\mathbf{\hat{G}}\mathbf{R}_\mathbf{t} = \mathbf{T} - \mathbf{g}\mathbf{t}^\top \bmod q\).
If \(\mathbf{\hat{G}}\) has full column rank over \(\mathbb{Z}_q\), this value uniquely determines \(\mathbf{R}_\mathbf{t}\).
The construction below does not require this global rank condition.
Instead, for each prescribed coordinate set \(U\) within the privacy threshold, it uses a unimodular minor of \(\mathbf{\hat{G}}_U\) to construct an integer right inverse.

Let \(U\subseteq[n]\) satisfy \(1\leq |U|\leq k_p\).
By \cite[Lemma~4]{arzaki2026SEC}, there exists a set \(J_U \subseteq [k-1]\) with \(|J_U| = |U|\), such that \(\mathbf{\hat{G}}_U[\cdot,J_U] \in \mathbb{Z}^{|U|\times|U|}\) is unimodular.
In particular, \(\det \left( \mathbf{\hat{G}}_U[\cdot,J_U] \right) \in \{-1,1\}\), so its inverse has integer entries.
Define \(\mathbf{K}_U \in \mathbb{Z}^{(k-1)\times|U|}\) by placing \(\mathbf{\hat{G}}_U[\cdot,J_U]^{-1}\) in the rows indexed by \(J_U\) and placing zero rows in all remaining positions.
Then \(\mathbf{\hat{G}}_U\mathbf{K}_U = \mathbf{I}_{|U|}\).
Thus, \(\mathbf{K}_U\) is an integer right inverse of \(\mathbf{\hat{G}}_U\).

For prescribed abort keys \(\mathbf{T}_U\in\mathbb{Z}_q^{|U|\times dp}\) and a target \(\mathbf{\breve{t}}\in\mathbb{Z}_q^{dp}\), define the particular completion
\begin{align}\label{Eq:Abort-Key-Particular-Completion}
	\mathbf{\breve{R}}_{\mathbf{t},0} &:= \left( \mathbf{K}_U \left( \mathbf{T}_U - (\mathbf{g})_U \mathbf{\breve{t}}^\top \right) \right) \bmod q,
\end{align}
and let \(\mathcal{K}_U := \{ \mathbf{Z} \in \mathbb{Z}_q^{(k-1)\times dp}:\allowbreak \mathbf{\hat{G}}_U\mathbf{Z} = \mathbf{O} \bmod q \}\).
For any \(\mathbf{N}\in\mathcal{K}_U\), set
\begin{align}
	\mathbf{\breve{R}}_\mathbf{t} &:= \left( \mathbf{\breve{R}}_{\mathbf{t},0} + \mathbf{N} \right) \bmod q, \label{Eq:Abort-Key-Mask-Completion} \\
	\mathbf{\breve{T}} &:= \left( \mathbf{g}\mathbf{\breve{t}}^\top + \mathbf{\hat{G}} \mathbf{\breve{R}}_\mathbf{t} \right) \bmod q. \label{Eq:Abort-Key-Public-Completion}
\end{align}

\begin{proof}[Proof of \Cref{Lemma:Algebraic-Abort-Key-Completion}]
	We first consider \(U\neq\emptyset\).
	The construction above gives an efficiently computable integer matrix \(\mathbf{K}_U\) satisfying \(\mathbf{\hat{G}}_U\mathbf{K}_U = \mathbf{I}_{|U|}\).
	Therefore, by \eqref{Eq:Abort-Key-Particular-Completion},
	\begin{align*}
		\mathbf{\hat{G}}_U \mathbf{\breve{R}}_{\mathbf{t},0} &= \left( \mathbf{\hat{G}}_U \mathbf{K}_U \left( \mathbf{T}_U - (\mathbf{g})_U \mathbf{\breve{t}}^\top \right) \right) \bmod q \\
		&= \left( \mathbf{T}_U - (\mathbf{g})_U \mathbf{\breve{t}}^\top \right) \bmod q.
	\end{align*}
	Moreover, since \(\mathbf{N}\in\mathcal{K}_U\), \(\mathbf{\hat{G}}_U\mathbf{N} = \mathbf{O}_{|U|\times dp} \bmod q\).
	It follows from \eqref{Eq:Abort-Key-Mask-Completion} and \eqref{Eq:Abort-Key-Public-Completion} that
	\begin{align*}
		\mathbf{\breve{T}}_U &= \left( (\mathbf{g})_U \mathbf{\breve{t}}^\top + \mathbf{\hat{G}}_U \mathbf{\breve{R}}_\mathbf{t} \right) \bmod q \\
		&= \left( (\mathbf{g})_U \mathbf{\breve{t}}^\top + \mathbf{\hat{G}}_U \left( \mathbf{\breve{R}}_{\mathbf{t},0} + \mathbf{N} \right) \right) \bmod q \\
		&= \left( (\mathbf{g})_U \mathbf{\breve{t}}^\top + \mathbf{T}_U - (\mathbf{g})_U \mathbf{\breve{t}}^\top \right) \bmod q \\
		&= \mathbf{T}_U.
	\end{align*}
	This proves that the prescribed abort keys are preserved.
	Now let \(T\subseteq[n]\) be any certified set, and let \(\bm{\lambda}_T\) be its deterministic reconstruction vector.
	Since \(\bm{\lambda}_T^\top\mathbf{G}_T = \mathbf{e}_1^\top\), we have \(\bm{\lambda}_T^\top (\mathbf{g})_T = 1\) and \(\bm{\lambda}_T^\top \mathbf{\hat{G}}_T = \mathbf{0}^\top\).
	Let \(\mathbf{\breve{t}}_i^\top := \mathbf{\breve{T}}[i,\cdot]\).
	Then
	\begin{align*}
		&\mathsf{SEC.Rec}_\mathsf{pp} \left( T, (\mathbf{\breve{t}}_i)_{i\in T} \right)^\top = \bm{\lambda}_T^\top \mathbf{\breve{T}}_T \\
		&= \bm{\lambda}_T^\top \left( (\mathbf{g})_T \mathbf{\breve{t}}^\top + \mathbf{\hat{G}}_T \mathbf{\breve{R}}_\mathbf{t} \right) \bmod q \\
		&= \left( \bm{\lambda}_T^\top (\mathbf{g})_T \right) \mathbf{\breve{t}}^\top \\
		&\quad + \left( \bm{\lambda}_T^\top \mathbf{\hat{G}}_T \right) \mathbf{\breve{R}}_\mathbf{t} \bmod q \\
		&= \mathbf{\breve{t}}^\top.
	\end{align*}
	Transposing gives \(\mathsf{SEC.Rec}_\mathsf{pp} \left( T, (\mathbf{\breve{t}}_i)_{i\in T} \right) = \mathbf{\breve{t}} \bmod q\).
	If \(\mathbf{T}_U\) contains the real abort keys of the parties in \(U\), then \(\mathbf{\breve{T}}_U=\mathbf{T}_U\) immediately gives \((\mathbf{\breve{t}}_i)_{i\in U} = (\mathbf{A}\mathbf{s}_i\bmod q)_{i\in U}\).
	For \(U=\emptyset\), the prescribed-coordinate condition is empty.
	We take \(\mathbf{\breve{R}}_{\mathbf{t},0} := \mathbf{O}\) and \(\mathcal{K}_U = \mathbb{Z}_q^{(k-1)\times dp}\).
	The reconstruction argument above remains unchanged.
\end{proof}

We next show how to sample uniformly from the affine completion space in \Cref{Definition:SimAK}.
Since \(q\) is prime, \(\mathbb{Z}_q\) is a field.
For \(U\neq\emptyset\), let \(\mathbf{B}_U \in \mathbb{Z}_q^{(k-1)\times(k-1-|U|)}\) be a matrix whose columns form a basis of \(\ker_{\mathbb{Z}_q}(\mathbf{\hat{G}}_U)\).
Sample \(\mathbf{W} \stackrel{\$}{\gets} \mathbb{Z}_q^{(k-1-|U|)\times dp}\) and set \(\mathbf{N} := \mathbf{B}_U\mathbf{W} \bmod q\).
Each column of \(\mathbf{N}\) belongs to \(\ker_{\mathbb{Z}_q}(\mathbf{\hat{G}}_U)\), so \(\mathbf{N}\in\mathcal{K}_U\).

Because the columns of \(\mathbf{B}_U\) form a basis, the map \(\mathbf{W} \mapsto \mathbf{B}_U\mathbf{W}\) is a bijection from \(\mathbb{Z}_q^{(k-1-|U|)\times dp}\) to \(\mathcal{K}_U\).
Therefore, \(\mathbf{N}\) is uniformly distributed over \(\mathcal{K}_U\).
Since \(\mathbf{\breve{R}}_\mathbf{t} = \mathbf{\breve{R}}_{\mathbf{t},0} + \mathbf{N} \bmod q\), translation by the fixed particular solution \(\mathbf{\breve{R}}_{\mathbf{t},0}\) is a bijection from \(\mathcal{K}_U\) onto the affine space \(\big\{ \mathbf{Z} \in \mathbb{Z}_q^{(k-1)\times dp}: \mathbf{\hat{G}}_U\mathbf{Z} = \mathbf{T}_U - (\mathbf{g})_U \mathbf{\breve{t}}^\top \bmod q \big\}\).
Hence, \(\mathsf{SimAK}\) samples uniformly from exactly the affine space specified in \Cref{Definition:SimAK}.

For \(U=\emptyset\), we take \(\mathbf{B}_U = \mathbf{I}_{k-1}\) and sample \(\mathbf{W} \stackrel{\$}{\gets} \mathbb{Z}_q^{(k-1)\times dp}\), so that \(\mathbf{N}=\mathbf{W}\) is uniform over the entire space.

The simulator uses deterministic polynomial-time procedures, with fixed tie-breaking rules, to select the unimodular minor, construct \(\mathbf{K}_U\), and compute the modular kernel basis \(\mathbf{B}_U\) from \((\mathsf{pp},U)\).
Thus, for fixed \((U,\mathbf{\breve{t}},\mathbf{T}_U)\), the simulator's only internal randomness is the uniform matrix \(\mathbf{W}\).

We conclude by recording how the preceding completion is used in \Cref{Definition:HC-Leak-AK}.
Ignoring values that are identically distributed in both branches, the real branch contains \(\big( \bm{A}, \bm{t}, (\bm{s}_i)_{i\in U_\mathrm{raw}}, (\bm{z}_{i,s})_{(s,i)\in\mathcal{S}}, (\bm{t}_i)_{i\in[n]} \big)\), where \(\bm{t} = \bm{A}\bm{s}\) and \(\bm{t}_i = \bm{A}\bm{s}_i\).
The simulated branch contains \(\big( \bm{A}, \bm{\breve{t}}, (\bm{s}_i)_{i\in U_\mathrm{raw}}, (\bm{z}_{i,s})_{(s,i)\in\mathcal{S}}, (\bm{\breve{t}}_i)_{i\in[n]} \big)\), where \(\bm{\breve{t}}\) is uniform and \((\bm{\breve{t}}_i)_{i\in[n]}\) is generated by \(\mathsf{SimAK}\).
By \Cref{Lemma:Algebraic-Abort-Key-Completion}, the simulated tuple satisfies \((\bm{\breve{t}}_i)_{i\in U_\mathrm{raw}} = (\bm{A}\bm{s}_i)_{i\in U_\mathrm{raw}}\) and, for every certified set \(T\), \(\mathsf{SEC.Rec}_\mathsf{pp} \left( T, (\bm{\breve{t}}_i)_{i\in T} \right) = \bm{\breve{t}}\).
Thus, the simulated branch preserves the public abort keys determined by the corrupted raw shares and remains globally consistent with the simulated verification target.

\begin{remark}[Algebraic versus distributional simulation]\label{Remark:Abort-Key-Algebraic-vs-Distributional}
	\Cref{Lemma:Algebraic-Abort-Key-Completion} and \Cref{Definition:SimAK} establish algebraic consistency of the simulated abort-key tuple.
	They show that the simulation preserves the prescribed corrupted abort keys and reconstructs to the simulated global target from every certified set.
	These properties do not by themselves imply that the simulated tuple is distributed indistinguishably from the real tuple.
	The real induced masking matrix has the structured form \(\mathbf{R}_\mathbf{t} = \mathbf{R}\mathbf{A}^\top\), whereas \(\mathsf{SimAK}\) samples uniformly from an affine completion space.
	The remaining analysis therefore proves that these two views are computationally indistinguishable in the presence of the corrupted raw shares and multi-query response leakage.
	This is established through the \(\mathsf{RA\text{-}MHMLWE}\) representation and reduction developed in the following subsections.
\end{remark}

\subsection{Conditional Distribution of the Hidden Coordinates}

We give the conditioning calculation underlying \Cref{Lemma:Conditional-Hidden-Coordinate-Precision}.
The calculation is performed for one column of the extended secret matrix \(\mathbf{X}\).
Since the \(D\) coefficient positions are sampled independently, the vectorized result follows by repeating the resulting Gaussian parameter matrix across the \(D\) positions.

Let \(\mathbf{x} \in \mathbb{Z}^k\) denote one column of \(\mathbf{X}\).
Its distribution is a centered discrete Gaussian governed by the positive-definite Gaussian parameter matrix \(\mathbf{\Sigma}_\mathbf{X} := \mathrm{diag} \left( \sigma_{\bm{s}}^2, \sigma_\mathbf{R}^2 \mathbf{I}_{k-1} \right)\).
Equivalently, up to normalization, its probability mass function has the form \(\Pr[\mathbf{x}] \propto \exp \left( -\frac{1}{2} \mathbf{x}^\top \mathbf{\Sigma}_\mathbf{X}^{-1} \mathbf{x} \right)\).

Under the unimodular transformation \(\mathbf{x}' := \mathbf{Q}_U^{-1}\mathbf{x}\), we have \(\mathbf{x}=\mathbf{Q}_U\mathbf{x}'\).
Therefore, the transformed discrete Gaussian is governed by
\begin{equation}\label{Eq:Transformed-X-Covariance}
	\mathbf{\Sigma}'_U := \mathbf{Q}_U^{-1} \mathbf{\Sigma}_\mathbf{X} \mathbf{Q}_U^{-\top},
\end{equation}
and its precision matrix is \({\mathbf{\Sigma}'_U}^{-1} = \mathbf{Q}_U^\top \mathbf{\Sigma}_\mathbf{X}^{-1} \mathbf{Q}_U\).

Partition the transformed coordinates as \(\mathbf{x}' = \begin{bmatrix} \mathbf{x}'_\mathsf{pub}\\ \mathbf{x}'_\mathsf{hid} \end{bmatrix}\), where \(\mathbf{x}'_\mathsf{pub} \in \mathbb{Z}^{|U|}\) and \(\mathbf{x}'_\mathsf{hid} \in \mathbb{Z}^{h_U}\).
Accordingly, partition the transformed Gaussian parameter matrix as
\begin{equation}\label{Eq:Transformed-X-Covariance-Partition}
	\mathbf{\Sigma}'_U = \begin{bmatrix} \mathbf{\Sigma}_{U,\mathsf{pub}} & \mathbf{\Sigma}_{U,\mathsf{pub,hid}} \\ \mathbf{\Sigma}_{U,\mathsf{hid,pub}} & \mathbf{\Sigma}_{U,\mathsf{hid}} \end{bmatrix}.
\end{equation}
Define the conditional Gaussian parameter matrix \(\mathbf{\Sigma}_{U,\mathsf{hid}\mid\mathsf{pub}} := \mathbf{\Sigma}_{U,\mathsf{hid}} - \mathbf{\Sigma}_{U,\mathsf{hid,pub}} \mathbf{\Sigma}_{U,\mathsf{pub}}^{-1} \mathbf{\Sigma}_{U,\mathsf{pub,hid}}\).
This is the Schur complement of the public block in \(\mathbf{\Sigma}'_U\).

\begin{lemma}[Conditional distribution of the hidden coordinates]\label{Lemma:Conditional-Hidden-Coordinates}
	Let \(\mathbf{x}' = \begin{bmatrix} \mathbf{x}'_\mathsf{pub}\\ \mathbf{x}'_\mathsf{hid} \end{bmatrix} = \mathbf{Q}_U^{-1}\mathbf{x}\) be distributed as above.
	For every fixed exposed value \(\mathbf{x}'_\mathsf{pub} = \mathbf{a} \in \mathbb{Z}^{|U|}\), the conditional distribution of \(\mathbf{x}'_\mathsf{hid}\) is a discrete Gaussian over \(\mathbb{Z}^{h_U}\) with conditional center \(\mathbf{\mu}_{U,\mathsf{hid}\mid\mathsf{pub}}(\mathbf{a}) := \mathbf{\Sigma}_{U,\mathsf{hid,pub}} \mathbf{\Sigma}_{U,\mathsf{pub}}^{-1} \mathbf{a}\) and Gaussian parameter matrix \(\mathbf{\Sigma}_{U,\mathsf{hid}\mid\mathsf{pub}} = \mathbf{\Sigma}_{U,\mathsf{hid}} - \mathbf{\Sigma}_{U,\mathsf{hid,pub}} \mathbf{\Sigma}_{U,\mathsf{pub}}^{-1} \mathbf{\Sigma}_{U,\mathsf{pub,hid}}\).
	Equivalently, its conditional precision is \(\mathbf{\Sigma}_{U,\mathsf{hid}\mid\mathsf{pub}}^{-1} = \left( {\mathbf{\Sigma}'_U}^{-1} \right)_{\mathsf{hid},\mathsf{hid}}\).
\end{lemma}

\begin{proof}
	Write the transformed precision matrix as \({\mathbf{\Sigma}'_U}^{-1} = \begin{bmatrix} \mathbf{J}_{U,\mathsf{pub}} & \mathbf{J}_{U,\mathsf{pub,hid}} \\ \mathbf{J}_{U,\mathsf{hid,pub}} & \mathbf{J}_{U,\mathsf{hid}} \end{bmatrix}\).
	Conditioning on \(\mathbf{x}'_\mathsf{pub}=\mathbf{a}\), the terms in the quadratic form that depend on \(\mathbf{x}'_\mathsf{hid}\) are
	\begin{align*}
		{\mathbf{x}'_\mathsf{hid}}^\top \mathbf{J}_{U,\mathsf{hid}} \mathbf{x}'_\mathsf{hid} + 2 \mathbf{a}^\top \mathbf{J}_{U,\mathsf{pub,hid}} \mathbf{x}'_\mathsf{hid}.
	\end{align*}
	Completing the square gives a discrete Gaussian over \(\mathbb{Z}^{h_U}\) whose conditional precision is \(\mathbf{J}_{U,\mathsf{hid}} = \left( {\mathbf{\Sigma}'_U}^{-1} \right)_{\mathsf{hid},\mathsf{hid}}\).
	By the standard block-inverse identity, \(\mathbf{J}_{U,\mathsf{hid}} = \big( \mathbf{\Sigma}_{U,\mathsf{hid}} - \mathbf{\Sigma}_{U,\mathsf{hid,pub}} \mathbf{\Sigma}_{U,\mathsf{pub}}^{-1} \mathbf{\Sigma}_{U,\mathsf{pub,hid}} \big)^{-1}\).
	Hence, \(\mathbf{J}_{U,\mathsf{hid}} = \mathbf{\Sigma}_{U,\mathsf{hid}\mid\mathsf{pub}}^{-1}\).
	The corresponding center is obtained by completing the square: \(\mathbf{\mu}_{U,\mathsf{hid}\mid\mathsf{pub}}(\mathbf{a}) = - \mathbf{J}_{U,\mathsf{hid}}^{-1} \mathbf{J}_{U,\mathsf{hid,pub}} \mathbf{a}\).
	The block-inverse identities also give \(- \mathbf{J}_{U,\mathsf{hid}}^{-1} \mathbf{J}_{U,\mathsf{hid,pub}} = \mathbf{\Sigma}_{U,\mathsf{hid,pub}} \mathbf{\Sigma}_{U,\mathsf{pub}}^{-1}\).
	Therefore, \(\mathbf{\mu}_{U,\mathsf{hid}\mid\mathsf{pub}}(\mathbf{a}) = \mathbf{\Sigma}_{U,\mathsf{hid,pub}} \mathbf{\Sigma}_{U,\mathsf{pub}}^{-1} \mathbf{a}\).
\end{proof}

\begin{proof}[Proof of \Cref{Lemma:Conditional-Hidden-Coordinate-Precision}]
	Partition \(\mathbf{Q}_U = \begin{bmatrix} \mathbf{Q}_{U,\mathsf{pub}} \ \mathbf{Q}_{U,\mathsf{hid}} \end{bmatrix}\) according to the public and hidden transformed coordinates.
	By \eqref{Eq:Transformed-X-Covariance}, the transformed joint precision is \({\mathbf{\Sigma}'_U}^{-1} = \mathbf{Q}_U^\top \mathbf{\Sigma}_\mathbf{X}^{-1} \mathbf{Q}_U\).
	Its hidden principal block is therefore \(\left( {\mathbf{\Sigma}'_U}^{-1} \right)_{\mathsf{hid},\mathsf{hid}} = \mathbf{Q}_{U,\mathsf{hid}}^\top \mathbf{\Sigma}_\mathbf{X}^{-1} \mathbf{Q}_{U,\mathsf{hid}}\).
	By \Cref{Lemma:Conditional-Hidden-Coordinates}, this hidden principal block is exactly the conditional precision: \(\mathbf{\Sigma}_{U,\mathsf{hid}\mid\mathsf{pub}}^{-1} = \mathbf{Q}_{U,\mathsf{hid}}^\top \mathbf{\Sigma}_\mathbf{X}^{-1} \mathbf{Q}_{U,\mathsf{hid}}\).
	Using submultiplicativity of the spectral norm,
	\begin{align*}
		\left\| \mathbf{\Sigma}_{U,\mathsf{hid}\mid\mathsf{pub}}^{-1} \right\|_2 &\leq \left\| \mathbf{Q}_{U,\mathsf{hid}}^\top \right\|_2 \left\| \mathbf{\Sigma}_\mathbf{X}^{-1} \right\|_2 \left\| \mathbf{Q}_{U,\mathsf{hid}} \right\|_2 \\
		&= \left\| \mathbf{\Sigma}_\mathbf{X}^{-1} \right\|_2 \left\| \mathbf{Q}_{U,\mathsf{hid}} \right\|_2^2.
	\end{align*}
	Since \(\mathbf{Q}_{U,\mathsf{hid}}\) is a column submatrix of \(\mathbf{Q}_U\), \(\left\| \mathbf{Q}_{U,\mathsf{hid}} \right\|_2 \leq \left\| \mathbf{Q}_U \right\|_2\).
	Therefore, \(\left\| \mathbf{\Sigma}_{U,\mathsf{hid}\mid\mathsf{pub}}^{-1} \right\|_2 \leq \left\| \mathbf{\Sigma}_\mathbf{X}^{-1} \right\|_2 \left\| \mathbf{Q}_U \right\|_2^2\).
\end{proof}

Since the columns of \(\mathbf{X}\) are independently distributed according to the same one-column discrete Gaussian, the conditioning calculation applies independently to all \(D\) coefficient positions.
Under the vectorization \(\mathbf{x}'_\mathsf{hid} = \mathrm{vec} \left( {\mathbf{X}'_\mathsf{hid}}^\top \right)\), the conditional Gaussian parameter matrix is \(\mathbf{\Sigma}_{U,\mathsf{hid}\mid\mathsf{pub}} \otimes \mathbf{I}_D\), and the corresponding conditional precision is \(\mathbf{\Sigma}_{U,\mathsf{hid}\mid\mathsf{pub}}^{-1} \otimes \mathbf{I}_D\).
Its spectral norm is \(\left\| \mathbf{\Sigma}_{U,\mathsf{hid}\mid\mathsf{pub}}^{-1} \otimes \mathbf{I}_D \right\|_2 = \left\| \mathbf{\Sigma}_{U,\mathsf{hid}\mid\mathsf{pub}}^{-1} \right\|_2\).
Thus, the one-column bound in \Cref{Lemma:Conditional-Hidden-Coordinate-Precision} also controls the vectorized hidden secret used in the \(\mathsf{RA\text{-}MHMLWE}\) representation.

\subsection{Construction of the Hidden-Coordinate Transformation}

We give the deterministic construction of the transformation matrix \(\mathbf{Q}_U\) used to separate the coordinates exposed by the corrupted raw-share view from the remaining hidden coordinates.
For every \(U\subseteq[n]\) with \(1\leq |U|\leq k_p\), the construction produces \(\mathbf{Q}_U \in GL_k(\mathbb{Z})\) such that \(\mathbf{G}_U\mathbf{Q}_U = \begin{bmatrix} \mathbf{I}_{|U|} \ \mathbf{O}_{|U|\times h_U} \end{bmatrix}\), where \(h_U:=k-|U|\).
It also establishes the pattern-dependent estimate in \eqref{Eq:Hidden-Coordinate-Transformation-Pattern-Bound} and the uniform estimate in \eqref{Eq:Uniform-Hidden-Coordinate-Transformation-Bound}.
For \(U=\emptyset\), we retain the convention \(\mathbf{Q}_\emptyset:=\mathbf{I}_k\).

Fix a non-empty admissible set \(U\subseteq[n]\), and write \(r:=|U|\).
We use the unit-valued unique-neighbor property established in the proof of \cite[Lemma~4]{arzaki2026SEC} to construct an \(r\times r\) unimodular submatrix of \(\mathbf{\hat{G}}_U\) by peeling.
The deterministic procedure is given in \cite[Algorithm~3]{arzaki2026SEC}.

By the unit-valued unique-neighbor property, the candidate set \(C\) in \cite[Algorithm~3]{arzaki2026SEC} is non-empty at every iteration.
Moreover, once a column \(j_t\) and its unique active row \(i_t\) are selected, the column \(j_t\) has no nonzero entry among the remaining active rows.
Hence, the selected columns are distinct.

Under the peeling row order \(I_U = (i_1,\dots,i_r)\), the matrix \(\mathbf{M}_U^\triangle := \mathbf{\hat{G}}_U[I_U,J_U]\) is triangular with diagonal entries in \(\{-1,1\}\).
Consequently, \(\left| \det \left( \mathbf{M}_U^\triangle \right) \right| = 1\).
Since \(\mathbf{M}_U\) differs from \(\mathbf{M}_U^\triangle\) only by a row permutation, \(|\det(\mathbf{M}_U)| = 1\).
Thus, \(\mathbf{M}_U \in GL_r(\mathbb{Z})\).
The deterministic choice \(j_t:=\min C\) and the fixed ordering conventions make \((\mathbf{M}_U,J_U,I_U)\) uniquely determined by \((\mathsf{pp},U)\).

Let \(\mathbf{P}_U \in GL_k(\mathbb{Z})\) be the permutation matrix that moves the columns of \(\mathbf{G}_U\) corresponding to \(J_U\) to its first \(r\) positions, in the peeling order, and places all remaining columns afterward in their original relative order.
Since the selected columns belong to \(\mathbf{\hat{G}}_U\), we may write \(\mathbf{G}_U\mathbf{P}_U = \begin{bmatrix} \mathbf{M}_U \ \mathbf{N}_U \end{bmatrix}\), where \(\mathbf{N}_U \in \mathbb{Z}^{r\times h_U}\).
Define \(\mathbf{V}_U := \begin{bmatrix} \mathbf{M}_U^{-1} & -\mathbf{M}_U^{-1}\mathbf{N}_U \\ \mathbf{O}_{h_U\times r} & \mathbf{I}_{h_U} \end{bmatrix}\) and \(\mathbf{Q}_U := \mathbf{P}_U\mathbf{V}_U\).
Since \(\mathbf{M}_U\) is unimodular, \(\mathbf{M}_U^{-1}\) has integer entries, and \(\det(\mathbf{V}_U) = \det(\mathbf{M}_U^{-1}) \in \{-1,1\}\).
Therefore, \(\mathbf{V}_U, \mathbf{Q}_U \in GL_k(\mathbb{Z})\).
Furthermore, \(\mathbf{G}_U\mathbf{Q}_U = \mathbf{G}_U\mathbf{P}_U\mathbf{V}_U = \begin{bmatrix} \mathbf{M}_U \ \mathbf{N}_U \end{bmatrix} \begin{bmatrix} \mathbf{M}_U^{-1} & -\mathbf{M}_U^{-1}\mathbf{N}_U \\ \mathbf{O}_{h_U\times r} & \mathbf{I}_{h_U} \end{bmatrix} = \begin{bmatrix} \mathbf{I}_r \ \mathbf{O}_{r\times h_U} \end{bmatrix}\).
Thus, this construction gives the deterministic transformation used in the hidden-coordinate analysis.

We now prove \eqref{Eq:Hidden-Coordinate-Transformation-Pattern-Bound}.
Since \(\mathbf{P}_U\) is a permutation matrix, \(\|\mathbf{Q}_U\|_2 = \|\mathbf{V}_U\|_2\).
Write \(\mathbf{V}_U = \begin{bmatrix} \mathbf{M}_U^{-1} & -\mathbf{M}_U^{-1}\mathbf{N}_U \\ \mathbf{O} & \mathbf{O} \end{bmatrix} + \begin{bmatrix} \mathbf{O} & \mathbf{O} \\ \mathbf{O} & \mathbf{I}_{h_U} \end{bmatrix}\).
The triangle inequality and submultiplicativity of the spectral norm give
\begin{align*}
	\|\mathbf{Q}_U\|_2 &\leq 1 + \left\| \mathbf{M}_U^{-1} \begin{bmatrix} \mathbf{I}_r \ -\mathbf{N}_U \end{bmatrix} \right\|_2 \\
	&\leq 1 + \|\mathbf{M}_U^{-1}\|_2 \left\| \begin{bmatrix} \mathbf{I}_r \ -\mathbf{N}_U \end{bmatrix} \right\|_2 \\
	&= 1 + \frac{ \sqrt{1+\|\mathbf{N}_U\|_2^2} }{ \sigma_{\min}(\mathbf{M}_U) }.
\end{align*}
The last equality follows from \(\left\| \begin{bmatrix} \mathbf{I}_r \ -\mathbf{N}_U \end{bmatrix} \right\|_2^2 = 1+\|\mathbf{N}_U\|_2^2\).
This proves \eqref{Eq:Hidden-Coordinate-Transformation-Pattern-Bound}.

Recall the quantities \(\delta_{k_p} := \max_{\substack{ U\subseteq[n]\\ 1\leq |U|\leq k_p }} \|\mathbf{G}_U\|_2\) and \(\gamma_{k_p} := \min_{\substack{ U\subseteq[n]\\ 1\leq |U|\leq k_p }} \sigma_{\min}(\mathbf{M}_U)\), where \(\mathbf{M}_U\) is the deterministic unimodular submatrix returned by \cite[Algorithm~3]{arzaki2026SEC}.
Since every \(\mathbf{M}_U\) is non-singular and the family of admissible sets is finite, \(\gamma_{k_p} > 0\).
The matrix \(\mathbf{N}_U\) is a column submatrix of \(\mathbf{G}_U\mathbf{P}_U\).
Therefore, \(\|\mathbf{N}_U\|_2 \leq \|\mathbf{G}_U\mathbf{P}_U\|_2 = \|\mathbf{G}_U\|_2 \leq \delta_{k_p}\).
Substituting this estimate into \eqref{Eq:Hidden-Coordinate-Transformation-Pattern-Bound} gives \(\|\mathbf{Q}_U\|_2 \leq 1 + \frac{ \sqrt{1+\delta_{k_p}^2} }{ \gamma_{k_p} }\) for every non-empty admissible set \(U\).
Recall that \(\beta_\mathbf{Q}^* := \max_{\substack{ U\subseteq[n]\\ |U|\leq k_p }} \|\mathbf{Q}_U\|_2\).
Since \(\|\mathbf{Q}_\emptyset\|_2=1\), we obtain \(\beta_\mathbf{Q}^* \leq 1 + \frac{ \sqrt{1+\delta_{k_p}^2} }{ \gamma_{k_p} }\), which proves \eqref{Eq:Uniform-Hidden-Coordinate-Transformation-Bound}.

For completeness, one may eliminate \(\gamma_{k_p}\) at the cost of a looser estimate.
Define \(\beta_{k_p} := \max_{\substack{ U\subseteq[n]\\ 1\leq |U|\leq k_p }} \|\mathbf{\hat{G}}_U\|_2\).
Fix a non-empty admissible set \(U\).
Since \(\mathbf{M}_U\) is unimodular, \(1 = |\det(\mathbf{M}_U)| = \prod_{j=1}^{|U|} \sigma_j(\mathbf{M}_U)\).
Moreover, because \(\mathbf{M}_U\) is a column submatrix of \(\mathbf{\hat{G}}_U\), \(\sigma_j(\mathbf{M}_U) \leq \|\mathbf{M}_U\|_2 \leq \|\mathbf{\hat{G}}_U\|_2 \leq \beta_{k_p}\).
The unimodularity of \(\mathbf{M}_U\) also implies \(\|\mathbf{M}_U\|_2\geq1\), and hence \(\beta_{k_p}\geq1\).
Therefore, \(\sigma_{\min}(\mathbf{M}_U) \geq \beta_{k_p}^{-(|U|-1)} \geq \beta_{k_p}^{-(k_p-1)}\).
Consequently, \(\gamma_{k_p} \geq \beta_{k_p}^{-(k_p-1)}\), and hence \(\beta_\mathbf{Q} \leq 1 + \beta_{k_p}^{k_p-1} \sqrt{1+\delta_{k_p}^2}\).

\subsection{Precision Analysis}

We give the detailed calculation underlying \eqref{Eq:Abort-Key-Hint-Precision-Uniform} and \eqref{Eq:Abort-Key-Combined-Precision}.
Recall that the centered multi-query response view is \(\mathbf{\widetilde{z}}_\mathcal{S} = \mathbf{M}_\mathsf{hid} \mathbf{x}'_\mathsf{hid} + \mathbf{r}_\mathcal{S}\), where \(\mathbf{M}_\mathsf{hid} = \left( \mathbf{M}_{i,s} \right)_{(s,i)\in\mathcal{S}} \in \mathbb{Z}^{|\mathcal{S}|D\times h_UD} \) is formed by block-row stacking, and \( \mathbf{M}_{i,s} := \mathbf{L}_s \left( \mathbf{G}'_\mathsf{hid}[i,\cdot] \otimes \mathbf{I}_D \right) \in \mathbb{Z}^{D\times h_UD}\).

The response masks \(\left( \mathbf{r}_{i,s} \right)_{(s,i)\in\mathcal{S}} \) are mutually independent, and each is governed by the Gaussian parameter matrix \(\sigma_{\bm{r}}^2\mathbf{I}_D\).
Therefore, the stacked perturbation vector \( \mathbf{r}_\mathcal{S} := \left( \mathbf{r}_{i,s} \right)_{(s,i)\in\mathcal{S}} \) is governed by \( \mathbf{\Sigma}_{\mathbf{r}_\mathcal{S}} = \sigma_{\bm{r}}^2 \mathbf{I}_{|\mathcal{S}|D}\), and its precision matrix is \( \mathbf{\Sigma}_{\mathbf{r}_\mathcal{S}}^{-1} = \frac{1}{\sigma_{\bm{r}}^2} \mathbf{I}_{|\mathcal{S}|D}\).

Fix \((s,i)\in\mathcal{S}\).
By the definition of \(\mathbf{M}_{i,s}\),
\begin{align*}
	&\mathbf{M}_{i,s}^\top \mathbf{M}_{i,s} \\
	&= \left( \mathbf{G}'_\mathsf{hid}[i,\cdot] \otimes \mathbf{I}_D \right)^\top \mathbf{L}_s^\top \mathbf{L}_s \left( \mathbf{G}'_\mathsf{hid}[i,\cdot] \otimes \mathbf{I}_D \right) \\
	&= \left( \mathbf{G}'_\mathsf{hid}[i,\cdot]^\top \otimes \mathbf{I}_D \right) \left( \mathbf{I}_1 \otimes \mathbf{L}_s^\top \mathbf{L}_s \right) \\
	&\quad\cdot \left( \mathbf{G}'_\mathsf{hid}[i,\cdot] \otimes \mathbf{I}_D \right) \\
	&= \left( \mathbf{G}'_\mathsf{hid}[i,\cdot]^\top \mathbf{G}'_\mathsf{hid}[i,\cdot] \right) \otimes \left( \mathbf{L}_s^\top \mathbf{L}_s \right).
\end{align*}
The final equality follows from the mixed-product rule for Kronecker products.

Because \(\mathbf{M}_\mathsf{hid}\) is obtained by vertically stacking the matrices \((\mathbf{M}_{i,s})_{(s,i)\in\mathcal{S}}\), we have
\begin{align*}
	\mathbf{M}_\mathsf{hid}^\top \mathbf{M}_\mathsf{hid} &= \sum_{(s,i)\in\mathcal{S}} \mathbf{M}_{i,s}^\top \mathbf{M}_{i,s} \\
	&= \sum_{s\in[Q_\mathsf{Sign}]} \left( (\mathbf{G}'_\mathsf{hid})_{V_s}^\top (\mathbf{G}'_\mathsf{hid})_{V_s} \right) \otimes \left( \mathbf{L}_s^\top \mathbf{L}_s \right),
\end{align*}
where \((\mathbf{G}'_\mathsf{hid})_{V_s}\) denotes the row restriction of \(\mathbf{G}'_\mathsf{hid}\) to the response-visibility set \(V_s\).
Using the triangle inequality, the multiplicativity of the spectral norm under Kronecker products, and \(\|\mathbf{L}_s\|_2 \leq \omega_c\), we obtain
\begin{align*}
	\left\| \mathbf{M}_\mathsf{hid}^\top \mathbf{M}_\mathsf{hid} \right\|_2 &\leq \sum_{s\in[Q_\mathsf{Sign}]} \left\| (\mathbf{G}'_\mathsf{hid})_{V_s}^\top (\mathbf{G}'_\mathsf{hid})_{V_s} \right\|_2 \left\| \mathbf{L}_s^\top \mathbf{L}_s \right\|_2 \\
	&= \sum_{s\in[Q_\mathsf{Sign}]} \left\| (\mathbf{G}'_\mathsf{hid})_{V_s} \right\|_2^2 \|\mathbf{L}_s\|_2^2 \\
	&\leq \omega_c^2 \sum_{s\in[Q_\mathsf{Sign}]} \left\| (\mathbf{G}'_\mathsf{hid})_{V_s} \right\|_2^2.
\end{align*}
Since \(\mathbf{\Sigma}_{\mathbf{r}_\mathcal{S}}^{-1} = \frac{1}{\sigma_{\bm{r}}^2} \mathbf{I}_{|\mathcal{S}|D}\), it follows that
\begin{align}\label{Eq:Abort-Key-Hint-Precision-Pattern}
	\left\| \mathbf{M}_\mathsf{hid}^\top \mathbf{\Sigma}_{\mathbf{r}_\mathcal{S}}^{-1} \mathbf{M}_\mathsf{hid} \right\|_2 \leq \frac{\omega_c^2}{\sigma_{\bm{r}}^2} \sum_{s\in[Q_\mathsf{Sign}]} \left\| (\mathbf{G}'_\mathsf{hid})_{V_s} \right\|_2^2.
\end{align}

We now derive the uniform estimate \eqref{Eq:Abort-Key-Hint-Precision-Uniform}.
Recall that \(\mathbf{G}\mathbf{Q}_U = \begin{bmatrix} \mathbf{G}'_\mathsf{pub} \ \mathbf{G}'_\mathsf{hid} \end{bmatrix}\).
Since row restriction and column restriction cannot increase the spectral norm,
\begin{align*}
	\left\| (\mathbf{G}'_\mathsf{hid})_{V_s} \right\|_2 &\leq \|\mathbf{G}'_\mathsf{hid}\|_2 \leq \|\mathbf{G}\mathbf{Q}_U\|_2 \\
	&\leq \|\mathbf{G}\|_2 \|\mathbf{Q}_U\|_2 \leq \|\mathbf{G}\|_2 \beta_\mathbf{Q}.
\end{align*}
Substituting this estimate into \eqref{Eq:Abort-Key-Hint-Precision-Pattern} gives
\begin{align*}
	\left\| \mathbf{M}_\mathsf{hid}^\top \mathbf{\Sigma}_{\mathbf{r}_\mathcal{S}}^{-1} \mathbf{M}_\mathsf{hid} \right\|_2 &\leq \frac{\omega_c^2}{\sigma_{\bm{r}}^2} \sum_{s\in[Q_\mathsf{Sign}]} \|\mathbf{G}\|_2^2 \beta_\mathbf{Q}^2 \\
	&= \frac{ \omega_c^2 Q_\mathsf{Sign} \|\mathbf{G}\|_2^2 \beta_\mathbf{Q}^2 }{ \sigma_{\bm{r}}^2 }.
\end{align*}
This proves \eqref{Eq:Abort-Key-Hint-Precision-Uniform}.

It remains to combine the response-hint precision with the conditional prior precision of the hidden transformed coordinates.
By \Cref{Lemma:Conditional-Hidden-Coordinate-Precision} and \eqref{Eq:Uniform-Hidden-Coordinate-Transformation-Bound},
\begin{align*}
	\left\| \mathbf{\Sigma}_{U,\mathsf{hid}\mid\mathsf{pub}}^{-1} \otimes \mathbf{I}_D \right\|_2 = \left\| \mathbf{\Sigma}_{U,\mathsf{hid}\mid\mathsf{pub}}^{-1} \right\|_2 \leq \left\| \mathbf{\Sigma}_{\mathbf{X}}^{-1} \right\|_2 \beta_\mathbf{Q}^2.
\end{align*}
Since \(\mathbf{\Sigma}_\mathbf{X} = \mathrm{diag} \left( \sigma_{\bm{s}}^2, \sigma_\mathbf{R}^2 \mathbf{I}_{k-1} \right)\), we have \(\left\| \mathbf{\Sigma}_{\mathbf{X}}^{-1} \right\|_2 = \max \left\{ \sigma_{\bm{s}}^{-2}, \sigma_\mathbf{R}^{-2} \right\}\).
Therefore, \( \left\| \mathbf{\Sigma}_{U,\mathsf{hid}\mid\mathsf{pub}}^{-1} \otimes \mathbf{I}_D \right\|_2 \leq \beta_\mathbf{Q}^2 \max \left\{ \sigma_{\bm{s}}^{-2}, \sigma_\mathbf{R}^{-2} \right\}\).
Using the triangle inequality together with \eqref{Eq:Abort-Key-Hint-Precision-Uniform}, we obtain
\begin{align*}
	&\left\| \mathbf{\Sigma}_{U,\mathsf{hid}\mid\mathsf{pub}}^{-1} \otimes \mathbf{I}_D + \mathbf{M}_\mathsf{hid}^\top \mathbf{\Sigma}_{\mathbf{r}_\mathcal{S}}^{-1} \mathbf{M}_\mathsf{hid} \right\|_2 \\
	&\leq \left\| \mathbf{\Sigma}_{U,\mathsf{hid}\mid\mathsf{pub}}^{-1} \otimes \mathbf{I}_D \right\|_2 + \left\| \mathbf{M}_\mathsf{hid}^\top \mathbf{\Sigma}_{\mathbf{r}_\mathcal{S}}^{-1} \mathbf{M}_\mathsf{hid} \right\|_2 \\
	&\leq \beta_\mathbf{Q}^2 \left( \max \left\{ \sigma_{\bm{s}}^{-2}, \sigma_\mathbf{R}^{-2} \right\} + \frac{ \omega_c^2 Q_\mathsf{Sign} \|\mathbf{G}\|_2^2 }{ \sigma_{\bm{r}}^2 } \right).
\end{align*}
This proves \eqref{Eq:Abort-Key-Combined-Precision}.

The resulting bounds depend only on the public parameter bounds \(Q_\mathsf{Sign}\), \(\omega_c\), \(\|\mathbf{G}\|_2\), \(\beta_\mathbf{Q}\), \(\sigma_{\bm{s}}\), \(\sigma_\mathbf{R}\), and \(\sigma_{\bm{r}}\).
They are therefore uniform over every admissible corrupted set \(U\) and every admissible multi-query response-visibility pattern \((V_s)_{s\in[Q_\mathsf{Sign}]}\).

\subsection{Proof of \texorpdfstring{\Cref{Lemma:Computational-Indistinguishability-Abort-Key-View}}{Lemma \getrefnumber{Lemma:Computational-Indistinguishability-Abort-Key-View}}}

We give the full proof of \Cref{Lemma:Computational-Indistinguishability-Abort-Key-View}.
The reduction proceeds through \(\mathsf{HC\text{-}Leak\text{-}AK} \to \mathsf{RA\text{-}MHMLWE}[h_U] \to \mathsf{RA\text{-}MLWE}[h_U] \to \mathsf{MLWE}\), where \(U := U_\mathrm{raw}\), \(h_U := k-|U|\), and \(D := d\ell\).
The first step uses the hidden-coordinate representation and the combined precision bound established in the preceding analyses.
The second step follows from the posterior-resampling reduction of \Cref{Lemma:Hardness-RA-MHMLWE}, and the final step applies \Cref{Lemma:Hardness-RA-MLWE}.

\begin{proof}[Proof of \Cref{Lemma:Computational-Indistinguishability-Abort-Key-View}]
	Fix an admissible multi-query leakage pattern \(\mathcal{U} = \left( U_\mathrm{raw}, (V_s)_{s\in[Q_\mathsf{Sign}]} \right)\) and a public challenge sequence \(\mathbf{c} = (c_s)_{s\in[Q_\mathsf{Sign}]}\).
	Write \(U:=U_\mathrm{raw}\), \(h_U:=k-|U|\), and \(D:=d\ell\).
	Fix an arbitrary value \(\mathbf{S}_U\) in the support of the exposed raw-share view.
	By \Cref{Lemma:Conditional-Hidden-Coordinates}, conditioned on \(\mathbf{X}'_\mathsf{pub}=\mathbf{S}_U\), the vectorized hidden secret \(\mathbf{x}'_\mathsf{hid} = \mathrm{vec} \left( {\mathbf{X}'_\mathsf{hid}}^\top \right) \in \mathbb{Z}^{h_UD}\) follows a discrete Gaussian governed by the Gaussian parameter matrix \(\mathbf{\Sigma}_{U,\mathsf{hid}\mid\mathsf{pub}} \otimes \mathbf{I}_D\), with a conditional center determined by \(\mathbf{S}_U\).
	After conditioning, this center is fixed public auxiliary information.

	As shown in the detailed \(\mathsf{RA\text{-}MHMLWE}\) representation, the centered response leakage is \(\mathbf{\widetilde{z}}_\mathcal{S} = \mathbf{M}_\mathsf{hid} \mathbf{x}'_\mathsf{hid} + \mathbf{r}_\mathcal{S}\), while the real and simulated hidden evaluated targets are, respectively, \(\mathbf{A}^{\langle h_U\rangle} \mathbf{x}'_\mathsf{hid}\) and \(\mathbf{u}_\mathsf{hid} \stackrel{\$}{\gets} \mathbb{Z}_q^{h_Udp}\).
	Hence, conditioned on \(\mathbf{S}_U\), the real-versus-simulated distinction has the form of an elliptic \(\mathsf{RA\text{-}MHMLWE}[h_U]\) instance with one stacked rectangular hint.
	When \(\mathcal{S}=\emptyset\), there is no hint, which is the case \(\nu=0\) of \Cref{Definition:RA-MHMLWE}.

	By \eqref{Eq:Abort-Key-Precision-via-Sigma-AK}, the combined conditional-prior and hint precision satisfies \(\big\| \mathbf{\Sigma}_{U,\mathsf{hid}\mid\mathsf{pub}}^{-1} \otimes \mathbf{I}_D + \mathbf{M}_\mathsf{hid}^\top \mathbf{\Sigma}_{\mathbf{r}_\mathcal{S}}^{-1} \mathbf{M}_\mathsf{hid} \big\|_2 \leq \frac{1}{2\sigma_\mathrm{AK}^2}\).
	Furthermore, since \(h_U\leq k\), the smoothing assumption \(\sigma_\mathrm{AK} \geq \sqrt{2}\eta_\epsilon \left( \mathbb{Z}^{kD} \right)\) implies \(\sigma_\mathrm{AK} \geq \sqrt{2}\eta_\epsilon \left( \mathbb{Z}^{h_UD} \right)\).
	Therefore, the hypotheses of \Cref{Lemma:Hardness-RA-MHMLWE} are satisfied for the realized hidden-coordinate instance.

	Applying \Cref{Lemma:Hardness-RA-MHMLWE} with \(\mathbf{S}_U\) and the conditional center as public auxiliary information replaces \(\mathbf{A}^{\langle h_U\rangle} \mathbf{x}'_\mathsf{hid}\) with \(\mathbf{u}_\mathsf{hid}\), while preserving the centered response leakage, with statistical loss at most \(2\epsilon\).
	The conditional Gaussian parameter matrix and the combined-precision bound are independent of the realized value of \(\mathbf{S}_U\).
	Thus, the reduction applies to the joint distribution obtained by first sampling the raw-share view and then sampling the hidden coordinates from their conditional distribution.
    The raw-share view is sampled exactly by drawing a fresh dealing matrix \(\mathbf{X}\), whose rows are spherical, and keeping only \(\mathbf{S}_U=\mathbf{G}_U\mathbf{X}\).

	The hidden-target replacement induces exactly the simulated abort-key distribution.
	Indeed, in the real branch, \(\mathbf{A}^{\langle h_U\rangle} \mathbf{x}'_\mathsf{hid} = \mathrm{vec} \left( \left( \mathbf{X}'_\mathsf{hid} \mathbf{A}^\top \right)^\top \right)\).
	In the uniform branch, write \(\mathbf{u}_\mathsf{hid} = \mathrm{vec} \left( \mathbf{U}_\mathsf{hid}^\top \right)\), where \(\mathbf{U}_\mathsf{hid} \stackrel{\$}{\gets} \mathbb{Z}_q^{h_U\times dp}\).
	Then \(\mathbf{\breve{Y}} := \mathbf{Q}_U \begin{bmatrix} \mathbf{S}_U\mathbf{A}^\top\\ \mathbf{U}_\mathsf{hid} \end{bmatrix}\) is uniform over \(\mathcal{Y}_U \left( \mathbf{S}_U\mathbf{A}^\top \right)\).
	By the affine-completion analysis, extracting the first row of \(\mathbf{\breve{Y}}\) and applying the public map \(\mathbf{G}\) produces exactly the uniform target and completed abort-key tuple generated by \(\mathsf{SimAK}\).
	Thus, the representation introduces no additional loss.
	Consequently, writing \(\mathsf{RA\text{-}MLWE}_{\sigma_\mathrm{AK}}[h_U]:=\mathsf{RA\text{-}MLWE}_{p,\ell-p,q,d,\sigma_\mathrm{AK}}[h_U]\) and \(\mathsf{MLWE}_{\sigma_\mathrm{AK}}:=\mathsf{MLWE}_{p,\ell-p,q,d,\sigma_\mathrm{AK}}\), there exists a PPT adversary \(\mathcal{B}\) against \(\mathsf{RA\text{-}MLWE}_{\sigma_\mathrm{AK}}[h_U]\) such that
	\begin{align}\label{Eq:Abort-Key-to-RA-MLWE}
		\mathsf{Adv}_{ \mathcal{A}, \mathsf{HC\text{-}Leak\text{-}AK} }^{ (\mathsf{pp},\mathcal{U},\mathbf{c}) } (\kappa) \leq \mathsf{Adv}_{\mathcal{B}}^{ \mathsf{RA\text{-}MLWE}_{\sigma_\mathrm{AK}}[h_U] } (\kappa) + 2\epsilon.
	\end{align}
	By \Cref{Lemma:Hardness-RA-MLWE}, there exists a PPT adversary \(\mathcal{B}'\) against \(\mathsf{MLWE}_{p,\ell-p,q,d,\sigma_\mathrm{AK}}\) such that
	\begin{align}\label{Eq:Abort-Key-RA-MLWE-to-MLWE}
		\mathsf{Adv}_{\mathcal{B}}^{ \mathsf{RA\text{-}MLWE}_{\sigma_\mathrm{AK}}[h_U] } (\kappa) \leq h_U \cdot \mathsf{Adv}_{\mathcal{B}'}^{ \mathsf{MLWE}_{\sigma_\mathrm{AK}} } (\kappa).
	\end{align}
	Combining \eqref{Eq:Abort-Key-to-RA-MLWE} and \eqref{Eq:Abort-Key-RA-MLWE-to-MLWE} gives \eqref{Eq:Abort-Key-View-Advantage-Bound}.
\end{proof}

\section{Hybrid Experiments and Proofs for \texorpdfstring{\Cref{Section:LaTSEC-Unforgeability}}{Section \getrefnumber{Section:LaTSEC-Unforgeability}}}\label{Appendix:LaTSEC-Unforgeability-Hybrids}

This section gives the complete experiments used in the hybrid argument of \Cref{Section:LaTSEC-Unforgeability}.
The main text describes the changes between consecutive hybrids and states the corresponding transition bounds.
In each figure below, any oracle or game step not explicitly modified is executed as in the preceding hybrid, and ``complete \(\dots\) as in \(\mathsf{Hyb}_j\)'' runs only the remaining steps of that oracle in \(\mathsf{Hyb}_j\), without repeating earlier ones.

\subsection{\texorpdfstring{\(\bm{\mathsf{Hyb}_1}\)}{Hyb1}: Programming Honest Commitments}

The complete modification defining \(\mathsf{Hyb}_1\) is given in \Cref{Figure:Hybrid1}.
In the first signing round, the challenger replaces each honest commitment hash with an independently sampled value in \(\{0,1\}^{2\kappa}\) and stores that value together with the corresponding opening \(\bm{w}_i\).
In the second signing round, before returning the opening to the adversary, the challenger programs \(H_\mathsf{cmt}\) at the associated input.
The random commitment has the same distribution as a fresh output of \(H_\mathsf{cmt}\).
Therefore, the transition changes the adversary's view only if the relevant oracle input was queried before the challenger programs it.
This event is recorded as \(\mathsf{Bad}_\mathsf{cmt}\).
All other signing-oracle behavior, including transcript maintenance, blame handling, and the third signing round, remains unchanged from \(\mathsf{Hyb}_0\).

\begin{figure*}[!t]
	\centering
	\begin{tcolorbox}[gamebox]
		\small
		\(\mathsf{Hyb}_1\): the game is identical to \(\mathsf{Hyb}_0\), except for the following modifications to \(\mathsf{OSign}_1\) and \(\mathsf{OSign}_2\).

		\smallskip
		\(\mathsf{OSign}_1(\mathsf{sid},\mathsf{act},\mathsf{msg})\)
		\hrule height 0.4pt
		\begin{algorithmic}[1]
			\State \textbf{assert} \{\(\mathsf{sid}\notin L_\mathsf{sid} \land \mathsf{act}\subseteq[n] \land |\mathsf{act}|\geq t_\mathsf{sig} \land \mathcal{P}(\mathsf{tspar},\mathsf{vk},\mathsf{sid},\mathsf{act},\mathsf{msg})=1\land (\mathsf{honest}\cap\mathsf{act}\neq\emptyset)\)\}
			\For{\(i \in \mathsf{honest}\cap\mathsf{act}\)}
				\State \textbf{assert} \{\(\mathsf{sid}\notin \mathsf{st}_i.\mathsf{session}\)\}
				\State \(\bm{r}_i \gets \mathcal{D}_{\sigma_{\bm{r}}}^\ell\).
				\State \(\bm{w}_i := \bm{A}\bm{r}_i\).
				\State \colorbox{lightgray}{\(\mathsf{cmt}_i\stackrel{\$}{\gets}\{0,1\}^{2\kappa}\).} \Comment{replace \(\mathsf{cmt}_i\) with random value}
				\State \(\mathsf{st}_i.\mathsf{session}[\mathsf{sid}] := (\mathsf{status}=\mathsf{running},\mathsf{rnd}=1,\mathsf{act},\mathsf{msg},\mathsf{open}=(\bm{r}_i,\bm{w}_i,\mathsf{cmt}_i),\mathsf{aux}=\emptyset)\)
				\State \(\mathsf{pm}_i^1 := \mathsf{cmt}_i\)
			\EndFor
			\State Initialize transcript table \(\mathsf{PM}\) and blame table \(\mathsf{BL}\) as in \Cref{Figure:Game-TS-UF-Oracle}.
			\State \(L_\mathsf{sig}:=L_\mathsf{sig}\cup\{\mathsf{msg}\}\)
			\State \(L_\mathsf{sid}[\mathsf{sid}]:=(\mathsf{running},1,\mathsf{act},\mathsf{msg},\mathsf{PM},\mathsf{BL})\)
			\State \Return \((\mathsf{pm}_i^1)_{i\in\mathsf{honest}\cap\mathsf{act}}\)
		\end{algorithmic}
		\smallskip

		\(\mathsf{OSign}_2(\mathsf{sid},(\mathsf{pm}_j^1)_{j \in \mathsf{corrupt}\cap\mathsf{act}})\)
		\hrule height 0.4pt
		\begin{algorithmic}[1]
			\State \textbf{assert} \{\(\mathsf{sid}\in L_\mathsf{sid}\)\}
			\State Fetch \((\mathsf{status},r,\mathsf{act},\mathsf{msg},\mathsf{PM},\mathsf{BL})\gets L_\mathsf{sid}[\mathsf{sid}]\).
			\If{\(\mathsf{status}=\mathsf{abort}\)}
				\State \Return \(\bot\)
			\EndIf
			\State \textbf{assert} \{\(r=1\)\}
			\State Record \((\mathsf{pm}_j^1)_{j\in\mathsf{corrupt}\cap\mathsf{act}}\) in \(\mathsf{PM}\).
			\For{\(i\in\mathsf{honest}\cap\mathsf{act}\)}
				\State Fetch \(\mathsf{open}=(\bm{r}_i,\bm{w}_i,\mathsf{cmt}_i)\) from \(\mathsf{st}_i.\mathsf{session}[\mathsf{sid}]\).
				\State Let \(x_i:=(\mathsf{sid},\mathsf{vk},\mathsf{act},i,\bm{w}_i)\).
				\If{\(x_i\) has already been queried to \(H_\mathsf{cmt}\)}
					\State \(\mathsf{Bad}_\mathsf{cmt}:=1\); abort the hybrid.
				\Else
					\State \colorbox{lightgray}{Program \(H_\mathsf{cmt}(x_i):=\mathsf{cmt}_i\).} \Comment{program random oracle}
				\EndIf
			\EndFor
			\For{\(i\in\mathsf{honest}\cap\mathsf{act}\)}
				\State \((\mathsf{st}_i,\mathsf{pm}_i^2)\gets
				\mathsf{LaTSEC.ShareSign}_2(\mathsf{st}_i,\mathsf{sid},\mathsf{act},(\mathsf{pm}_u^1)_{u\in\mathsf{act}})\).
			\EndFor
			\State Set \(\mathsf{pm}_i^2:=\bm{w}_i\) for every \(i\in\mathsf{honest}\cap\mathsf{act}\).
			\State Update \(\mathsf{PM}\), \(\mathsf{BL}\), and \(L_\mathsf{sid}\), including abort handling, exactly as in \Cref{Figure:Game-TS-UF-Oracle}.
			\State \Return \((\mathsf{pm}_i^2)_{i\in\mathsf{honest}\cap\mathsf{act}}\).
		\end{algorithmic}
	\end{tcolorbox}
	\caption{\(\mathsf{Hyb}_1\) of the unforgeability proof of our scheme; the differences from the previous hybrid are \colorbox{lightgray}{highlighted}.
	For conciseness, we omit some routine assertions in \Cref{Algorithm:LaTSEC-ShareSign2}.}
	\label{Figure:Hybrid1}
\end{figure*}

Conditioned on \(\neg\mathsf{Bad}_\mathsf{cmt}\), the programmed random-oracle value is distributed exactly as the value that would have been returned by an honest evaluation of \(H_\mathsf{cmt}\).
Moreover, programming occurs before the opening \(\bm{w}_i\) is revealed, so the subsequent commitment-opening check remains unchanged.
Hence, conditioned on \(\neg\mathsf{Bad}_\mathsf{cmt}\), the adversary's view in \(\mathsf{Hyb}_1\) is identically distributed to its view in \(\mathsf{Hyb}_0\).

For completeness, the transition bound follows from the entropy of the programmed input.
By \Cref{Lemma:LaTSEC-Systematic-MLWE-MinEntropy}, each value \(\bm{w}_i=\bm{A}\bm{r}_i\) has min-entropy at least \(d-1\), except with probability \(2^{-d+1}\) over the choice of \(\bm{A}'\).
Thus, for each oracle query, the probability of guessing the programmed opening is at most \(2^{-d+1}\).
A union bound over at most \(Q_H\) oracle queries, \(n\) honest active signers per signing session, and \(Q_\mathsf{Sign}\) signing sessions, together with the possible failure of the min-entropy event, gives \(\Pr[\mathsf{Bad}_\mathsf{cmt}] \leq n(Q_H+1)Q_\mathsf{Sign}\cdot 2^{-d+1}\).
This proves the transition bound in \eqref{Eq:Hybrid0-Hybrid1}.

\subsection{\texorpdfstring{\(\bm{\mathsf{Hyb}_2}\)}{Hyb2}: Extraction of Corrupted Openings}

The complete modification defining \(\mathsf{Hyb}_2\) is given in \Cref{Figure:Hybrid2}.
The challenger searches the \(H_\mathsf{cmt}\)-table during \(\mathsf{OSign}_2\) for openings of all commitments supplied by corrupted active signers.
If every corrupted opening can be extracted, the challenger combines these values with the stored honest openings, reconstructs the aggregate commitment, computes the common challenge, and precomputes the honest third-round responses.
The table \(\mathsf{Programmed}\) records the extracted corrupted openings, the resulting challenge, and the precomputed honest responses.
All of these values remain internal during \(\mathsf{OSign}_2\): the only values returned to the adversary are the honest second-round openings, exactly as in \(\mathsf{Hyb}_1\).
The stored responses are returned only in \(\mathsf{OSign}_3\), after the challenger has verified that the corrupted openings agree with the extracted values.
Thus, the modification advances the challenger's internal computation without advancing the adversary's view.

\begin{figure*}[!t]
	\centering
	\begin{tcolorbox}[gamebox]
		\small
		\(\mathsf{Hyb}_2\): the game is identical to \(\mathsf{Hyb}_1\), except that the challenger extracts corrupted openings in \(\mathsf{OSign}_2\) and stores precomputed honest third-round responses in \(\mathsf{Programmed}\).
		\(\mathsf{Programmed}\) is initially empty and we write \(\mathsf{Programmed}[\mathsf{sid},\bm{w}] = \bot\) when no entry is defined.

		\smallskip
		\(\mathsf{OSign}_2(\mathsf{sid},(\mathsf{pm}_j^1)_{j\in\mathsf{corrupt}\cap\mathsf{act}})\)
		\hrule height 0.4pt
		\begin{algorithmic}[1]
			\State Run \(\mathsf{OSign}_2\) as in \(\mathsf{Hyb}_1\), up to the point after \(H_\mathsf{cmt}\) has been programmed for all honest openings.
			\For{\(j\in\mathsf{corrupt}\cap\mathsf{act}\)}
				\State Search the \(H_\mathsf{cmt}\)-table for \(\widehat{\bm{w}}_j\in\mathcal{R}_q^p\) such that \(H_\mathsf{cmt}(\mathsf{sid},\mathsf{vk},\mathsf{act},j,\widehat{\bm{w}}_j)=\mathsf{pm}_j^1\).
				\If{no such \(\widehat{\bm{w}}_j\) exists}
					\State Complete \(\mathsf{OSign}_2\) exactly as in \(\mathsf{Hyb}_1\) and return its output.
				\EndIf
			\EndFor
			\State \colorbox{lightgray}{Define \(\widetilde{\bm{w}}_u := \begin{cases} \bm{w}_u, & u\in\mathsf{honest}\cap\mathsf{act},\\ \widehat{\bm{w}}_u, & u\in\mathsf{corrupt}\cap\mathsf{act}.\end{cases}\)}
			\State \colorbox{lightgray}{Compute \(\bm{w}:=\mathsf{SEC.Rec}_\mathsf{pp}\left(\mathsf{act},(\widetilde{\bm{w}}_u)_{u\in\mathsf{act}}\right)\).}
			\State \colorbox{lightgray}{Set \(c:=H_c(\mathsf{sid},\mathsf{vk},\mathsf{act},\mathsf{msg},\bm{w})\).}
			\For{\(i\in\mathsf{honest}\cap\mathsf{act}\)}
				\State \colorbox{lightgray}{\(\bm{z}_i:=c\bm{s}_i+\bm{r}_i\)}
			\EndFor
			\State \colorbox{lightgray}{\(\mathsf{Programmed}[\mathsf{sid},\bm{w}]:=\left((\widehat{\bm{w}}_j)_{j\in\mathsf{corrupt}\cap\mathsf{act}},c,(\bm{z}_i)_{i\in\mathsf{honest}\cap\mathsf{act}}\right)\).}
			\State Complete \(\mathsf{OSign}_2\) exactly as in \(\mathsf{Hyb}_1\) and return its output.
		\end{algorithmic}

		\smallskip
		\(\mathsf{OSign}_3(\mathsf{sid},(\mathsf{pm}_j^2)_{j\in\mathsf{corrupt}\cap\mathsf{act}})\)
		\hrule height 0.4pt
		\begin{algorithmic}[1]
			\State Let \((\mathsf{pm}_u^2)_{u\in\mathsf{act}}\) be obtained from the honest stored openings and the corrupted openings provided to \(\mathsf{OSign}_3\).
			\State Verify the supplied openings and update the transcript, blame table, and abort status as in \(\mathsf{Game}_\mathsf{TS\text{-}UF}\).
			\If{\(\exists j\in\mathsf{corrupt}\cap\mathsf{act}\) such that \(H_\mathsf{cmt}\left(\mathsf{sid},\mathsf{vk},\mathsf{act},j,\mathsf{pm}_j^2\right) \neq \mathsf{pm}_j^1\)}
				\State Follow the ordinary abort behavior of \(\mathsf{Game}_\mathsf{TS\text{-}UF}\) and return.
			\EndIf
			\State Form \(\bm{w}:=\mathsf{SEC.Rec}_\mathsf{pp}\left(\mathsf{act},(\mathsf{pm}_u^2)_{u\in\mathsf{act}}\right)\).
			\If{\(\mathsf{Programmed}[\mathsf{sid},\bm{w}] =\bot\)}
				\State \(\mathsf{Bad}_\mathsf{open}:=1\); abort the hybrid.
			\EndIf
			\State Parse \(\mathsf{Programmed}[\mathsf{sid},\bm{w}] = \left((\widehat{\bm{w}}_j)_{j\in\mathsf{corrupt}\cap\mathsf{act}}, c, (\bm{z}_i)_{i\in\mathsf{honest}\cap\mathsf{act}} \right)\).
			\If{\((\mathsf{pm}_j^2)_{j\in\mathsf{corrupt}\cap\mathsf{act}} \neq (\widehat{\bm{w}}_j)_{j\in\mathsf{corrupt}\cap\mathsf{act}}\)}
				\State \(\mathsf{Bad}_\mathsf{open}:=1\); abort the hybrid.
			\EndIf
			\State \Return \((\bm{z}_i)_{i\in\mathsf{honest}\cap\mathsf{act}}\), updating the transcript exactly as in \(\mathsf{Game}_\mathsf{TS\text{-}UF}\).
		\end{algorithmic}
	\end{tcolorbox}
	\caption{\(\mathsf{Hyb}_2\) of the unforgeability proof.
	The challenger extracts corrupted openings in \(\mathsf{OSign}_2\) and stores precomputed honest third-round responses in \(\mathsf{Programmed}\); the differences from \(\mathsf{Hyb}_1\) are \colorbox{lightgray}{highlighted}.}
	\label{Figure:Hybrid2}
\end{figure*}

We justify that the preceding extraction and precomputation preserve the transcript distribution unless \(\mathsf{Bad}_\mathsf{open}\) occurs.
If every corrupted opening later supplied in \(\mathsf{OSign}_3\) matches the value extracted during \(\mathsf{OSign}_2\), then the reconstructed commitment \(\bm{w}\), the challenge \(c\), and every honest response \(\bm{z}_i=c\bm{s}_i+\bm{r}_i\) are exactly the values that would have been computed in \(\mathsf{Hyb}_1\).
If no opening was extracted for some corrupted commitment and that commitment later fails verification, both hybrids follow the same abort behavior.
Consequently, conditioned on \(\neg\mathsf{Bad}_\mathsf{open}\), the adversary's view is identical in \(\mathsf{Hyb}_1\) and \(\mathsf{Hyb}_2\).

The bad event can occur in two ways.
First, a corrupted signer may later supply a valid opening for a commitment value for which no preimage appeared in the \(H_\mathsf{cmt}\)-table during \(\mathsf{OSign}_2\).
For a fixed commitment value in \(\{0,1\}^{2\kappa}\), each fresh oracle query finds such a preimage with probability \(2^{-2\kappa}\).
Taking a union bound over at most \(t_\mathsf{uf}\) corrupted active signers, \(Q_\mathsf{Sign}\) signing sessions, and \(Q_H\) oracle queries gives a probability of at most \(t_\mathsf{uf}Q_\mathsf{Sign}Q_H\cdot 2^{-2\kappa}\).
Second, a corrupted signer may open a commitment to a value different from the one extracted during \(\mathsf{OSign}_2\).
This requires a collision in \(H_\mathsf{cmt}\), whose probability among at most \(Q_H\) oracle queries is at most \(Q_H^2\cdot 2^{-2\kappa}\).
Taking a union bound over these events gives \( \Pr[\mathsf{Bad}_\mathsf{open}] \leq \left( Q_H^2 + t_\mathsf{uf}Q_\mathsf{Sign}Q_H \right) 2^{-2\kappa}\).
This proves the transition bound in \eqref{Eq:Hybrid1-Hybrid2}.

\subsection{\texorpdfstring{\(\bm{\mathsf{Hyb}_3}\)}{Hyb3}: Programming the Challenge Oracle}

The complete experiment defining \(\mathsf{Hyb}_3\) is given in \Cref{Figure:Hybrid3}.
As in \(\mathsf{Hyb}_2\), the challenger extracts the corrupted commitment openings during \(\mathsf{OSign}_2\) and combines them with the stored honest openings to reconstruct the aggregate commitment \(\bm{w}\).
The challenger then samples \(c\stackrel{\$}{\gets}\mathcal{C}\) and programs \(H_c\) at \((\mathsf{sid},\mathsf{vk},\mathsf{act},\mathsf{msg},\bm{w})\).
If that point has already been queried, the challenger records \(\mathsf{Bad}_c\) and aborts the hybrid.
Otherwise, the challenger computes the honest responses \(\bm{z}_i:=c\bm{s}_i+\bm{r}_i\) and stores them in \(\mathsf{Programmed}\) as in \(\mathsf{Hyb}_2\).
This modification remains internal during \(\mathsf{OSign}_2\): only the honest second-round openings are returned at that point, while the stored responses are returned later by \(\mathsf{OSign}_3\).
The latter oracle is therefore unchanged from \(\mathsf{Hyb}_2\).

\begin{figure*}[!t]
	\centering
	\begin{tcolorbox}[gamebox]
		\small
		\(\mathsf{Hyb}_3\): the game is identical to \(\mathsf{Hyb}_2\), except that the challenge \(c\) is sampled uniformly from \(\mathcal{C}\) and the random oracle \(H_c\) is programmed at \((\mathsf{sid},\mathsf{vk},\mathsf{act},\mathsf{msg},\bm{w})\).
		\(\mathsf{Programmed}\) is maintained as in \(\mathsf{Hyb}_2\).

		\smallskip
		\(\mathsf{OSign}_2(\mathsf{sid},(\mathsf{pm}_j^1)_{j\in\mathsf{corrupt}\cap\mathsf{act}})\)
		\hrule height 0.4pt
		\begin{algorithmic}[1]
			\State Run \(\mathsf{OSign}_2\) as in \(\mathsf{Hyb}_1\), up to the point after \(H_\mathsf{cmt}\) has been programmed for all honest openings.
			\For{\(j\in\mathsf{corrupt}\cap\mathsf{act}\)}
				\State Search the \(H_\mathsf{cmt}\)-table for \(\widehat{\bm{w}}_j\in\mathcal{R}_q^p\) such that \(H_\mathsf{cmt}(\mathsf{sid},\mathsf{vk},\mathsf{act},j,\widehat{\bm{w}}_j)=\mathsf{pm}_j^1\).
				\If{no such \(\widehat{\bm{w}}_j\) exists}
					\State Complete \(\mathsf{OSign}_2\) exactly as in \(\mathsf{Hyb}_1\) and return its output.
				\EndIf
			\EndFor
			\State Define \(\widetilde{\bm{w}}_u := \begin{cases} \bm{w}_u, & u\in\mathsf{honest}\cap\mathsf{act},\\ \widehat{\bm{w}}_u, & u\in\mathsf{corrupt}\cap\mathsf{act}.\end{cases}\)
			\State Compute \(\bm{w}:=\mathsf{SEC.Rec}_\mathsf{pp}\left(\mathsf{act},(\widetilde{\bm{w}}_u)_{u\in\mathsf{act}}\right)\).
			\State \colorbox{lightgray}{Sample \(c \stackrel{\$}{\gets}\mathcal{C}\).}
			\State Let \(y:=(\mathsf{sid},\mathsf{vk},\mathsf{act},\mathsf{msg},\bm{w})\).
			\If{\(y\) has already been queried to \(H_c\)}
				\State \(\mathsf{Bad}_c := 1\); abort the hybrid.
			\Else
				\State \colorbox{lightgray}{Program \(H_c(y):= c\).} \Comment{program random oracle}
			\EndIf
			\For{\(i\in\mathsf{honest}\cap\mathsf{act}\)}
				\State \(\bm{z}_i:=c\bm{s}_i+\bm{r}_i\)
			\EndFor
			\State \(\mathsf{Programmed}[\mathsf{sid},\bm{w}]:=\left((\widehat{\bm{w}}_j)_{j\in\mathsf{corrupt}\cap\mathsf{act}},c,(\bm{z}_i)_{i\in\mathsf{honest}\cap\mathsf{act}}\right)\).
			\State Complete \(\mathsf{OSign}_2\) exactly as in \(\mathsf{Hyb}_2\) and return its output.
		\end{algorithmic}

		\smallskip
		\(\mathsf{OSign}_3(\mathsf{sid},(\mathsf{pm}_j^2)_{j\in\mathsf{corrupt}\cap\mathsf{act}})\) is unchanged from \(\mathsf{Hyb}_2\) in \Cref{Figure:Hybrid2}.
	\end{tcolorbox}
	\caption{\(\mathsf{Hyb}_3\) of the unforgeability proof.
	The challenger samples \(c\) uniformly from \(\mathcal{C}\) and programs \(H_c\) at the reconstructed aggregate commitment; the differences from \(\mathsf{Hyb}_2\) are \colorbox{lightgray}{highlighted}.}
	\label{Figure:Hybrid3}
\end{figure*}

Conditioned on \(\neg\mathsf{Bad}_c\), the programmed value has the same distribution as a fresh evaluation of \(H_c\), and hence the adversary's view is identical in \(\mathsf{Hyb}_2\) and \(\mathsf{Hyb}_3\).
The min-entropy analysis bounding \(\Pr[\mathsf{Bad}_c]\) is given in Appendix~\ref{Appendix:LaTSEC-Unforgeability-MinEntropy}, and the resulting transition bound is stated in \eqref{Eq:Hybrid2-Hybrid3}.

\subsection{\texorpdfstring{\(\bm{\mathsf{Hyb}_4}\)}{Hyb4}: Rewriting the Honest Openings}

The complete experiment defining \(\mathsf{Hyb}_4\) is given in \Cref{Figure:Hybrid4}.
The only modification from \(\mathsf{Hyb}_3\) is the highlighted rewriting of each honest second-round opening.
All other oracle and game steps are executed as in \(\mathsf{Hyb}_3\).

\begin{figure*}[!t]
	\centering
	\begin{tcolorbox}[gamebox]
		\small
		\(\mathsf{Hyb}_4\): the game is identical to \(\mathsf{Hyb}_3\), except that the honest second-round openings are rewritten as \(\bm{w}_i=\bm{A}\bm{z}_i-c\bm{t}_i\).
		\(\mathsf{Programmed}\) is maintained as in \(\mathsf{Hyb}_3\).

		\smallskip
		\(\mathsf{OSign}_2(\mathsf{sid},(\mathsf{pm}_j^1)_{j\in\mathsf{corrupt}\cap\mathsf{act}})\)
		\hrule height 0.4pt
		\begin{algorithmic}[1]
			\State Run \(\mathsf{OSign}_2\) as in \(\mathsf{Hyb}_1\), up to the point after \(H_\mathsf{cmt}\) has been programmed for all honest openings.
			\For{\(j\in\mathsf{corrupt}\cap\mathsf{act}\)}
				\State Search the \(H_\mathsf{cmt}\)-table for \(\widehat{\bm{w}}_j\in\mathcal{R}_q^p\) such that \(H_\mathsf{cmt}(\mathsf{sid},\mathsf{vk},\mathsf{act},j,\widehat{\bm{w}}_j)=\mathsf{pm}_j^1\).
				\If{no such \(\widehat{\bm{w}}_j\) exists}
					\State Complete \(\mathsf{OSign}_2\) exactly as in \(\mathsf{Hyb}_1\) and return its output.
				\EndIf
			\EndFor
			\State Define \(\widetilde{\bm{w}}_u := \begin{cases} \bm{w}_u, & u\in\mathsf{honest}\cap\mathsf{act},\\ \widehat{\bm{w}}_u, & u\in\mathsf{corrupt}\cap\mathsf{act}.\end{cases}\)
			\State Compute \(\bm{w}:=\mathsf{SEC.Rec}_\mathsf{pp}\left(\mathsf{act},(\widetilde{\bm{w}}_u)_{u\in\mathsf{act}}\right)\).
			\State Sample \(c \stackrel{\$}{\gets}\mathcal{C}\).
			\State Let \(y:=(\mathsf{sid},\mathsf{vk},\mathsf{act},\mathsf{msg},\bm{w})\).
			\If{\(y\) has already been queried to \(H_c\)}
				\State \(\mathsf{Bad}_c := 1\); abort the hybrid.
			\Else
				\State Program \(H_c(y):= c\). \Comment{program random oracle}
			\EndIf
			\For{\(i\in\mathsf{honest}\cap\mathsf{act}\)}
				\State \(\bm{z}_i:=c\bm{s}_i+\bm{r}_i\)
				\State \colorbox{lightgray}{Rewrite honest opening \(\bm{w}_i\) as \(\bm{w}_i := \bm{A}\bm{z}_i-c\bm{t}_i\).} \Comment{equal to \(\bm{A}\bm{r}_i\)}
			\EndFor
			\State \(\mathsf{Programmed}[\mathsf{sid},\bm{w}]:=\left((\widehat{\bm{w}}_j)_{j\in\mathsf{corrupt}\cap\mathsf{act}},c,(\bm{z}_i)_{i\in\mathsf{honest}\cap\mathsf{act}}\right)\).
			\State Complete \(\mathsf{OSign}_2\) exactly as in \(\mathsf{Hyb}_3\) and return its output.
		\end{algorithmic}

		\smallskip
		\(\mathsf{OSign}_3(\mathsf{sid},(\mathsf{pm}_j^2)_{j\in\mathsf{corrupt}\cap\mathsf{act}})\) is unchanged from \(\mathsf{Hyb}_3\) in \Cref{Figure:Hybrid3}.
	\end{tcolorbox}
	\caption{\(\mathsf{Hyb}_4\) of the unforgeability proof.
	The honest openings are rewritten as \(\bm{w}_i = \bm{A}\bm{z}_i - c\bm{t}_i\); the differences from \(\mathsf{Hyb}_3\) are \colorbox{lightgray}{highlighted}.}
	\label{Figure:Hybrid4}
\end{figure*}

\subsection{\texorpdfstring{\(\bm{\mathsf{Hyb}_5}\)}{Hyb5}: Simulation of the Public Abort-Key View}

The complete experiment defining \(\mathsf{Hyb}_5\) is given in \Cref{Figure:Hybrid5}.
At initialization, the challenger replaces the real global verification key and public abort-key tuple by a uniform target and the corresponding completion generated by \(\mathsf{SimAK}\).
The honest response values remain unchanged, but each honest opening is computed using the simulated abort key as \(\bm{w}_i := \bm{A}\bm{z}_i - c\bm{\breve{t}}_i\).
Since these openings depend on the sampled challenge, programming \(H_\mathsf{cmt}\) at the honest openings is deferred until the challenge, responses, and openings have been determined.
All remaining signing-oracle bookkeeping is performed as in \(\mathsf{Hyb}_4\).

\begin{figure*}[!t]
	\centering
	\begin{tcolorbox}[gamebox]
		\small
		\(\mathsf{Hyb}_5\): the game is identical to \(\mathsf{Hyb}_4\), except that the real global verification key \(\bm{t}=\bm{A}\bm{s}\) and the real public abort-key tuple \((\bm{t}_i)_{i\in[n]}\) are jointly replaced by a uniform global verification key \(\bm{\breve{t}}\) and a consistent abort-key completion \((\bm{\breve{t}}_i)_{i\in[n]}\).
		The response values remain real, while the honest openings are defined by \(\bm{w}_i:=\bm{A}\bm{z}_i-c\bm{\breve{t}}_i\).
		The table \(\mathsf{Programmed}\) is maintained as in \(\mathsf{Hyb}_4\).

		\smallskip
		\colorbox{lightgray}{Initialization}
		\hrule height 0.4pt
		\begin{algorithmic}[1]
			\State Sample \(\bm{\breve{t}} \stackrel{\$}{\gets} \mathcal{R}_q^p\), and let \(\mathbf{\breve{t}}\in\mathbb{Z}_q^{dp}\) be its coefficient embedding.
			\State Compute \(\mathbf{\breve{T}}\gets\mathsf{SimAK}_\mathsf{pp}\left(\mathsf{corrupt},\mathbf{\breve{t}},\mathbf{T}_\mathsf{corrupt}\right)\) as in \Cref{Definition:SimAK}.
			\State Let \((\bm{\breve{t}}_i)_{i\in[n]}\) be the ring-valued tuple corresponding to the rows of \(\mathbf{\breve{T}}\).
			\State Use \(\bm{\breve{t}}\) as the public global verification key and \((\bm{\breve{t}}_i)_{i\in[n]}\) as the public abort-key tuple.
		\end{algorithmic}

		\smallskip
		\(\mathsf{OSign}_2(\mathsf{sid},(\mathsf{pm}_j^1)_{j\in\mathsf{corrupt}\cap\mathsf{act}})\)
		\hrule height 0.4pt
		\begin{algorithmic}[1]
			\State Run the common \(\mathsf{OSign}_2\) prefix as in \(\mathsf{Hyb}_4\), except that programming \(H_\mathsf{cmt}\) at the honest openings is deferred.
			\State Extract the corrupted openings from the \(H_\mathsf{cmt}\)-table as in \(\mathsf{Hyb}_4\), obtaining \((\widehat{\bm{w}}_j)_{j\in\mathsf{corrupt}\cap\mathsf{act}}\); if extraction fails, complete \(\mathsf{OSign}_2\) exactly as in \(\mathsf{Hyb}_1\) and return its output.
			\State Sample \(c\stackrel{\$}{\gets}\mathcal{C}\).
			\For{\(i\in\mathsf{honest}\cap\mathsf{act}\)}
				\State \(\bm{z}_i:=c\bm{s}_i+\bm{r}_i\)
				\State \colorbox{lightgray}{\(\bm{w}_i:=\bm{A}\bm{z}_i-c\bm{\breve{t}}_i\)}
				\State \colorbox{lightgray}{Store \(\bm{w}_i\) as the honest opening and set \(\mathsf{pm}_i^2:=\bm{w}_i\).}
				\State Let \(x_i:=(\mathsf{sid},\mathsf{vk},\mathsf{act},i,\bm{w}_i)\).
				\If{\(x_i\) has already been queried to \(H_\mathsf{cmt}\)}
					\State \(\mathsf{Bad}_\mathsf{cmt}:=1\); abort the hybrid.
				\Else
					\State Program \(H_\mathsf{cmt}(x_i):=\mathsf{pm}_i^1\).
				\EndIf
			\EndFor
			\State Define \(\widetilde{\bm{w}}_u:=\begin{cases}\bm{w}_u, & u\in \mathsf{honest}\cap\mathsf{act}, \\ \widehat{\bm{w}}_u, & u\in \mathsf{corrupt}\cap\mathsf{act}.\end{cases}\)
			\State Compute \(\bm{w}:=\mathsf{SEC.Rec}_\mathsf{pp}\left(\mathsf{act},(\widetilde{\bm{w}}_u)_{u\in\mathsf{act}}\right)\).
			\State Let \(y:=(\mathsf{sid},\mathsf{vk},\mathsf{act},\mathsf{msg},\bm{w})\).
			\If{\(y\) has already been queried to \(H_c\)}
				\State \(\mathsf{Bad}_c:=1\); abort the hybrid.
			\Else
				\State Program \(H_c(y):=c\).
			\EndIf
			\State \(\mathsf{Programmed}[\mathsf{sid},\bm{w}]:=\left((\widehat{\bm{w}}_j)_{j\in\mathsf{corrupt}\cap\mathsf{act}},c,(\bm{z}_i)_{i\in\mathsf{honest}\cap\mathsf{act}}\right)\).
			\State Complete \(\mathsf{OSign}_2\) exactly as in \(\mathsf{Hyb}_4\) and return its output.
		\end{algorithmic}

		\smallskip
		\(\mathsf{OSign}_3(\mathsf{sid},(\mathsf{pm}_j^2)_{j\in\mathsf{corrupt}\cap\mathsf{act}})\) is unchanged from \(\mathsf{Hyb}_4\).
	\end{tcolorbox}

	\caption{\(\mathsf{Hyb}_5\) of the unforgeability proof.
	The real global verification key and public abort-key tuple are jointly replaced by a uniform target and the corresponding algebraic completion generated by \(\mathsf{SimAK}\).
	The response values remain real, while each honest opening is defined as \(\bm{w}_i=\bm{A}\bm{z}_i-c\bm{\breve{t}}_i\); the differences from \(\mathsf{Hyb}_4\) are \colorbox{lightgray}{highlighted}.}
	\label{Figure:Hybrid5}
\end{figure*}

By \Cref{Lemma:Algebraic-Abort-Key-Completion}, the completed tuple preserves the abort keys indexed by \(\mathsf{corrupt}\) and reconstructs to \(\bm{\breve{t}}\) from every certified set.
Thus, the simulated public material is algebraically consistent with both the corrupted raw-share view and the uniform global verification key.
The computational indistinguishability of the real and simulated public views is established by \Cref{Lemma:Computational-Indistinguishability-Abort-Key-View}, whose full proof is given in Appendix~\ref{Appendix:LaTSEC-Abort-Key-Simulation}.
The resulting transition bound is stated in \eqref{Eq:Hybrid4-Hybrid5}.
\Cref{Figure:Hybrid5} specifies the hybrid, whose challenger knows \(\bm{s}_i\) and \(\bm{r}_i\).
The reduction \(\mathcal{B}_\mathrm{AK}\) of \Cref{Section:LaTSEC-Initial-Hybrids} simulates it without them: once extraction succeeds, it takes \(c\) and each \(\bm{z}_i\) from the next unused leakage slot, and when extraction fails, it samples a fresh \(\bm{r}_i\).
Leakage slots left unused are ignored.

\subsection{Reduction to \texorpdfstring{\(\bm{\mathsf{SelfTargetMSIS}}\)}{SelfTargetMSIS}}

We give the complete simulation and extraction argument underlying \Cref{Lemma:Hybrid5-to-STMSIS}.

\begin{proof}[Proof of \Cref{Lemma:Hybrid5-to-STMSIS}]
	The adversary \(\mathcal{B}_\mathsf{STMSIS}\) receives \(\bm{A}_\mathsf{ST} \stackrel{\$}{\gets} \mathcal{R}_q^{p\times(\ell-p+1)}\) from the \(\mathsf{SelfTargetMSIS}\) challenger.
	It parses \(\bm{A}_\mathsf{ST} = \begin{bmatrix} \bm{A}' \ \bm{a} \end{bmatrix}\), where \(\bm{A}'\in\mathcal{R}_q^{p\times(\ell-p)}\) and \(\bm{a}\in\mathcal{R}_q^p\), and sets \(\bm{A} := \begin{bmatrix} \bm{I}_p \ \bm{A}' \end{bmatrix}\) and \(\bm{\breve{t}} := -\bm{a}\).
	Thus, \(\bm{A}_\mathsf{ST} = \begin{bmatrix} \bm{A}' \ -\bm{\breve{t}} \end{bmatrix}\).
	Since \(\bm{A}_\mathsf{ST}\) is uniform, the pair \((\bm{A}',\bm{\breve{t}})\) consists of mutually independent uniform values and has exactly the distribution prescribed in \(\mathsf{Hyb}_5\).

	The reduction samples \(\bm{s} \gets \mathcal{D}_{\bm{s}}\) and generates \((\bm{s}_i)_{i\in[n]} \gets \mathsf{SEC.Deal}_\mathsf{pp}(\bm{s})\).
	Let \(U_\mathrm{raw} := \mathsf{corrupt}\).
	For every \(i\in U_\mathrm{raw}\), it computes \(\bm{t}_i^\mathrm{raw} := \bm{A}\bm{s}_i\), and lets \(\mathbf{T}_{U_\mathrm{raw}}\) be the row-stacked coefficient embedding of \((\bm{t}_i^\mathrm{raw})_{i\in U_\mathrm{raw}}\).
	Writing \(\mathbf{\breve{t}}\in\mathbb{Z}_q^{dp}\) for the coefficient embedding of \(\bm{\breve{t}}\), the reduction computes \(\mathbf{\breve{T}} \gets \mathsf{SimAK}_\mathsf{pp} \left( U_\mathrm{raw}, \mathbf{\breve{t}}, \mathbf{T}_{U_\mathrm{raw}} \right)\) and lets \((\bm{\breve{t}}_i)_{i\in[n]}\) be the corresponding ring-valued abort-key tuple.
	By \Cref{Lemma:Algebraic-Abort-Key-Completion}, \((\bm{\breve{t}}_i)_{i\in U_\mathrm{raw}} = (\bm{A}\bm{s}_i)_{i\in U_\mathrm{raw}}\), and, for every certified set \(T\), \(\mathsf{SEC.Rec}_\mathsf{pp} \left( T, (\bm{\breve{t}}_i)_{i\in T} \right) = \bm{\breve{t}}\).
	Hence, the global verification target, raw-share view, and public abort-key tuple have exactly the initialization distribution of \(\mathsf{Hyb}_5\).

	The reduction runs \(\mathcal{A}\) with this public material and simulates the signing interaction as in \(\mathsf{Hyb}_5\).
	In particular, for each honest active signer \(i\), it computes \(\bm{z}_i := c\bm{s}_i+\bm{r}_i\) and \(\bm{w}_i := \bm{A}\bm{z}_i-c\bm{\breve{t}}_i\), programs \(H_\mathsf{cmt}\) and \(H_c\), maintains \(\mathsf{Programmed}\), and performs the signing-oracle bookkeeping exactly as prescribed by \(\mathsf{Hyb}_5\).

	To simulate \(H_c\), the reduction records and returns all values programmed during signing.
	For every unprogrammed query \((\mathsf{sid},\mathsf{vk},\mathsf{act},\mathsf{msg},\bm{w})\), it defines \(\mathsf{msg}_\mathsf{ST} := \mathsf{Encode} \left( \mathsf{sid}, \mathsf{vk}, \mathsf{act}, \mathsf{msg} \right)\) for a fixed injective encoding \(\mathsf{Encode}\) and answers with \(\mathsf{G} \left( \bm{w}, \mathsf{msg}_\mathsf{ST} \right)\).
	Thus, the simulated challenge oracle has the distribution required in \(\mathsf{Hyb}_5\).

	Suppose that \(\mathcal{A}\) outputs an accepting forgery \((\mathsf{sid}^*,\mathsf{act}^*,\mathsf{msg}^*,c^*,\bm{z}^*)\).
	Define \(\mathsf{msg}_\mathsf{ST}^* := \mathsf{Encode} \big( \mathsf{sid}^*, \mathsf{vk}, \mathsf{act}^*, \mathsf{msg}^* \big)\) and \(\bm{z}_\mathsf{ST}^* := \begin{bmatrix} \bm{z}^*\\ c^* \end{bmatrix}\).
	Verification gives \(c^* = H_c \left( \mathsf{sid}^*, \mathsf{vk}, \mathsf{act}^*, \mathsf{msg}^*, \bm{w}^* \right)\), where \(\bm{w}^* = \bm{A}\bm{z}^* - c^*\bm{\breve{t}}\).
	By the relation in \Cref{Section:LaTSEC-STMSIS-Reduction}, \(\bm{w}^* = \begin{bmatrix} \bm{I}_p \ \bm{A}_\mathsf{ST} \end{bmatrix} \bm{z}_\mathsf{ST}^*\).

	By freshness, the forgery point is not one of the signing-query points at which \(H_c\) was programmed.
	Its value is therefore obtained through \(\mathsf{G}\), and hence \(\mathsf{G} \left( \begin{bmatrix} \bm{I}_p \ \bm{A}_\mathsf{ST} \end{bmatrix} \bm{z}_\mathsf{ST}^*, \mathsf{msg}_\mathsf{ST}^* \right) = c^*\).

	It remains to verify the norm condition.
	Since \(\|c^*\|_\infty=1\) and \(\|c^*\|_1=\omega_c\), the coefficients of \(c^*\) belong to \(\{-1,0,1\}\), with exactly \(\omega_c\) nonzero coefficients.
	Therefore, \(\|c^*\|_2^2 = \omega_c\), and \(\|\bm{z}_\mathsf{ST}^*\|_2^2 = \|\bm{z}^*\|_2^2+\|c^*\|_2^2 \leq B_2^2+\omega_c \leq B_\mathsf{STMSIS}^2\).
	Moreover, \(\bm{z}_\mathsf{ST}^*\neq\bm{0}\) because \(\omega_c>0\) implies \(c^*\neq0\).
	Thus, \(0 < \|\bm{z}_\mathsf{ST}^*\|_2 \leq B_\mathsf{STMSIS}\).

	Consequently, \(\left( \mathsf{msg}_\mathsf{ST}^*, \bm{z}_\mathsf{ST}^* \right)\) is a valid solution to \(\mathsf{SelfTargetMSIS}_{p,\ell-p+1,q,d,\mathcal{C},B_\mathsf{STMSIS}}\).
	Whenever \(\mathcal{A}\) wins \(\mathsf{Hyb}_5\), the reduction therefore solves its \(\mathsf{SelfTargetMSIS}\) instance.
	This proves \Cref{Lemma:Hybrid5-to-STMSIS}.
\end{proof}

\subsection{Detailed Derivation of the Unforgeability Bound}

\begin{proof}[Proof of \Cref{Theorem:LaTSEC-TS-UF}]
	By definition, \(\mathsf{Hyb}_0\) is the original threshold-unforgeability experiment.
	Hence, \(\Pr[\mathsf{Hyb}_0=1] = \mathsf{Adv}_\mathcal{A}^\mathsf{TS\text{-}UF}(\kappa)\).
	By \eqref{Eq:Hybrid0-Hybrid1}, \(\left| \Pr[\mathsf{Hyb}_0=1] - \Pr[\mathsf{Hyb}_1=1] \right| \leq n(Q_H+1)Q_\mathsf{Sign}\cdot 2^{-d+1}\).
	By \eqref{Eq:Hybrid1-Hybrid2}, \(\left| \Pr[\mathsf{Hyb}_1=1] - \Pr[\mathsf{Hyb}_2=1] \right| \leq \left( Q_H^2 + t_\mathsf{uf}Q_\mathsf{Sign}Q_H \right)2^{-2\kappa}\).
	By \eqref{Eq:Hybrid2-Hybrid3}, \(\left| \Pr[\mathsf{Hyb}_2=1] - \Pr[\mathsf{Hyb}_3=1] \right| \leq (Q_H+1)Q_\mathsf{Sign} \cdot 2^{-d+1}\).
	By \eqref{Eq:Hybrid3-Hybrid4}, \(\Pr[\mathsf{Hyb}_3=1] = \Pr[\mathsf{Hyb}_4=1]\).
	By \eqref{Eq:Hybrid4-Hybrid5-via-HCLeakAK}, \(\left| \Pr[\mathsf{Hyb}_4=1] - \Pr[\mathsf{Hyb}_5=1] \right| \leq \mathsf{Adv}_{\mathcal{B}_\mathrm{AK}}^{ \mathsf{HC\text{-}Leak\text{-}AK} }(\kappa)\).
	Finally, by \Cref{Lemma:Hybrid5-to-STMSIS}, \(\Pr[\mathsf{Hyb}_5=1] \leq \mathsf{Adv}_{\mathcal{B}_\mathsf{STMSIS}}^\mathsf{STMSIS}(\kappa)\).
	Combining these bounds by the triangle inequality and collecting the two terms proportional to \(2^{-d+1}\) gives \eqref{Eq:LaTSEC-TS-UF-Modular-Bound}.
	The running-time claims follow because each reduction invokes \(\mathcal{A}\) once and performs only polynomially many additional random-oracle simulations, table operations, Gaussian samples, and algebraic computations.
\end{proof}

\begin{proof}[Proof of \Cref{Corollary:LaTSEC-TS-UF-MLWE-MSIS}]
	By \eqref{Eq:Hybrid4-Hybrid5} and the smoothing assumption \(\sigma_\mathrm{AK} \geq \sqrt{2}\eta_\epsilon \left(\mathbb{Z}^{kD} \right)\), there exists a PPT adversary \(\mathcal{B}_\mathsf{MLWE}\) against \(\mathsf{MLWE}_{p,\ell-p,q,d,\sigma_\mathrm{AK}}\) such that \(\mathsf{Adv}_{\mathcal{B}_\mathrm{AK}}^{\mathsf{HC\text{-}Leak\text{-}AK} }(\kappa) \leq k \cdot \mathsf{Adv}_{\mathcal{B}_\mathsf{MLWE}}^{\mathsf{MLWE}_{ p,\ell-p,q,d,\sigma_\mathrm{AK}}}(\kappa) + 2\epsilon\).
	By \Cref{Lemma:Hardness-STMSIS}, there exists a PPT adversary \(\mathcal{B}_\mathsf{MSIS}\) against \(\mathsf{MSIS}_{p,\ell-p+1,q,d,2B_\mathsf{STMSIS}}\) such that \(\mathsf{Adv}_{\mathcal{B}_\mathsf{STMSIS}}^\mathsf{STMSIS}(\kappa) \leq \sqrt{ Q_H \cdot \mathsf{Adv}_{\mathcal{B}_\mathsf{MSIS}}^{\mathsf{MSIS}_{p,\ell-p+1,q,d,2B_\mathsf{STMSIS}}}(\kappa) } + \frac{Q_H}{|\mathcal{C}|}\).
	Substituting these two inequalities into \eqref{Eq:LaTSEC-TS-UF-Modular-Bound} gives \eqref{Eq:LaTSEC-TS-UF-MLWE-MSIS}.

	If \(\mathsf{MLWE}_{p,\ell-p,q,d,\sigma_\mathrm{AK}}\) and \(\mathsf{MSIS}_{p,\ell-p+1,q,d,2B_\mathsf{STMSIS}}\) are hard, then the corresponding computational advantages are negligible.
	Moreover, by assumption, \(Q_H, Q_\mathsf{Sign}, n, k, t_\mathsf{uf}\) are polynomially bounded in \(\kappa\), and the parameter choices ensure that \(2^{-d+1}\), \(2^{-2\kappa}\), and \(Q_H/|\mathcal{C}|\) are negligible.
	Thus, the right-hand side of \eqref{Eq:LaTSEC-TS-UF-MLWE-MSIS} consists of negligible computational and oracle-programming terms, together with the explicit statistical loss \(2\epsilon\).
	If \(2\epsilon\) is also negligible, then \(\mathsf{Adv}_\mathcal{A}^{ \mathsf{TS\text{-}UF} }(\kappa) \leq \mathsf{negl}(\kappa)\).
	The running-time claims follow by composing the reductions from \(\mathsf{HC\text{-}Leak\text{-}AK}\) to repeated-\(\bm{A}\) MLWE, from repeated-\(\bm{A}\) MLWE to ordinary MLWE, and from \(\mathsf{SelfTargetMSIS}\) to \(\mathsf{MSIS}\), together with the polynomial-time hybrid simulations.
	In particular, the factor \(2\) in \(\mathsf{Time}(\mathcal{B}_\mathsf{MSIS})\) is inherited from the reduction in \Cref{Lemma:Hardness-STMSIS}.
\end{proof}

\section{Full Empirical Profiles}\label{Appendix:LaTSEC-Detailed-Empirical-Profiles}

We provide the complete empirical profiles supporting the concrete-instantiation methodology in \Cref{Section:EmpiricalProfiles-InstantiationMethod}.
For each selected \(\mathsf{SEC}\) generator matrix, we evaluate the additional quantities required by the hidden-coordinate and abort-key analyses.
For every \(U\subseteq[n]\), let \(\mathbf{\hat{G}}_U\) and \(\mathbf{G}_U\) denote the corresponding row restrictions.
We compute \(\beta_{k_p} := \max_{\substack{ U\subseteq[n]\\ 1\leq|U|\leq k_p }} \|\mathbf{\hat{G}}_U\|_2\), \(\delta_{k_p} := \max_{\substack{ U\subseteq[n]\\ 1\leq|U|\leq k_p }} \|\mathbf{G}_U\|_2\), whose maxima are attained at \(|U|=k_p\) because a row restriction cannot increase the spectral norm, and \(\gamma_{k_p} := \min_{\substack{ U\subseteq[n]\\ 1\leq |U|\leq k_p }} \sigma_{\min}(\mathbf{M}_U)\), where \(\mathbf{M}_U\) is the deterministic unimodular submatrix returned by \cite[Algorithm~3]{arzaki2026SEC}.
For the concrete parameter selection, we use the explicit value \(\beta_\mathbf{Q} := 1 + \frac{ \sqrt{1+\delta_{k_p}^2} }{ \gamma_{k_p} }\), which is the right-hand side of the uniform transformation bound in \eqref{Eq:Uniform-Hidden-Coordinate-Transformation-Bound}.
All extrema are evaluated exhaustively over their indicated subset domains.
The resulting values for the complete empirical range are reported in \Cref{Table:LaTSEC-Empirical-Matrix-Parameters}.

\begin{table}[!t]
	\caption{Exhaustively evaluated matrix parameters of the \(\mathsf{SEC}\) instances of \cite[Table~4]{arzaki2026SEC}}
	\label{Table:LaTSEC-Empirical-Matrix-Parameters}
	\centering
	\footnotesize
	\begin{tabular}{@{}cccccccc@{}}
		\toprule
		No. & \(n\) & \(k=m+1\) & \(k_p\) & \(\beta_{k_p}\) & \(\delta_{k_p}\) & \(\gamma_{k_p}\) & \(\beta_\mathbf{Q}\) \\
		\midrule
		1 & \(10\) & \(6\) & \(2\) & \(1.732\) & \(2.236\) & \(0.618\) & \(4.963\) \\
		2 & \(10\) & \(6\) & \(2\) & \(2.236\) & \(2.646\) & \(0.618\) & \(5.576\) \\
		3 & \(15\) & \(8\) & \(3\) & \(2.646\) & \(3.059\) & \(0.445\) & \(8.232\) \\
		4 & \(20\) & \(10\) & \(3\) & \(2.000\) & \(2.646\) & \(0.445\) & \(7.355\) \\
		5 & \(20\) & \(10\) & \(3\) & \(2.646\) & \(3.162\) & \(0.445\) & \(8.452\) \\
		6 & \(25\) & \(13\) & \(3\) & \(2.000\) & \(2.646\) & \(0.445\) & \(7.355\) \\
		7 & \(25\) & \(13\) & \(3\) & \(2.646\) & \(3.059\) & \(0.445\) & \(8.232\) \\
		8 & \(30\) & \(15\) & \(3\) & \(2.000\) & \(2.646\) & \(0.445\) & \(7.355\) \\
		9 & \(30\) & \(15\) & \(3\) & \(2.646\) & \(3.059\) & \(0.445\) & \(8.232\) \\
		10 & \(30\) & \(20\) & \(4\) & \(2.236\) & \(3.000\) & \(0.347\) & \(10.105\) \\
		11 & \(30\) & \(20\) & \(4\) & \(2.665\) & \(3.317\) & \(0.295\) & \(12.744\) \\
		12 & \(35\) & \(18\) & \(4\) & \(2.665\) & \(3.245\) & \(0.295\) & \(12.512\) \\
		13 & \(40\) & \(20\) & \(5\) & \(2.940\) & \(3.646\) & \(0.193\) & \(20.635\) \\
		14 & \(45\) & \(23\) & \(5\) & \(2.868\) & \(3.635\) & \(0.193\) & \(20.582\) \\
		15 & \(50\) & \(25\) & \(5\) & \(2.940\) & \(3.580\) & \(0.193\) & \(20.305\) \\
		16 & \(55\) & \(37\) & \(6\) & \(2.893\) & \(3.788\) & \(0.152\) & \(26.759\) \\
		17 & \(60\) & \(40\) & \(6\) & \(2.893\) & \(3.788\) & \(0.152\) & \(26.759\) \\
		\bottomrule
	\end{tabular}
\end{table}

Across the evaluated instances, the overall growth of \(\beta_\mathbf{Q}\) as \(k_p\) increases is driven primarily by the deterioration of the minimum singular value \(\gamma_{k_p}\), with additional variation arising from \(\delta_{k_p}\).
The concrete parameter sets use the profiles for \(n=40\) and \(n=60\), whose corresponding transformation bounds are \(20.635\) and \(26.759\), respectively.

\section{Detailed Size Estimates}\label{Appendix:LaTSEC-Detailed-Size-Estimates}

We give the calculations underlying the size estimates reported in \Cref{Table:LaTSEC-Selected-Sizes}.
We represent the uniform block of \(\bm{A}\) by a \(256\)-bit seed and do not apply bit dropping to the verification target or public abort keys.
Charging \(\lceil\lg q\rceil\) bits per uniform coefficient gives \(|\mathsf{vk}| = 256+pd\lceil\lg q\rceil\), \(|\mathsf{ak}_i| = pd\lceil\lg q\rceil\), and \(|\mathsf{ak}| = n|\mathsf{ak}_i|\).
For non-uniform components, we report ideal entropy-coded sizes.
Arithmetic coding approaches the entropy of a stationary memoryless source up to a constant coding overhead \cite[Theorem 2]{saarinen2018arithmetic}.

Let \(D:=d\ell\).
Each party stores \(\mathsf{sk}_i:=\bm{s}_i\in\mathcal{R}^\ell\).
By \cite[Eq.~(7)]{arzaki2026SEC}, each coefficient satisfies \(\mathrm{Var}(\mathbf{s}_i[j]) \leq \sigma_\mathbf{s}^2 + c_\mathrm{exp}\sigma_\mathbf{R}^2\), because \(\beta_\mathbf{g}=1\) and every row of \(\mathbf{\hat{G}}\) has \(c_\mathrm{exp}\) nonzero entries in \(\{-1,1\}\).
We therefore use the effective coefficient scale \(\sigma_{\mathsf{sk}_i} := \sqrt{ \sigma_\mathbf{s}^2 + c_\mathrm{exp}\sigma_\mathbf{R}^2 }\).
Since \(\sigma_\mathbf{s}=\sigma_\mathbf{R}\) in every selected instance, \(\sigma_{\mathsf{sk}_i} = \sigma_\mathbf{s} \sqrt{1+c_\mathrm{exp}}\).
Using a variance-matched discrete Gaussian as the ideal source model, whose per-coordinate entropy is approximated by \(\lg\sigma + \frac{1}{2}\lg(2\pi e)\) bits \cite[Section 5.1]{espitau2022shorter}, we estimate \(|\mathsf{sk}_i| \approx D \left( \lg\sigma_{\mathsf{sk}_i} + \frac{1}{2}\lg(2\pi e) \right) \approx D \left( \lg\sigma_\mathbf{s} + \frac{1}{2}\lg(1+c_\mathrm{exp}) + 2.047 \right)\) bits.

For a response vector in \(\mathbb{Z}^D\) with Euclidean bound \(B\), define \(\mathsf{Size}(B) := D \big( \lg\frac{B}{\sqrt{D}} + \frac{1}{2}\lg(2\pi e) \big)\).
This agrees with the dominant \(D\)-linear term in the logarithmic volume of the Euclidean ball, since \(\lg \mathrm{Vol} \left( \mathcal{B}_D(B) \right) = \mathsf{Size}(B) - \frac{1}{2}\lg(\pi D) + O(D^{-1})\).
Writing \(b_c := \left\lceil \omega_c + \lg \binom{d}{\omega_c} \right\rceil\) for the challenge encoding length, the bounds \(\|\bm{z}_i\|_2 \leq B_1\) and \(\|\bm{z}\|_2 \leq B_2\) give \(|\mathsf{sig}_i| \approx b_c+\mathsf{Size}(B_1)\) and \(|\mathsf{sig}| \approx b_c+\mathsf{Size}(B_2)\).

The per-party signing communication consists of a \(2\kappa\)-bit commitment, an opening \(\bm{w}_i\in\mathcal{R}_q^p\), and the encoded response \(\bm{z}_i\).
Since the common challenge is not retransmitted by each participant, \(\mathsf{Comm}_i \approx 2\kappa + pd\lceil\lg q\rceil + \mathsf{Size}(B_1)\).
The public abort key \(\mathsf{ak}_i\) is long-term storage and is not included in the per-signing communication.
All reported values use the underlying full-precision parameters and exclude implementation-specific entropy-coding, framing, finite-block, and byte-alignment overheads.

\section{Storage Comparison with sDKG and Hermine}\label{Appendix:Storage-Comparison}

We compare the per-participant long-term secret-share storage \(|\mathsf{sk}_i|\) and the participant-specific public verification material \(|\mathsf{ak}_i|\) of \(\mathsf{LaTSEC}\) (\Cref{Table:LaTSEC-Selected-Sizes}) with those of the replicated and ramp sDKG constructions \cite{del2024shortly} and of Hermine \cite{borin2026hermine}, all at \(\kappa=128\).
None of these works gives parameters for \(n\in\{40,60\}\), so their values are extrapolations of the sharing structure, not security-validated parameter sets.
Since \(\mathsf{LaTSEC}\) is ramp, no single threshold \(t\) matches it; as in \cite[Appendix J.2]{arzaki2026SEC}, we use the privacy-matched profile \(t_\mathsf{pm}:=k_p+1\), which tolerates the same number of corruptions, and the quorum-matched profile \(t_\mathsf{qm}:=k_c\), which has the same minimum signing-set size, giving \((n,t)\in\{(40,6),(40,26),(60,7),(60,46)\}\).
Neither is like-for-like: privacy matching lets the baseline sign with fewer parties, and quorum matching gives it a stronger corruption threshold.

\paragraph{\textbf{Secret-share storage}}

The sub-share counts and the ideal entropy-coded size of one secret sub-share are those of \cite[Appendix J.2]{arzaki2026SEC}.
Replicated sDKG stores \(\binom{n-1}{t-1}\) sub-shares of \(6.024\) KiB each, using its first \(\kappa=128\) set, with \((p,\ell)=(4,4)\) and \(\sigma_{\bm{s}}=2^{10}\).
Ramp sDKG stores \(\mu=\kappa=128\) sub-shares of \(7.529\) KiB each, using its second set, with \((p,\ell)=(5,5)\) and \(\sigma_\mathsf{share}=2^{10}\); this estimate ignores the larger dependent bucket and is optimistic.
In Hermine, participant \(i\) stores \(u_i(n,t)\) Vandermonde sub-shares, between \(L(n,t)\) and \(U(n,t)\), of at most \(512\cdot4\cdot(13+\frac12\lg7+2.047)\) bits each, using its \(\lceil\lg q\rceil=48\) family with \((p,\ell)=(4,4)\).
\Cref{Table:Storage-Comparison-sk} reports the resulting storage and its ratio to \(|\mathsf{sk}_i|\) of \(\mathsf{LaTSEC}\), which is \(13.883\) KiB for \(n=40\) and \(14.273\) KiB for \(n=60\).
For ramp sDKG, the storage is \(963.768\) KiB, that is, \(69.421\) and \(67.524\) times that of \(\mathsf{LaTSEC}\) for \(n=40\) and \(n=60\), respectively.

\begin{table*}[!t]
	\caption{Estimated per-participant long-term secret-share storage at \(\kappa=128\), compared with \(\mathsf{LaTSEC}\)}
	\label{Table:Storage-Comparison-sk}
	\centering
	\footnotesize
	\setlength{\tabcolsep}{4pt}
	\begin{tabular}{@{}ccl rr rrrr@{}}
		\toprule
		& & & \multicolumn{2}{c}{Replicated sDKG} & \multicolumn{4}{c}{Hermine (KiB)} \\
		\cmidrule(lr){4-5}\cmidrule(l){6-9}
		\(n\) & \(t\) & Profile & \(|\mathsf{sk}_i|\) & Ratio & Min. & Avg. & Max. & Min.\ ratio \\
		\midrule
		\(40\) & \(6\) & Privacy & \(3.307\) GiB & \(2.50\times10^{5}\) & \(353.692\) & \(429.365\) & \(522.312\) & \(25.477\) \\
		\(40\) & \(26\) & Quorum & \(8.665\times10^{4}\) GiB & \(6.54\times10^{9}\) & \(1{,}949.417\) & \(2{,}391.942\) & \(2{,}932.350\) & \(140.418\) \\
		\(60\) & \(7\) & Privacy & \(258.831\) GiB & \(1.90\times10^{7}\) & \(699.158\) & \(1{,}075.881\) & \(1{,}151.554\) & \(48.985\) \\
		\(60\) & \(46\) & Quorum & \(7.639\times10^{7}\) GiB & \(5.61\times10^{12}\) & \(4{,}441.709\) & \(7{,}192.278\) & \(7{,}896.371\) & \(311.197\) \\
		\bottomrule
	\end{tabular}
	\par\smallskip
	\parbox{\textwidth}{\scriptsize Ratios divide by \(|\mathsf{sk}_i|\) of \(\mathsf{LaTSEC}\) (\(13.883\) and \(14.273\) KiB for \(n=40\) and \(60\)). The sDKG values exclude indexing and framing overheads and the seed-and-tree compression proposed in \cite{del2024shortly}; the Hermine values measure the initial sharing before any refresh.}
\end{table*}

\paragraph{\textbf{Abort-key storage}}

Each sub-share in these constructions carries its own public verification value in \(\mathcal{R}_q^p\), and, as for \(\mathsf{LaTSEC}\), we charge \(pd\lceil\lg q\rceil\) bits per value without bit dropping.
In sDKG, this is the auxiliary key \(\bm{b}_\mathsf{sub}:=\begin{bmatrix}\bm{A}'\ \bm{I}_p\end{bmatrix}\begin{bmatrix}\bm{s}_\mathsf{sub}\\ \bm{e}_\mathsf{sub}\end{bmatrix}\), of \(4\cdot512\cdot45\) bits \(=11.25\) KiB for the replicated set and \(5\cdot512\cdot48\) bits \(=15\) KiB for the ramp set.
In Hermine, it is the indexed partial verification key \(\mathsf{vk}_i[\mathsf{idx}]:=2(\bm{A}'\bm{s}_\mathsf{idx}+\bm{e}_\mathsf{idx})\), which Hermine does not compress \cite[Figure 6, line 11]{borin2026hermine}, of \(4\cdot512\cdot48\) bits \(=12\) KiB.
Participant \(i\) is associated with \(\binom{n-1}{t-1}\) such values in replicated sDKG, \(\mu=128\) in ramp sDKG, and \(u_i(n,t)\) in Hermine.
The globally distinct collection has \(\binom{n}{t-1}\) values in replicated sDKG and \(\nu\mu\) in ramp sDKG; for the latter we choose the smallest bucket count \(\nu:=k_p+1\) for which no set of \(k_p\) parties covers all buckets of a parallel sharing, which is favorable to sDKG.
In Hermine, \(\mathsf{vk}_i[\mathsf{idx}]\) depends only on \(\mathsf{idx}\), so one copy per distinct index suffices; the number \(D(n,t)\) of distinct values satisfies \(D(n,1)=1\), \(D(n,n)=n\) (Hermine's \(t=n\) optimization), \(D(n,t)=0\) for \(t\leq0\) or \(t>n\), and, for \(2\leq t\leq n-1\) with \(n_\mathsf{L}:=\lfloor n/2\rfloor\) and \(n_\mathsf{R}:=\lceil n/2\rceil\),
\begin{align*}
	D(n,t)=\sum_{t_\mathsf{L}=\max\{0,t-n_\mathsf{R}\}}^{\min\{t,n_\mathsf{L}\}}\left(D(n_\mathsf{L},t_\mathsf{L})+D(n_\mathsf{R},t-t_\mathsf{L})\right).
\end{align*}
\Cref{Table:Storage-Comparison-ak} reports the results against \(|\mathsf{ak}_i|\) and \(|\mathsf{ak}|=n|\mathsf{ak}_i|\) of \(\mathsf{LaTSEC}\).
For ramp sDKG, the per-participant bundle is \(1920\) KiB, that is, \(83.934\) and \(80.000\) times \(|\mathsf{ak}_i|\) for \(n=40\) and \(n=60\), and the distinct collection is \(11{,}520\) and \(13{,}440\) KiB, that is, \(12.590\) and \(9.333\) times \(|\mathsf{ak}|\).

\begin{table*}[!t]
	\caption{Estimated participant-specific public verification material at \(\kappa=128\), compared with \(\mathsf{LaTSEC}\)}
	\label{Table:Storage-Comparison-ak}
	\centering
	\footnotesize
	\setlength{\tabcolsep}{3pt}
	\begin{tabular}{@{}ccl rrrr rrrrr@{}}
		\toprule
		& & & \multicolumn{4}{c}{Replicated sDKG} & \multicolumn{5}{c}{Hermine} \\
		\cmidrule(lr){4-7}\cmidrule(l){8-12}
		\(n\) & \(t\) & Profile & \(|\mathsf{aux}_i|\) & Ratio & Total & Ratio & Min.\ \(|\mathsf{vk}_i|\) (MiB) & Ratio & \(D(n,t)\) & Total (MiB) & Ratio \\
		\midrule
		\(40\) & \(6\) & Privacy & \(6.177\) GiB & \(2.83\times10^{5}\) & \(7.060\) GiB & \(8.09\times10^{3}\) & \(1.008\) & \(45.115\) & \(2{,}582\) & \(30.258\) & \(33.862\) \\
		\(40\) & \(26\) & Quorum & \(158.0\) TiB & \(7.42\times10^{9}\) & \(421.5\) TiB & \(4.95\times10^{8}\) & \(5.555\) & \(248.656\) & \(16{,}888\) & \(197.906\) & \(221.482\) \\
		\(60\) & \(7\) & Privacy & \(483.414\) GiB & \(2.11\times10^{7}\) & \(537.127\) GiB & \(3.91\times10^{5}\) & \(1.992\) & \(85.000\) & \(9{,}646\) & \(113.039\) & \(80.383\) \\
		\(60\) & \(46\) & Quorum & \(1.393\times10^{5}\) TiB & \(6.23\times10^{12}\) & \(5.573\times10^{5}\) TiB & \(4.16\times10^{11}\) & \(12.656\) & \(540.000\) & \(76{,}744\) & \(899.344\) & \(639.533\) \\
		\bottomrule
	\end{tabular}
	\par\smallskip
	\parbox{\textwidth}{\scriptsize Per-participant ratios divide by \(|\mathsf{ak}_i|\) of \(\mathsf{LaTSEC}\) (\(22.875\) and \(24.000\) KiB for \(n=40\) and \(60\)); total ratios divide by \(|\mathsf{ak}|\) (\(915\) and \(1440\) KiB). Replicated sDKG uses its first \(\kappa=128\) set (\(p=4\), \(\lceil\lg q\rceil=45\)); Hermine uses \(p=4\) and \(\lceil\lg q\rceil=48\). Values exclude indexing and framing overheads.}
\end{table*}

At \(\kappa=256\), the sDKG sizes scale by \(7/4\) for the replicated set and by \(14/5\) (secret shares) or \(21/8\) (auxiliary keys) for the ramp construction, and the Hermine sizes by \(2\) (secret shares, \(\ell=8\)) and \(7/4\) (partial verification keys, \(p=7\)); the corresponding \(\mathsf{LaTSEC}\) sizes grow by about \(1.7\) (\Cref{Table:LaTSEC-Selected-Sizes}).
In all cases, the multiplicity of sub-shares, rather than the size of an individual sub-share, accounts for the gap: each \(\mathsf{LaTSEC}\) participant holds exactly one share and one abort key.

\section{Extended Benchmarks}\label{Appendix:Extended-Benchmarks}

This appendix gives the measurement method behind \Cref{Section:Implementation-Running-Time}, the running times of all algorithms with their dispersion, their dependence on the size of the signing set, and the cost of encoding and decoding.

\subsection{Method}

All measurements use the Rust implementation, built with \texttt{rustc}~1.100.0 in release mode (optimization level~3) for the baseline \texttt{x86-64} target, that is, without \texttt{target-cpu=native} and without vector instructions beyond SSE2.
The implementation is single-threaded.
The runs use one core of an AMD Ryzen AI 9 HX 370 (a Zen~5 core, at most 5.16~GHz) under Linux~7.2, with the process pinned to that core, the \texttt{amd-pstate-epp} driver, the \texttt{powersave} governor, and the energy-performance preference \texttt{balance\_performance}; frequency boosting is enabled.
Each run records these settings together with its timings.

For each parameter set and signing-set size, key generation is run \(11\) times.
Then five all-honest signing sessions are run and discarded, and signer sets are drawn uniformly at random until \(100\) sessions have completed for \(|\mathsf{act}|=t_\mathsf{sig}\), or \(30\) for the larger sizes.
Every algorithm is timed with the monotonic clock around its call.
The time of \(\mathsf{ShareSign}_1\) therefore includes parsing \(\mathsf{msg}\) and evaluating \(\mathcal{P}\), and \(\mathsf{PartialVerify}\) and \(\mathsf{Combine}\) evaluate \(\mathcal{P}\) again; computing \(\bm{\lambda}_\mathsf{act}\) is the initiator's work and is reported separately.
The parties exchange values in memory, so encoding and decoding are not part of these times; they are measured separately (\Cref{Table:LaTSEC-Bench-Codec}).
The time of a signing round is the total over the signers of a session divided by \(|\mathsf{act}|\), and a stand-alone \(\mathsf{PartialVerify}\) is timed once for every signer of every session.
We report medians over the sessions (over the calls, for \(\mathsf{PartialVerify}\)). At \(|\mathsf{act}|=t_\mathsf{sig}\), the interquartile range of the times of \(\mathsf{ShareSign}_1\), \(\mathsf{ShareSign}_3\), \(\mathsf{Combine}\), and \(\mathsf{Verify}\) is at most \(11\%\) of the median, and that of \(\mathsf{PartialVerify}\), whose cost varies with the signer set, at most \(16\%\); \(\mathsf{ShareSign}_2\), which takes a few microseconds, varies more.

\subsection{Running Times}

\Cref{Table:LaTSEC-Bench-Times} gives the medians at \(|\mathsf{act}|=t_\mathsf{sig}\).
Of the \(603\) signer sets drawn for it, three were not certified and were refused before the first round; none was certified but outside the reconstruction bounds, and all \(600\) sessions signed and verified.
Computing \(\bm{\lambda}_\mathsf{act}\) with the exact integral LLL takes \(0.18\)~ms for \(26\) of \(40\) parties and \(0.57\)~ms for \(46\) of \(60\), and evaluating \(\mathcal{P}\) on a supplied vector takes between \(1\) and \(2.4\)~\(\mu\)s; neither depends on \(\kappa\).
A signer spends \(7.6\)~ms in total in the three rounds at \(n=60\) and \(\kappa=128\), and \(14.7\)~ms at \(\kappa=256\).

\begin{table}[!t]
	\caption{Median running times of \(\mathsf{LaTSEC}\) in ms on one core, at \(|\mathsf{act}|=t_\mathsf{sig}\)}
	\label{Table:LaTSEC-Bench-Times}
	\centering
	\footnotesize
	\setlength{\tabcolsep}{2.5pt}
	\begin{tabular}{@{}ccrrrrrrrr@{}}
		\toprule
		\(n\) & \(\kappa\) & \(\mathsf{Keygen}\) & \(\mathsf{SS}_1\) & \(\mathsf{SS}_2\) & \(\mathsf{SS}_3\) & \(\mathsf{PV}\) & \(\mathsf{Comb}\) & \(\mathsf{Vfy}\) & \(\bm{\lambda}_\mathsf{act}\) \\
		\midrule
		\(40\) & \(128\) & \(63.3\) & \(4.0\) & \({<}0.01\) & \(2.3\) & \(1.0\) & \(18.1\) & \(0.76\) & \(0.18\) \\
		\(40\) & \(192\) & \(81.6\) & \(4.3\) & \({<}0.01\) & \(3.2\) & \(1.5\) & \(28.6\) & \(1.2\) & \(0.18\) \\
		\(40\) & \(256\) & \(145\) & \(5.5\) & \({<}0.01\) & \(4.1\) & \(2.2\) & \(43.5\) & \(1.7\) & \(0.18\) \\
		\midrule
		\(60\) & \(128\) & \(109\) & \(3.3\) & \({<}0.01\) & \(4.3\) & \(1.3\) & \(32.6\) & \(0.76\) & \(0.57\) \\
		\(60\) & \(192\) & \(167\) & \(4.3\) & \({<}0.01\) & \(6.0\) & \(2.0\) & \(52.4\) & \(1.2\) & \(0.56\) \\
		\(60\) & \(256\) & \(220\) & \(7.1\) & \({<}0.01\) & \(7.6\) & \(2.8\) & \(76.2\) & \(1.8\) & \(0.56\) \\
		\bottomrule
	\end{tabular}
	\par\smallskip
	\parbox{\columnwidth}{\scriptsize \(|\mathsf{act}|=26\) for \(n=40\) and \(46\) for \(n=60\). Notation as in \Cref{Table:LaTSEC-Timings}; \(\bm{\lambda}_\mathsf{act}\) is the initiator's computation of the reconstruction vector. \(\mathsf{Keygen}\) covers all \(n\) parties and is the median of \(11\) runs.}
\end{table}

\subsection{Dependence on the Signing Set}

\Cref{Table:LaTSEC-Bench-Scaling} varies \(|\mathsf{act}|\) from \(t_\mathsf{sig}\) to \(n\) for \(\kappa=128\) and \(256\); the values for \(\kappa=192\) lie in between and are in the artifact.
The times of \(\mathsf{ShareSign}_1\) and \(\mathsf{Verify}\) remain approximately constant over the measured sizes.
\(\mathsf{ShareSign}_3\) grows with it, because a signer checks the commitment of every active signer, hashing \(|\mathsf{act}|\) openings of \(pd\lceil\lg q\rceil\) bits each.
\(\mathsf{Combine}\) runs one partial verification per signer and grows accordingly, from \(32.6\) to \(42.2\)~ms at \(n=60\) and \(\kappa=128\).
Larger signing sets admit lighter and shorter reconstruction vectors: the average weight of \(\bm{\lambda}_\mathsf{act}\) falls from \(10.3\) to \(6\) for \(n=40\) and from about \(23\) to \(18\) for \(n=60\), and \(\|\bm{\lambda}_\mathsf{act}\|_2\) from \(3.3\) to \(2.4\) and from \(5.6\) to \(4.2\).
The signature, whose width follows \(\|\bm{\lambda}_\mathsf{act}\|_2\), becomes slightly shorter, by up to \(0.8\%\).
With all \(n\) parties there is a single signer set, so its vector is computed once; it has weight \(6\) for \(n=40\) and \(18\) for \(n=60\).

\begin{table}[!t]
	\caption{Median running times in ms and signature sizes for signing sets of \(t_\mathsf{sig}\) to \(n\) parties}
	\label{Table:LaTSEC-Bench-Scaling}
	\centering
	\footnotesize
	\setlength{\tabcolsep}{2.2pt}
	\begin{tabular}{@{}cccrrrrrrr@{}}
		\toprule
		\(n\) & \(\kappa\) & \(|\mathsf{act}|\) & \(wt_H\) & \(\|\cdot\|_2\) & \(\bm{\lambda}_\mathsf{act}\) & \(\mathsf{SS}_3\) & \(\mathsf{PV}\) & \(\mathsf{Comb}\) & \(|\mathsf{sig}|\) \\
		\midrule
		\(40\) & \(128\) & \(26\) & \(10.3\) & \(3.3\) & \(0.18\) & \(2.3\) & \(1.00\) & \(18.1\) & \(36.68\) \\
		\(40\) & \(128\) & \(33\) & \(7.9\) & \(2.8\) & \(0.40\) & \(2.8\) & \(0.95\) & \(22.9\) & \(36.52\) \\
		\(40\) & \(128\) & \(40\) & \(6.0\) & \(2.4\) & -- & \(3.3\) & \(0.91\) & \(27.6\) & \(36.40\) \\
		\(40\) & \(256\) & \(26\) & \(10.3\) & \(3.2\) & \(0.18\) & \(4.1\) & \(2.2\) & \(43.5\) & \(64.15\) \\
		\(40\) & \(256\) & \(33\) & \(7.6\) & \(2.7\) & \(0.39\) & \(4.7\) & \(2.0\) & \(53.3\) & \(63.84\) \\
		\(40\) & \(256\) & \(40\) & \(6.0\) & \(2.4\) & -- & \(5.6\) & \(2.0\) & \(65.1\) & \(63.66\) \\
		\midrule
		\(60\) & \(128\) & \(46\) & \(23.1\) & \(5.6\) & \(0.57\) & \(4.3\) & \(1.3\) & \(32.6\) & \(37.84\) \\
		\(60\) & \(128\) & \(53\) & \(19.2\) & \(4.4\) & \(1.00\) & \(4.6\) & \(1.2\) & \(36.4\) & \(37.58\) \\
		\(60\) & \(128\) & \(60\) & \(18.0\) & \(4.2\) & -- & \(5.2\) & \(1.2\) & \(42.2\) & \(37.54\) \\
		\(60\) & \(256\) & \(46\) & \(23.2\) & \(5.7\) & \(0.56\) & \(7.6\) & \(2.8\) & \(76.2\) & \(66.15\) \\
		\(60\) & \(256\) & \(53\) & \(18.4\) & \(4.3\) & \(1.1\) & \(8.7\) & \(2.6\) & \(90.1\) & \(65.68\) \\
		\(60\) & \(256\) & \(60\) & \(18.0\) & \(4.2\) & -- & \(9.0\) & \(2.5\) & \(98.2\) & \(65.74\) \\
		\bottomrule
	\end{tabular}
	\par\smallskip
	\parbox{\columnwidth}{\scriptsize \(wt_H\) and \(\|\cdot\|_2\) are the mean weight and Euclidean norm of \(\bm{\lambda}_\mathsf{act}\); \(|\mathsf{sig}|\) is the mean encoded signature size in KiB. \(100\) sessions for \(|\mathsf{act}|=t_\mathsf{sig}\) and \(30\) otherwise.}
\end{table}

\subsection{Encoding and Decoding}

\Cref{Table:LaTSEC-Bench-Codec} gives the median cost, at \(|\mathsf{act}|=t_\mathsf{sig}\), of the encodings of \Cref{Section:Implementation-Instantiation} for the three objects that are transmitted in a session.
At \(|\mathsf{act}|=t_\mathsf{sig}\), the median time to encode or decode a partial response is at most \(0.2\)~ms, below \(5\%\) of the time of \(\mathsf{ShareSign}_3\), which computes it, and that of an opening is at most \(0.07\)~ms, below \(2\%\) of the time of \(\mathsf{ShareSign}_1\), which computes it.
A signature takes longer to encode than to decode because the encoder searches for the Golomb--Rice parameter that gives the shortest code, while the decoder only confirms it.
The decoders reject an input before reading it if it is longer than any encoding of a response within the verification bound, so their work is bounded by the parameter set: a signature has at most \(77{,}034\) bytes for the largest set, \(1.5\%\) above the estimate of \Cref{Table:LaTSEC-Selected-Sizes}.

\begin{table}[!t]
	\caption{Median encoding and decoding times in ms}
	\label{Table:LaTSEC-Bench-Codec}
	\centering
	\footnotesize
	\setlength{\tabcolsep}{4pt}
	\begin{tabular}{@{}ccrrrrrr@{}}
		\toprule
		& & \multicolumn{2}{c}{\(\bm{z}_i\)} & \multicolumn{2}{c}{\(\mathsf{sig}\)} & \multicolumn{2}{c}{\(\bm{w}_i\)} \\
		\cmidrule(lr){3-4}\cmidrule(lr){5-6}\cmidrule(l){7-8}
		\(n\) & \(\kappa\) & enc. & dec. & enc. & dec. & enc. & dec. \\
		\midrule
		\(40\) & \(128\) & \(0.11\) & \(0.11\) & \(0.47\) & \(0.15\) & \(0.04\) & \(0.04\) \\
		\(40\) & \(192\) & \(0.15\) & \(0.15\) & \(0.66\) & \(0.20\) & \(0.05\) & \(0.05\) \\
		\(40\) & \(256\) & \(0.17\) & \(0.17\) & \(0.84\) & \(0.25\) & \(0.06\) & \(0.06\) \\
		\midrule
		\(60\) & \(128\) & \(0.11\) & \(0.11\) & \(0.49\) & \(0.16\) & \(0.04\) & \(0.03\) \\
		\(60\) & \(192\) & \(0.14\) & \(0.14\) & \(0.68\) & \(0.21\) & \(0.05\) & \(0.05\) \\
		\(60\) & \(256\) & \(0.20\) & \(0.20\) & \(0.86\) & \(0.25\) & \(0.07\) & \(0.06\) \\
		\bottomrule
	\end{tabular}
\end{table}

\end{document}